\documentstyle[emulateapj,danonecolfloat]{article}
\lefthead{Adelberger et al.}

\righthead{Galaxies and Intergalactic Matter at $z\sim 3$}

\submitted{Received 2002 January; Accepted 2002 October}

\begin{document}
\newcommand{\lya}{Lyman~$\alpha$}
\newcommand{\lyb}{Lyman~$\beta$}
\newcommand{\degpoint}{\mbox{$^\circ\mskip-7.0mu.\,$}}
\newcommand{\minpoint}{\mbox{$'\mskip-4.7mu.\mskip0.8mu$}}
\newcommand{\secpoint}{\mbox{$''\mskip-7.6mu.\,$}}
\newcommand{\sqdeg}{\mbox{${\rm deg}^2$}}
\newcommand{\squig}{\sim\!\!}
\newcommand{\subsun}{\mbox{$_{\twelvesy\odot}$}}
\newcommand{\et}{{\it et al.}~}
\newcommand{\Rs}{{\cal R}}

\newcommand{\papertwop}{Adelberger et al. 2003}
\newcommand{\paperthreep}{Adelberger 2003}
\newcommand{\papertwo}{Adelberger et al. (2003)}
\newcommand{\paperthree}{Adelberger (2003)}
 
\def\ltsima{$\; \buildrel < \over \sim \;$}
\def\simlt{\lower.5ex\hbox{\ltsima}}
\def\gtsima{$\; \buildrel > \over \sim \;$}
\def\simgt{\lower.5ex\hbox{\gtsima}}
\def\propsima{$\; \buildrel \propto \over \sim \;$}
\def\simprop{\lower.5ex\hbox{\propsima}}
\def\arcs{$''~$}
\def\arcm{$'~$}
 
\twocolumn[
\title{GALAXIES AND INTERGALACTIC MATTER AT REDSHIFT $Z\sim 3$: OVERVIEW\altaffilmark{1}}

\author{\sc Kurt L. Adelberger\altaffilmark{2}}
\affil{Center for Astrophysics, 60 Garden Street, Cambridge, MA 02138}

\author{\sc Charles C. Steidel\altaffilmark{3} and Alice E. Shapley}
\affil{Palomar Observatory, Caltech 105--24, Pasadena, CA 91125}

\author{\sc Max Pettini}
\affil{Institute of Astronomy, Madingley Road, Cambridge CB3 0HA, UK}

\altaffiltext{1}{Based in part on observations obtained at the W.M. Keck
Observatory, which is operated jointly by the California Institute of
Technology, the University of California, and NASA, and was
made possible by a gift from the W.M. Keck Foundation.}
\altaffiltext{2}{Harvard Society Junior Fellow}
\altaffiltext{3}{Packard Fellow}

\begin{abstract}
We present the first results from a survey of the
relative spatial distributions of galaxies, intergalactic
neutral hydrogen, and intergalactic metals at high redshift.
We obtained high-resolution spectra of 8 bright QSOs
at $3.1<z<4.1$ and spectroscopic redshifts for 431 Lyman-break galaxies (LBGs)
at slightly lower redshifts.  Comparing the locations of galaxies
to the absorption lines in the QSO spectra shows that
the intergalactic medium contains less
neutral hydrogen than the global average within
$r \simlt 0.5h^{-1}$ comoving Mpc of LBGs and more
than average at slightly larger distances 
$1 \simlt r \simlt 5 h^{-1}$ comoving Mpc.
The intergalactic medium within
the largest overdensities at $z \sim 3$, which will presumably
evolve into the intracluster medium by $z \sim 0$, is
rich in neutral hydrogen and CIV.
The lack of HI absorption at small distances
from LBGs appears unlikely to be produced solely by the Lyman continuum
radiation they emit; 
it may show that the galaxies'
supernovae-driven winds 
maintain their measured outflow
velocities of $\sim 600 {\rm km\,s}^{-1}$ for 
a few hundred million years and drive away
nearby intergalactic gas. 
We present correlation functions of galaxies with Lyman-$\alpha$
forest flux decrements, with CIV systems, and with other galaxies.
We describe the association of galaxies with damped
Lyman-$\alpha$ systems and with intergalactic HeII opacity.
A strong observed correlation of galaxies with intergalactic
metals supports the idea that Lyman-break
galaxies' winds have enriched their surroundings.
\end{abstract}
]

\section{INTRODUCTION}
\label{sec:intro}
The large-scale distribution of matter in the universe
is well understood only at the earliest times, a few hundred
thousand years after the big bang, when primordial photons
strained against the opacity of matter and drove
small acoustic waves throughout space.
But though this era was appealingly simple, elegantly predicted,
magnificently observed, and so on, it was nevertheless brief.
Particles slowed as the universe
cooled.  Electrons became unable to outrun the Coulombic pull
of nuclei and vanished into atomic orbits.
The universe became transparent; the photons of the microwave
background hurtled
through history;
and the baryons left behind, freed from the regulating pressure of light,
began to evolve according to their own complicated rules.
Many acoustic waves had not even completed their first oscillation
when the universe ceased to be described 
by the simple physics of linear fluid dynamics.
Baryons fell towards
overdensities in the matter distribution.  They crashed together and became shock heated.
They cooled. Nuclear reactions ignited. Black holes formed.
Intense radiation filled the universe.
The final result of this 
chaos is known
only because we can survey the wreckage that
surrounds us.  A small fraction of baryons 
were crushed
into stars that now huddle together in galaxies, burning through their
measure of fuel, dying one by one.  Most of the remainder were blasted
to $T\sim 10^5$--$10^7$ K, temperatures they had not experienced since
shortly after the big bang, and were left to drift for
a sterile eternity in the vast stretches of intergalactic space.
Few goals in cosmology are more fundamental than understanding the
physical processes that transformed the large-scale distribution of baryons
in this way.

Many basic observations suggest that supernovae played a major role.
The disruption of star formation by supernova explosions is
the favored explanation for why so few baryons are found in stars today
(e.g., White \& Rees 1978, 
Springel \& Hernquist 2002).  
Numerical simulations cannot easily reproduce the large disk
galaxies that we observe around us 
(e.g., Weil, Eke, \& Efstathiou 1998)
or the high temperature of intergalactic gas at redshift $z\sim 3$
(e.g., Cen \& Bryan 2001)
unless they include
substantial heat input from supernovae.
The shape of galaxy clusters'
X-ray luminosity/temperature relationship 
differs from naive self-similar expectations 
in a way that suggests that supernovae may have
imparted $\sim 1$ keV of energy to each
of the young universe's nucleons  
(Kaiser 1991; Ponman, Cannon, \& Navarro 1999).
It is difficult to explain why the soft X-ray background
is so faint and so dominated by AGN without asserting that
supernovae blew apart dense clumps of baryons that would
otherwise have produced copious free-free emission
(e.g., Pen 1999).  
The scarcity of faint galaxies in the local universe relative
to naive expectations from cold dark matter models
is often attributed to the destruction
of low-mass galaxies by numerous concurrent supernovae
(e.g., Cole et al. 1994).

These examples are only a few among many.  We are unable to account
for much of what we observe around us without invoking the
indistinct notion of strong supernova ``feedback,'' and our
understanding of the evolving universe will remain seriously
incomplete until we comprehend quantitatively how
this feedback works.
Here is what we know:
a large fraction of the gas in starburst galaxies
at low and high redshift appears to be flowing outwards
rapidly enough to escape the galaxies' gravitational pull
(e.g., Heckman et al. 2000; Pettini et al. 2001, 2002);
galaxies that experienced intense bursts of star formation 
in the past now contain little interstellar gas
(e.g., Mayall 1958; Roberts 1972);
and metals produced
by stars can be found far from known galaxies 
(e.g.,
de Young 1978; Cowie et al. 1995; Mushotzky \& Loewenstein 1997; 
Ellison et al. 2000).
Most attempts to understand how supernovae affect the universe
are founded on the picture that these observations inspire:
the numerous supernova explosions in a young
galaxy create an enormous blast wave that rips through the galaxy
and lays waste to its surroundings.  
But working through the details of this picture
remains challenging even after 30 years of theoretical studies
(e.g., Mathews \& Baker 1971;
Larson 1974; Ozernoi \& Chernomordik 1978; Ostriker \& Cowie 1981;
Dekel \& Silk 1986;
Ikeuchi \& Ostriker 1986; Voit 1996; Nath \& Trentham 1997;
Mac Low \& Ferrara 1999; Aguirre et al. 2001;
Cen \& Bryan 2001; Madau, Ferrara, \& Rees 2001;
Scannapieco \& Broadhurst 2001; Croft et al. 2002). 
It is still unclear, for example,
which sorts of galaxies were responsible for seeding the
intergalactic medium with metals, or what effect blast waves
have on galaxy formation and evolution, or even whether
realistic blast waves would be physically capable of fulfilling
the large role that they are assigned in the standard lore.

These blast waves, or ``superwinds,'' are not
easy to study theoretically.
Supernovae themselves are not well understood, and
treating the propagation of their numerous overlapping
shock waves into a galaxy's inhomogeneous
surroundings is difficult enough on its own; but the central problem
is that we have little idea of the characteristic energy scale
to associate with the winds that supernovae drive.
The energy released by a single supernova, $\sim 10^{51}$ erg,
is no mystery,
but the number of supernovae in a young
galaxy is not easy to estimate theoretically or observationally,
and it is unclear in any case how large a fraction of the energy released
by supernovae is imparted to nascent winds.  
Much of it may be harmlessly radiated 
away by the dense gas it heats.  
Physical arguments and numerical simulations are at present
incapable of estimating a priori the energy of a galaxy's
superwind to within even an order of magnitude.  Yet the energy of
the winds is largely what determines how large an impact they
have on the evolving baryonic universe.
Current theoretical approaches usually amount to little more
than treating the winds' energy
as a free parameter that can be adjusted until supernovae have
the desired impact on the rest of the universe---but
the circular logic is unsettling to skeptics who wonder
if supernovae may be merely a convenient scapegoat for
the failures of popular cosmogonic models.  
The role of supernovae is probably the biggest remaining
gap in our attempt to account for the evolution of
the baryonic universe since the time of recombination, and
this is unlikely to change until observations
can provide direct constraints on the strength of
supernovae-driven winds.

Inspired in part by these considerations, we began in the spring
of 1999 a systematic survey of starburst galaxies
and the intergalactic material near them. 
Our strategy was to apply QSO absorption-line and
faint-galaxy techniques to the same volumes of
space.  Comparing the locations 
of galaxies and intergalactic absorbing
gas would let us map the relative
spatial distributions 
of diffuse and collapsed baryons throughout large
volumes of the universe.  We felt that these maps
would test our understanding of the evolving baryonic
universe in a number of ways.  Because one of the most
robust ways to measure a blast wave's energy is to see
how far out of its galaxy's potential it has managed to
climb, searching for disturbances to the intergalactic
gas near galaxies seemed a particularly promising
way to measure at last the strength of supernovae-driven winds.

We chose to conduct the survey at redshift $z \sim 3$.
Our reasons were primarily practical.
First, experience had shown us that galaxies at $z \sim 3$ were
easy to find in deep images and easy to study spectroscopically.
Second, intergalactic gas at $z\sim 3$ produces a wealth of strong
absorption lines that 
are easily detected from the ground in optical high-resolution spectra of background QSOs.
At higher redshifts the galaxy spectroscopy becomes prohibitively
difficult (e.g., Steidel et al. 1999) and background QSOs become increasingly rare;
at lower redshifts the strongest intergalactic absorption lines
can only be detected through difficult space-based spectroscopy.
Similar surveys can be constructed at very low redshifts, $z\sim 0$
(e.g., Norman et al. 1996), but it is difficult to cover representative
comoving volumes, and in any case simple arguments suggest that
the strongest supernova-driven winds likely existed at high redshift,
when galaxies were less massive and star-formation rates were
far higher (e.g., Adelberger \& Steidel 2000, \S 4).

This paper, the first in a series, describes the survey
and its early results. 
It compares
the relative spatial distributions of Lyman-break galaxies (LBGs), intergalactic metals,
and intergalactic neutral hydrogen
and emphasizes the
evidence that star formation in the LBGs directly influences the
properties of the nearby IGM. Paper II in the series (\papertwop)
sharpens these conclusions, focusing on the unexpectedly strong
larger-scale correlations between galaxies and HI in the IGM
and the implications for the physics of the IGM. Paper III (\paperthreep)
examines in more theoretical detail the possibility that superwinds from star-forming
galaxies are responsible for the observed correlations between the galaxies and the
IGM.  We tacitly assume throughout the series that supernovae would be the
primary source of any explosive energy release in young galaxies, but
AGN are an equally plausible candidate.  The source of the energy
makes little difference to our conclusions.

\S2 presents the observational
strategy and some aspects of the data reduction.
\S3 describes the characteristics of intergalactic gas
within $\sim 5h^{-1}$ comoving Mpc of galaxies.  This
is farther than the galaxies' winds are likely to propagate,
but the discussion sets the stage for \S4 where we describe
the state of the intergalactic medium nearest the galaxies.
It appears to be rarefied and metal-enriched, observations
that may suggest the galaxies' winds have propagated out
to comoving radii $r\sim 0.5h^{-1}$ Mpc.  In \S5
we show the two-dimensional correlation functions of
galaxies with galaxies, with intergalactic HI, and with 
intergalactic CIV.  These data show the connection of
galaxies and intergalactic gas on spatial scales between
those of \S3 and \S4.  \S5 contains
a brief and largely empirical summary.

\section{DATA}
\label{sec:data}
\subsection{Observations}
Our strategy was to use the Lyman-break technique to
locate star-forming galaxies at $z\simeq 3.0\pm 0.25$
in fields surrounding background QSOs whose spectra
were suitable for measuring Lyman-$\alpha$ absorbing gas
at $z\sim 3$.  Two criteria were used to select fields.
First, we wanted the QSO to lie at $3.3\simlt z\simlt 3.6$
and to be bright enough for high-resolution spectroscopy.
The lower redshift limit was chosen to maximize the number of
Lyman-break galaxies at redshifts where HI absorption
was probed by absorption in the QSO spectra; the upper limit
was chosen to minimize the impact of Lyman-continuum
and (especially) Lyman-$\beta$ absorption at higher redshifts
upon the $z\sim 3$ Lyman-$\alpha$ forest in each QSO
spectrum.  Second, we wanted the fields to have as little $100\mu$m
cirrus flux from the Galaxy as possible, to minimize the attenuation
of Lyman-break galaxies' light from dust in our own galaxy.

We were able to obtain data in five fields satisfying these
criteria (Table~\ref{tab:fields}); one field contains two QSOs.  Our sample
was augmented to six fields by including data from the ``SSA22'' region
where previous Lyman-break observations (Steidel \et 1998)
had discovered a QSO at $z=3.352$.

\begin{deluxetable}{lrrcccccrr
}\tablewidth{0pc}
\scriptsize
\tablecaption{QSO/LBG Fields}
\tablehead{
	\colhead{QSO} &
	\colhead{$\alpha_{2000}$\tablenotemark{a}} &
	\colhead{$\delta_{2000}$\tablenotemark{a}} &
	\colhead{$z_{\rm QSO}$\tablenotemark{a}} &
	\colhead{$G_{\rm QSO}$\tablenotemark{b}} &
	\colhead{$R_{\rm spec}$\tablenotemark{c}} &
	\colhead{$t_{\rm spec}$\tablenotemark{d}} &
	\colhead{$\Delta\Omega$\tablenotemark{e}} &
	\colhead{$N_1$\tablenotemark{f}} &
	\colhead{$N_2$\tablenotemark{g}}
}
\startdata
Primary: & & & & & & & & & \\
Q0256-0000  & 02$^{\rm h}$59$^{\rm m}$05.6 & +00$^o11'22''$ & 3.364 & 18.2 & 44000 & 41800 & 8.5\arcmin$\times$ 8.5\arcmin  & 45 & 38 \\
Q0302-0019   & 03$^{\rm h}$04$^{\rm m}$49.9 & -00$^o08'13''$ & 3.281 & 17.8 & 44000 & 21000 & 6.5\arcmin$\times$ 6.9\arcmin\tablenotemark{h} & 47 & 23 \\
Q0933+2845  & 09$^{\rm h}$33$^{\rm m}$37.3 & +28$^o45'32''$ & 3.428 & 17.5 & 33000 & 28800 & 8.9\arcmin$\times$ 9.3\arcmin\tablenotemark{h} & 65 & 32 \\
Q1422+2309\tablenotemark{i}  & 14$^{\rm h}$24$^{\rm m}$38.1 & +22$^o56'01''$ & 3.620 & 16.5 & 44000 & 52800 & 7.3\arcmin$\times$ 15.6\arcmin & 111 & 62 \\
Q1422+2309b & 14$^{\rm h}$24$^{\rm m}$40.6 & +22$^o55'43''$ & 3.629 & 23.4 & 6000 & 43600 &                         &    &    \\
SSA22D13    & 22$^{\rm h}$17$^{\rm m}$22.3 & +00$^o16'41''$ & 3.352 & 21.6 & 6000 & 24600 & 8.7\arcmin$\times$ 17.4\arcmin & 79 & 54 \\
Q2233+1341   & 22$^{\rm h}$36$^{\rm m}$27.2 & +13$^o57'13''$ & 3.210 & 20.0 & 44000 & 68400 & 9.2\arcmin$\times$ 9.3\arcmin  & 47 & 29 \\
Supplemental: & & & & & & & & & \\
Q0000-2620   & 00$^{\rm h}$03$^{\rm m}$22.9 & -26$^o03'17''$ & 4.098 & 19.4 & 48000 & 12500 & 3.7\arcmin$\times$ 5.1\arcmin & 18 & 1 \\
Q0201+1120   & 02$^{\rm h}$03$^{\rm m}$46.7 & +11$^o34'45''$ & 3.610 & 20.1 & 44000 & 40500 & 8.7\arcmin$\times$ 8.7\arcmin & 19 & 11 \\
\enddata
\tablenotetext{a}{QSO coordinates}
\tablenotetext{b}{QSO AB magnitude}
\tablenotetext{c}{Spectral resolution}
\tablenotetext{d}{Spectral integration time (sec)}
\tablenotetext{e}{Angular size of surrounding region with known $z\sim 3$ galaxies}
\tablenotetext{f}{Number of Lyman-break galaxies with spectroscopic redshifts}
\tablenotetext{g}{Number of Lyman-break galaxies with $z_{{\rm Ly}\beta} < z < z_{\rm QSO}-0.05$ and $2.6<z<3.4$}
\tablenotetext{h}{Region with spectroscopic follow-up; images cover a larger area}
\tablenotetext{i}{Sum of lensed components A and C (see, e.g., Rauch, Sargent, \& Barlow 1999)}
\label{tab:fields}
\end{deluxetable}

High resolution ($R\sim 40000$; see table~\ref{tab:fields}) 
spectra of Q0256-0000, Q0302-0019, Q0933+2845, Q1422+2309\footnote{The beautiful spectrum
of this QSO was taken and generously shared by W. Sargent}, and Q2233+1341
were obtained with the HIRES echelle spectrograph (Vogt 1994) on Keck I
between 1996 and 2000. 
Two overlapping echelle grating angles were chosen
to provide complete wavelength coverage through the $z\sim 3$ Lyman-$\alpha$
forest.  SSA22D13 was observed with the Echellette Spectrograph and Imager (ESI; Sheinis et al. 2000) 
on Keck II in echellette
mode ($R\sim 6000$)
in June and August 2000.  An ESI spectrum was also obtained
for Q1422+2309b, a faint QSO that we discovered $\sim 40''$ from Q1422+2309.
The HIRES spectra were reduced in the usual
way with Tom Barlow's Makee package.  Their continua were fit using
an interactive program kindly provided by R. Simcoe.
The ESI spectra were reduced and continuum fit
with the ``Dukee'' suite of custom IRAF scripts and C programs
written by KLA and M. Hunt.

$U_nG{\cal R}$ images were obtained in each field, with COSMIC (Kells \et 1998)
on the Palomar 5m Hale Telescope for Q0256-0000, Q0933+2845, Q2233+1341, and two adjacent pointings in SSA22,
and with the
Prime Focus Imager on the
William Herschel Telescope for Q0302-0019 and Q1422+2309.  The Palomar images
of Q0933+2845 and WHT images of Q0302-0019 were supplemented 
with data obtained for another project
with MOSAIC on the Kitt Peak 4m Mayall telescope.
The images were reduced
and photometric catalogs were constructed as described in
Steidel \et (1999).  The typical reduced image depths ($1\sigma$) were
$\sim$ 29.1, 29.2, 28.6 AB magnitudes per arcsec$^2$
in $U_n$,$G$, and ${\cal R}$, and approximately 2.2 objects
per square arcminute in each field were found to
satisfy the Lyman-break selection criteria
\begin{equation}
U_n-G\geq G-{\cal R}+1.0,\quad\quad G-{\cal R}\leq 1.2,\quad\quad {\cal R}\leq 25.5.
\label{eq:lbg_sel_crit}
\end{equation}
In Q1422 exceptionally
deep data let us extend our sample to ${\cal R}=26$.

\begin{figure}[htb]
\centerline{\epsfxsize=9cm\epsffile{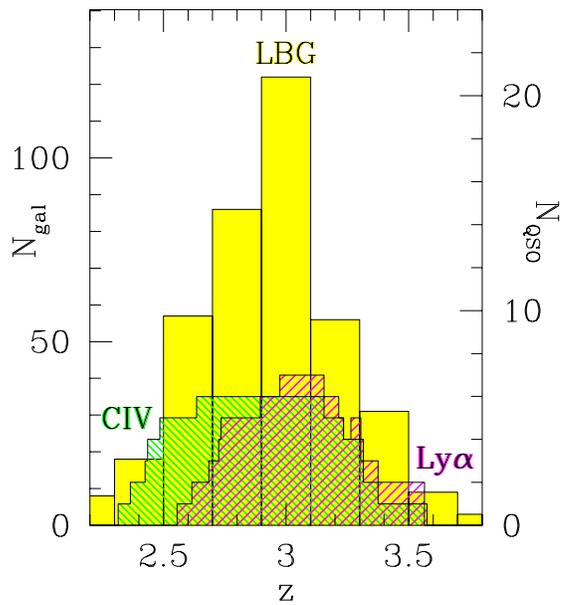}}
\figcaption[f1.eps]{
The range of redshifts probed in our survey.  The solid histogram
in the background shows the distribution of galaxy redshifts
in the primary sample;
refer to left axis.  The cross-hatched histograms in the foreground
show the number of primary-sample QSOs that could be used to
detect CIV and Lyman-$\alpha$ at different redshifts; refer to right axis.
The Lyman-$\alpha$ histogram conservatively assumes that we could not
detect Lyman-$\alpha$ in parts of the spectra that were contaminated
by Lyman-$\beta$ from gas at higher redshifts.  The CIV histogram
assumes that we could not detect CIV in parts of the spectra
that were contaminated by Lyman-$\alpha$ from gas at higher redshifts.
Neither assumption is strictly true.
\label{fig:zranges}
}
\end{figure}

Low resolution ($R \sim 600$)
spectra for a subset of these Lyman-break candidates were
obtained with LRIS (Oke \et 1995) on the Keck I and II telescopes
between 1995 and 2001.  Each field was observed through at least
4 multislit masks that accommodated $\sim 20$ objects each.
Spectra were obtained for Lyman-break candidates throughout
the regions observed in $U_nG{\cal R}$, except in the cases
of Q0302 and Q0933 where spectra were
obtained only in a $\sim 7'-9'$ region surrounding the QSO rather than
throughout the larger imaged fields.
The field sizes and number of useful redshifts obtained
in each field are listed in Table~\ref{tab:fields}.
Figure~\ref{fig:zranges} shows the redshift histogram for our
galaxy sample together with the range of redshifts where our
QSO spectra could be used to detect intergalactic CIV and HI.

These six fields make up the primary sample
used in most of our analysis below.  In a few cases
we also used data in fields surrounding
two additional QSOs, Q0000-2620 and Q0201+1120.  $U_n G {\cal R}$ images
for Q0000-2620 were obtained with EMMI on the 3.6m NTT telescope
in 1994; images for Q0201+1120 were obtained with COSMIC on the Palomar 5m Hale
telescope in 1995.  Our echelle spectrum of Q0000-2620 is a combination
of a HIRES spectrum taken by W. Sargent and
the UVES (Dekker et al. 2000) spectrum
made public after the instrument's commissioning.  P. Molaro kindly 
provided us with the reduced UVES spectrum.
A HIRES spectrum of Q0201+1120 was obtained in 1999
and reduced as described above;
this spectrum is the same as the one analyzed by
Ellison et al. (2001).
LRIS on Keck I was used in 1995 and 1996 to measure redshifts for a few objects 
in each field whose $U_n G {\cal R}$
colors satisfied equation~\ref{eq:lbg_sel_crit}.
We excluded these two fields from most of the analysis for a number of reasons.
Their images are shallow.
Few spectroscopic redshifts were measured.
The $U_nG{\cal R}$ filters used for observing Q0000-2620 differed somewhat
from those used in the rest of the observations, leading to significant
changes in the redshift distribution of spectroscopically observed
galaxies.  Moreover Q0000-2620 is at so high a redshift that most of the interesting
CIV lines are buried in the Lyman-$\alpha$ forest and most of the
Lyman-$\alpha$ forest close to our galaxies is badly affected by absorption from
Lyman series lines from gas at higher redshifts.  The spectrum of Q0201+1120 contains large gaps between
echelle orders in the red that prevented us from detecting CIV systems in a
uniform way.  
Nevertheless in two cases below---figures~\ref{fig:dgaldlya} and~\ref{fig:page1to3}---we
were able to make some use of these data.

\subsection{Redshifts}

\subsubsection{Initial wavelength calibration}
Of the many steps required to reduce our data,
one of the most crucial is the wavelength calibration,
the estimate of which wavelengths of
light were cast on each pixel by the optics of the spectrograph.
This section describes how we calibrated our LRIS data.
The HIRES and ESI calibrations were similar.

The data for a single multislit mask typically consisted
of three 30 minute exposures followed immediately by a brief
spectrum of an arc lamp at the same telescope position
as the final exposure.  To correct for gravitational flexure
of the instrument during the series of exposures, spectra from
the earlier exposures were shifted in the wavelength direction
as required to make their sky lines overlap with the sky lines
in the last exposure.  The required shifts often approached
but rarely exceeded 3\AA ($\sim 1$ pixel).  
The data were then summed and a wavelength
was assigned to each pixel based on a 5th order polynomial
fit to the wavelength vs.~pixel number relationship in the
arc-lamp exposure.  Occasionally the wavelengths assigned to 
sky lines by this procedure were systematically incorrect by $\simlt 2$\AA;
in these cases each spectrum was shifted by a fixed wavelength
increment. 
In the resulting spectra, the mean difference between the true
and estimated sky-line wavelengths was negligible and the
$1\sigma$ scatter was $\sim 0.25$\AA.
The wavelengths assigned to each pixel were finally adjusted
by small amounts using standard formulae to correct for
the motion of the earth around the sun and for
the decrease in wavelength due to the index of refraction
of air.  After applying this procedure (or minor variants) to our
data from LRIS, HIRES, and ESI, we were left with spectra
in a common vacuum-heliocentric frame that allowed them to
be intercompared.

Although this approach is standard, readers should be aware
that it is not perfect.  For example, assigning correct wavelengths to
sky lines in our low-resolution spectra
does not guarantee that the wavelength assignment
will be correct for the galaxies: if differential atmospheric
refraction or errors in slitmask alignment,
guiding, or astrometry caused the centroid of a galaxy's light
to be displaced by $0\secpoint 1$ from the center of
its $1\secpoint 4$ slit, the wavelength solution for the galaxy
would differ by $\sim 1$\AA\ from the wavelength solution for
the sky lines.  We made no attempt to correct
for this difficult problem.  It may contribute significantly
to our derived uncertainties in the galaxy redshifts, which
we will shortly discuss. 

\subsubsection{Correction for galactic winds}
\label{sec:dvsys}
\begin{figure}[htb]
\centerline{\epsfxsize=9cm\epsffile{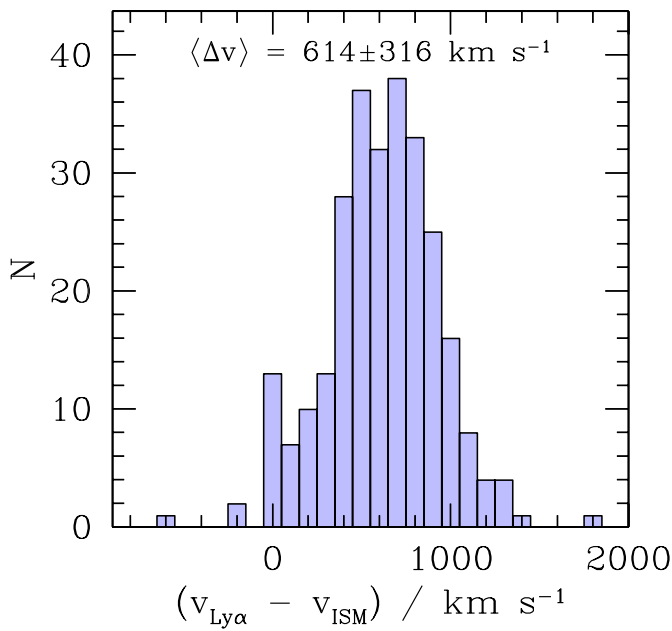}}
\figcaption[f2.eps]{
The distribution of velocity differences between Lyman-$\alpha$ emission and
interstellar absorption in the spectra of Lyman-break galaxies.
\label{fig:dz_emabs}
}
\end{figure}

\begin{figure}[htb]
\centerline{\epsfxsize=9cm\epsffile{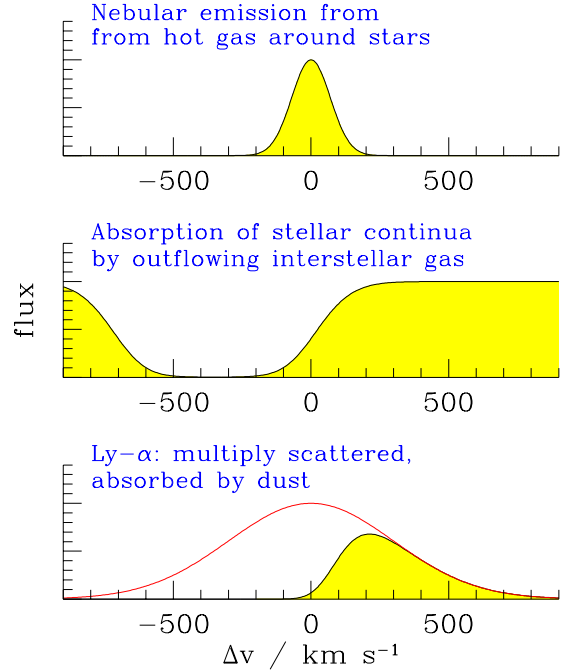}}
\figcaption[f3.eps]{Schematic view of the different redshifts in the spectrum of
a single Lyman-break galaxy.  Narrow ($\sigma_{1D}\sim 70$ km s$^{-1}$) nebular emission
lines have been detected through the near-IR spectroscopy of 27 Lyman-break
galaxies.  These emission lines presumably lie near the galaxy's systemic redshift.
Wide interstellar absorption lines are usually seen blueward of the nebular
lines.  Lyman-$\alpha$ emission of varying widths lies redward.
The relative redshifts can be understood through a simple wind model for
the gas in Lyman-break galaxies.  Absorption of stellar continua is produced
by outflowing (i.e., blueshifted) gas lying between the observer and the stars.
Resonant scattering sends Lyman-$\alpha$ photons on long random walks
through the dusty outflow.
The few photons that escape without being absorbed by dust
are those that were scattered along unusually short paths out
of the galaxy; most of these photons
scattered off the redshifted back of the outflow
and along the steep velocity gradient towards us.
\label{fig:schematicwinds}
}
\end{figure}

The $\sim 30$ km s$^{-1}$ velocity of the earth about the
sun perturbs our redshifts by $\Delta z\sim 0.0001$.
The gravitational flexure of the spectrograph and 
the slowing of light by the earth's atmosphere perturb them by
larger amounts $\Delta z\sim 0.001$.  Larger still
are the perturbations due to the chaotic motions of material
within Lyman-break galaxies themselves.
The various emission and absorption lines within the spectrum of
a single Lyman-break galaxy seldom have the same redshift.
Absorption lines from the cool interstellar gas tend to have the
lowest redshifts;
nebular emission lines from the hot gas close to stars tend to have slightly
higher redshifts; and Lyman-$\alpha$ almost always has the highest redshift
(e.g. Pettini et al. 2002; Pettini et al. 2001).
The redshift range spanned by the interstellar lines and Lyman-$\alpha$
often exceeds $\Delta z=0.01$ ($\sim 750 {\rm km s}^{-1}$; see Fig.~\ref{fig:dz_emabs}).
This reflects velocity differences within the galaxy itself---the 
observed redshift differences are qualitatively consistent with the idea that a
typical Lyman-break galaxy is expelling some fraction of its interstellar
gas in a supernovae-driven wind (figure~\ref{fig:schematicwinds};
see also Tenorio-Tagle et al. 1999)---but if the redshift differences
were erroneously assumed to be caused by the Hubble flow, the implied 
comoving distance between the reddest and bluest features
would be 
$\sim 7h^{-1}$ Mpc ($z=3$, $\Omega_M=0.3$, $\Omega_\Lambda=0.7$).
The challenge is to estimate from our data the redshift of each Lyman-break
galaxy's stars, to pick out where each Lyman-break galaxy lies
within the $\sim 7h^{-1}$ Mpc range that our observations would seem to allow.


Our approach was guided by the assumption that
the redshift of a galaxy's nebular lines (e.g., [OII]$\lambda 3727$, H$\beta$, 
[OIII]$\lambda\lambda 4959,5007$) ought to be nearly equal to the redshift of its stars,
that the gas responsible for nebular emission should always lie close to 
hot stars.  Because nebular lines are redshifted to the near-IR, where
spectroscopy is difficult, we were unable to measure their
redshifts for the vast majority of galaxies in our sample.  Instead we searched
for correlations between nebular-line redshifts and UV spectral characteristics
among the 27 Lyman-break galaxies 
that have measured nebular redshifts, then used these correlations to
estimate the systemic redshift of each galaxy in our larger sample
from its rest-frame UV spectrum.  The near-IR data for 14 of these 27 galaxies
were taken from Pettini et al. (2001); our NIRSPEC (McLean et al 1998)
spectra of the remaining galaxies will be presented elsewhere.

The following four relationships were found to hold for the 27 galaxies
with nebular redshifts.  Each resulted from a singular-value decomposition
solution of a linear least-squares equation (e.g. Press et al. 1994, \S 15.4).

Among the galaxies with detectable Lyman-$\alpha$ emission, the velocity
of Lyman-$\alpha$ relative to the nebular lines roughly satisfied
\begin{equation}
v_{{\rm Ly}\alpha} \simeq 670-8.9W_\lambda \quad {\rm km\,s}^{-1}
\label{eq:dv_vs_ew}
\end{equation}
where $W_\lambda$ is the rest-frame equivalent width of Lyman-$\alpha$ in 
\AA\footnote{Here $W_\lambda$ refers to the equivalent width of only the
part of the Lyman-$\alpha$ line that is observed in emission.  It is
therefore non-negative by definition.  Because the lines often have
P-Cygni profiles, the values $W_\lambda$ we measure from our spectra
according to this definition will generally differ from the equivalent
widths one would derive from (e.g.) a narrow-band image; we would
record a positive (emission) value for $W_\lambda$ if we saw
a weak Lyman-$\alpha$ emission line at the red edge of a deep
Lyman-$\alpha$ absorption trough, while narrow-band photometry would
reveal only that Lyman-$\alpha$ appeared dominantly in absorption.}.
The rms scatter about this mean relationship was $170 {\rm km\,s}^{-1}$.
The sense of this correlation is consistent with the idea that dust absorption
of resonantly scattered photons is responsible for driving Lyman-$\alpha$ to the red;
the reddest Lyman-$\alpha$ lines ought to be the weakest.

Averaging equation~\ref{eq:dv_vs_ew} over the distribution of $W_\lambda$ among
galaxies with detectable Lyman-$\alpha$ emission but no detectable absorption lines
leads to the mean relationship
\begin{equation}
v_{{\rm Ly}\alpha}\simeq 310\quad {\rm km\,s}^{-1}.
\label{eq:dv_eonlyavg}
\end{equation}
This equation can provide an estimate of a Lyman-$\alpha$--emitting galaxy's systemic redshift
when $W_\lambda$ has not been measured, but the redshift's precision ($\sigma_v\sim 250$ km s$^{-1}$)
suffers. 
Note that equation~\ref{eq:dv_eonlyavg}
is applicable only to data of quality similar to ours, since the adopted distribution
of $W_\lambda$ was appropriate for galaxies with detectable Lyman-$\alpha$ emission
and no detectable interstellar metal lines, and this classification is affected
by the signal-to-noise ratio of our typical spectra.

Among galaxies that had both measurable Lyman-$\alpha$ and interstellar-absorption
redshifts, the mean of the Lyman-$\alpha$ and interstellar redshifts
was larger than the nebular redshift by an amount
\begin{equation}
v_{\rm mid} \simeq -0.014\Delta v - 7.0W_\lambda + 280\quad {\rm km\,s}^{-1}
\label{eq:dv_eamidew}
\end{equation}
if $W_\lambda$ was used in the fit or
\begin{equation}
v_{\rm mid} \simeq -0.114\Delta v + 230\quad {\rm km\,s}^{-1}
\label{eq:dv_eamid}
\end{equation}
if $W_\lambda$ was ignored.  Here $\Delta v$ is the observed velocity
difference between Lyman-$\alpha$ and the absorption lines.
The rms scatter about equations~\ref{eq:dv_eamidew} and~\ref{eq:dv_eamid}
was $150$ and $210 {\rm km\,s}^{-1}$ respectively.
The physical origins of correlations~\ref{eq:dv_eamidew} and~\ref{eq:dv_vs_ew}
are similar.  

The mean velocity difference between the interstellar lines and
nebular lines for galaxies with measurable interstellar redshifts
was
\begin{equation}
v_{\rm abs} \simeq -150\quad  {\rm km\,s}^{-1}
\label{eq:dv_abs}
\end{equation}
with an rms scatter of $160 {\rm km\,s}^{-1}$.  
The qualitative picture of figure~\ref{fig:schematicwinds} suggests
that taking account of the velocity widths of the interstellar absorption lines
might improve the accuracy of equation~\ref{eq:dv_abs}, but in practice
the resolution of galaxy spectra, set by the seeing to $\sim 300$--600 km s$^{-1}$ (FWHM),
was unusably coarse.

Two anomalous galaxies were excluded when calculating each
of these fits and their scatter.  The large residuals of the
excluded galaxies could be traced back to their unusual spectra.
One had two Lyman-$\alpha$ emission lines separated by $\sim 1000$ km s$^{-1}$;
the other
had interstellar absorption at two redshifts
separated by more than $1500{\rm km\,s}^{-1}$.  As far as we could
tell nothing about the UV spectra of these objects would have allowed
us to predict their nebular redshifts with much precision.  
We chose to give up on the 2 pathological cases and optimize our
fits for the 25 normal galaxies where we had some chance of success.
Readers should be aware that for perhaps 1 galaxy among 10 the equations
above will lead to an estimated redshift that is incorrect by
several times the quoted rms.

Equations~\ref{eq:dv_vs_ew} or~\ref{eq:dv_eonlyavg}, \ref{eq:dv_eamidew} or~\ref{eq:dv_eamid}, and~\ref{eq:dv_abs}
were used to estimate the systemic redshifts of galaxies in our
sample that had, respectively, detectable Ly-$\alpha$ emission only,
both detectable Ly-$\alpha$ emission and interstellar absorption,
and only detectable interstellar absorption.  Lyman-$\alpha$ equivalent widths
were measured only for galaxies within $60''$ of a QSO sightline, because
only for these galaxies did the moderately improved precision of equations~\ref{eq:dv_vs_ew}
and~\ref{eq:dv_eamidew}
relative to~\ref{eq:dv_eonlyavg} and~\ref{eq:dv_eamid} make a significant difference.

We can check the accuracy of our redshift estimates by noting
that that the cross-correlation function of galaxies and
intergalactic material will be isotropic in an isotropic
universe.  On average the intergalactic medium ought to be
symmetric when reflected in the $z$ direction about the
true location of each Lyman-break galaxy.  The same is not
true for reflection about the measured interstellar or
Lyman-$\alpha$ redshift of the galaxies, because these redshifts
are influenced only by material that lies between the galaxies' stars
and us, and so the symmetry is broken. 

\begin{figure}[htb]
\centerline{\epsfxsize=9cm\epsffile{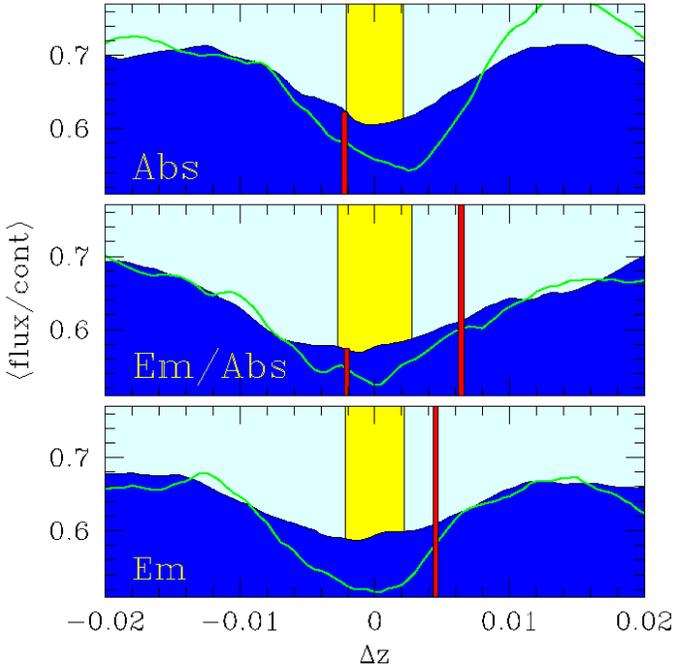}}
\figcaption[f4.eps]{The mean transmissivity of all parts of the QSO spectra
that lie within $4h^{-1}$ (solid line) or $6h^{-1}$ (shaded region)
comoving Mpc of a Lyman-break galaxy.  The height of these curves at
$\Delta z=0$ shows our best estimate from the redshift assignment
described in \S~\ref{sec:dvsys}; the height of the
curves at other values of $\Delta z$ shows the mean transmissivity
we would have measured had we applied an additional redshift
adjustment of $\Delta z$ to each galaxy.  Large systematic errors in our
redshift estimates would show up as asymmetries about $\Delta z=0$
in this plot, but in fact $\Delta z=0$ lies near the minimum of
each curve.  The sample was divided into classes
defined by the type of features we could
detect in each object's spectrum.  The mean redshift of
the detected features relative to our assigned redshift is marked
with vertical bars, short for absorption lines
and tall for emission lines.  The size of our estimated uncertainty ($\pm 1\sigma$)
in a single galaxy's redshift is indicated by the rectangular strip
surrounding $\Delta z=0$.
\label{fig:ztweak}
}
\end{figure}

Figure~\ref{fig:ztweak} shows the mean Ly-$\alpha$ transmissivity
of the intergalactic medium for all pixels of our QSO spectra
that lay within an apparent distance of $4h^{-1}$ and $6h^{-1}$ comoving
Mpc of a Lyman-break galaxy.  Also shown is the mean transmissivity we
would have measured had our estimated galaxy redshifts
been systematically higher or lower by an amount ranging from $\Delta z=-0.02$
to $\Delta z=+0.02$.  The three panels correspond (from top to bottom)
to galaxies with only detectable absorption, both detectable absorption
and emission, and only detectable emission.
In each case the curves of mean transmissivity vs. $\Delta z$
are roughly symmetrical about $\Delta z=0$, which suggests that
the systemic redshifts assigned through
equations~\ref{eq:dv_vs_ew}, \ref{eq:dv_eamid}, and~\ref{eq:dv_abs}
are accurate on average.  
The mean redshift offset of the emission and/or
absorption lines for each class is marked by the vertical bars.
(Readers may find other pertinent data in 
Figure~\ref{fig:hi_vs_rthet}, which is discussed 
in a different context below).

Further evidence of the procedure's accuracy comes from recent work
by Shapley et al (2003, in preparation), who used the above equations
to shift the spectra of $\sim 800$ LBGs into a common rest-frame.
Adding these shifted spectra together revealed weak
stellar photospheric lines with velocity $v=0\pm 30{\rm km s}^{-1}$, showing
that on average the equations successfully estimate the redshift
of the stars in Lyman-break galaxies.

\section{GALAXIES AND INTERGALACTIC GAS AT LARGE SEPARATIONS}
\label{sec:large}
We are now ready to discuss the relative spatial distributions
of galaxies and intergalactic material at $z\sim 3$.  This section
will concentrate on the properties of the intergalactic gas
that lies within a few comoving Mpc of the galaxies in our sample.
Simple arguments based on energetics suggest that a wind
driven by the supernovae in a Lyman-break galaxy
will be unable to propagate more
than $\sim 1h^{-1}$ comoving Mpc from its source (e.g., \paperthreep), and so data
averaged over several Mpc will not tell us much directly relevant
to the observed galaxies' impact on their surroundings.
But galaxies ought to form preferentially where the large-scale
density of dark matter is high (e.g., Kaiser 1984), and we can expect that
the bright galaxies we observe will lie near numerous fainter galaxies and
near any sites of previous star formation.
The cumulative effect 
of winds from the seen and unseen galaxies is something the results
of this section could in principle reveal.

\subsection{HI}
\label{sec:largehi}
\begin{figure}[htb]
\centerline{\epsfxsize=9cm\epsffile{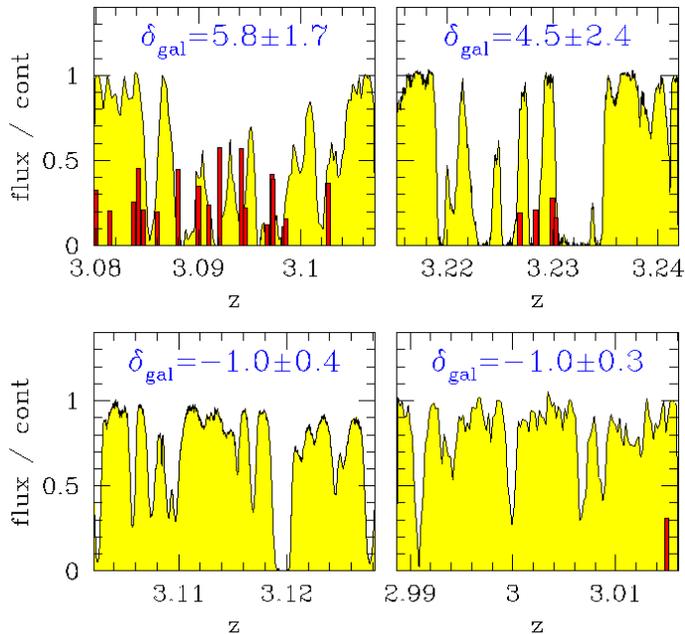}}
\figcaption[f5.eps]{  Intergalactic absorption in the $\sim 10\times 10\times 10h^{-3}$ comoving
Mpc$^3$ cells that contain the two most significant galaxy overdensities (top panel)
and underdensities (bottom panel).  The shaded curves show the Lyman-$\alpha$ forest
absorption along segments of our QSO spectra that pass through the overdensities.
Each segment is $\sim 17h^{-1}$ comoving Mpc long for $\Omega_M=0.3$, $\Omega_\Lambda=0.7$.
Galaxy redshifts are indicated with vertical
lines. The height of each line is proportional to the galaxy's distance
to the QSO sightline;
a line reaching ${\rm flux}/{\rm cont}=1$ represents a galaxy $10'$ away 
($\sim 12.9h^{-1}$ comoving Mpc for $\Omega_M=0.3$, $\Omega_\Lambda=0.7$).
The quoted overdensities $\delta_{\rm gal}$ are measured in a $10h^{-1}$ Mpc
cube at the center of the panel.
The galaxy overdensity in the upper left panel is the ``spike'' in SSA22
(e.g., Steidel et al. 1998), the largest galaxy
overdensity in the entire Lyman-break galaxy sample and our best candidate
for a proto-Abell cluster.  These data suggest that the intercluster medium
contained a relatively large amount of neutral hydrogen at early times.
\label{fig:clustvoid}
}
\end{figure}

\begin{figure}[htb]
\centerline{\epsfxsize=9cm\epsffile{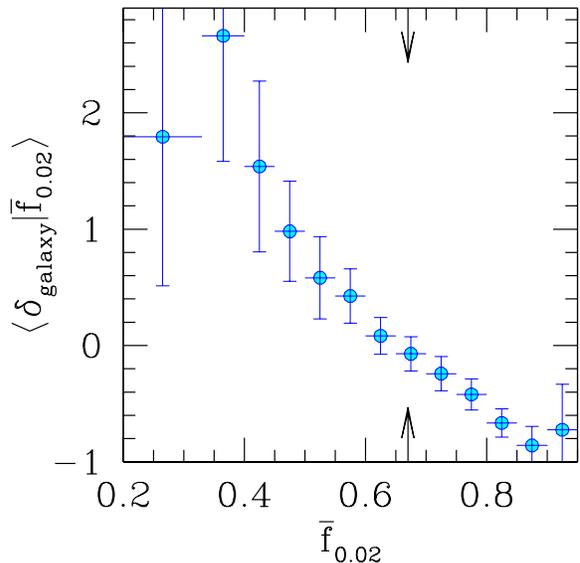}}
\figcaption[f6.eps]{The mean galaxy density in a $\sim 13h^{-1}$ comoving Mpc cubical cell as a function
of the mean Lyman-$\alpha$ forest transmissivity along the $\Delta z=0.02$ sightline skewer
through the cell.  Galaxy overdensities contain large amounts of neutral hydrogen;
galaxy voids contain less.  Arrows indicate the mean value of $\bar f_{0.02}$
among all sightline segments in our sample.
\label{fig:dgdflux}
}
\end{figure}

One of the most striking aspects of our data is the fact that
the transmissivity of the Lyman-$\alpha$ forest tends to be low
in volumes that contain a large number of Lyman-break galaxies.
Figure~\ref{fig:clustvoid} shows the Lyman-$\alpha$ forest along
skewers through the $\sim 10\times 10\times 10 h^{-3}$ comoving
Mpc$^{3}$ cubes that contain the two most significant galaxy overdensities
and the two most significant galaxy underdensities in our sample.  The
mean transmissivity is much lower through the galaxy overdensities.
The trend is observed throughout the range of galaxy densities, not merely
at the extremes.  This can be seen in figure~\ref{fig:dgdflux}, which
shows the mean galaxy overdensity in 3-dimensional cells surrounding
the QSO sightlines in our primary sample as a function of the mean flux $\bar f_{0.02}$ on
the $\Delta z=0.02$ sightline segment enclosed by the cell.  Each
cell was a right rectangular parallelepiped with depth $\Delta z=0.02$
and transverse dimensions equal to its CCD image's (table~\ref{tab:fields}); in comoving units
each was roughly a cube with side-length $13h^{-1}$ comoving Mpc
($\Omega_M=0.3$, $\Omega_\Lambda=0.7$).  The mean galaxy overdensity
associated with Lyman-$\alpha$ forest spectral segments with mean
transmissivity in the range $\bar f_{0.02}\pm\delta\bar f$ was calculated by summing
the observed number of galaxies in every cell with a mean transmissivity in that
range, then dividing by the sum of the number of galaxies that would have been
expected in those cells if Lyman-break galaxies were distributed uniformly
in space:
\begin{equation}
\langle\delta_{\rm gal}\rangle\equiv\frac{\sum_i^{\rm cells} N_i}{\sum_i^{\rm cells} \mu_{g,i}} - 1,
\label{eq:dgdf_estimator}
\end{equation}
where $N_i$ is the observed number of galaxies in the $i$th cell
and $\mu_{g,i}$ is the expected number in the absence of clustering.
$\mu_{g,i}$ was estimated by scaling our average selection function by the number of
galaxies with spectroscopic redshifts in the appropriate 
field.  See appendix~\ref{sec:apdensest} for a justification for this estimator.
In order to (slightly) reduce the shot noise, the centers of adjacent
cells were separated by $\Delta z=0.004$, so that we oversampled
our data by a factor of 5.
The QSO spectra were used over
the redshift range
$(1+z_q)\lambda_{{\rm Ly}\beta}/\lambda_{{\rm Ly}\alpha}-1<\lambda/\lambda_{{\rm Ly}\alpha}-1<z_q-0.05$,
where $z_q$ is the QSO redshift, in order to avoid contamination of the Lyman-$\alpha$
forest by material associated with the QSO or by Lyman-$\beta$ absorption from
gas at higher redshifts.
Segments of the QSO spectra with $2.91 < \lambda/\lambda_{{\rm Ly}\alpha}-1 < 2.98$
in SSA22 and with $3.214 < \lambda/\lambda_{{\rm Ly}\alpha}-1 < 3.264$ in Q0933
were removed from the analysis to avoid damped Lyman-$\alpha$ systems.


Some may be surprised that the correlation between HI and galaxy
density does not break down at large values of 
$\langle\delta_{\rm gal}\rangle$.  The largest
galaxy overdensities in our sample are
likely to evolve into rich clusters by $z=0$; this conclusion
follows from the comoving volume
between $z=2.7$ and $z=3.3$
that we have observed, $\sim 3.7\times 10^5 h^{-3}$ Mpc$^3$ for
$\Omega_M=0.3$, $\Omega_\Lambda=0.7$, which is large enough to contain
$\sim 3$ structures that are destined to evolve into clusters with X-ray
temperatures $kT>2.5$keV at $z\sim 0$.  We have no better candidates for
the protoclusters than the two overdensities shown in figure~\ref{fig:clustvoid}
or (more generally) than the large galaxy overdensities associated with
the lowest $\bar f_{0.02}$ bins in figure~\ref{fig:dgdflux}.
A number of arguments
suggest that the intracluster medium was heated at early times,
before the cluster itself had formed, and this has led some to speculate
that young clusters at $z\sim 3$ might contain far less neutral hydrogen
than the universal average (e.g. Theuns, Mo, \& Schaye 2001).  Figures~\ref{fig:clustvoid}
and~\ref{fig:dgdflux} show that
the opposite is true.  The large Lyman-$\alpha$ opacities of
(presumed) intracluster media at early times do not imply that preheating has not
happened, however, as we will argue in \paperthree:
putative shocks traveling outward
from high-redshift galaxies and heating the young intracluster medium could easily
increase the mean HI content of the volumes they affect.  
See \papertwo\ for an extended discussion of Figure~\ref{fig:dgdflux}
and its implications.

\subsection{CIV}
\label{sec:largeciv}
\begin{figure}[htb]
\centerline{\epsfxsize=9cm\epsffile{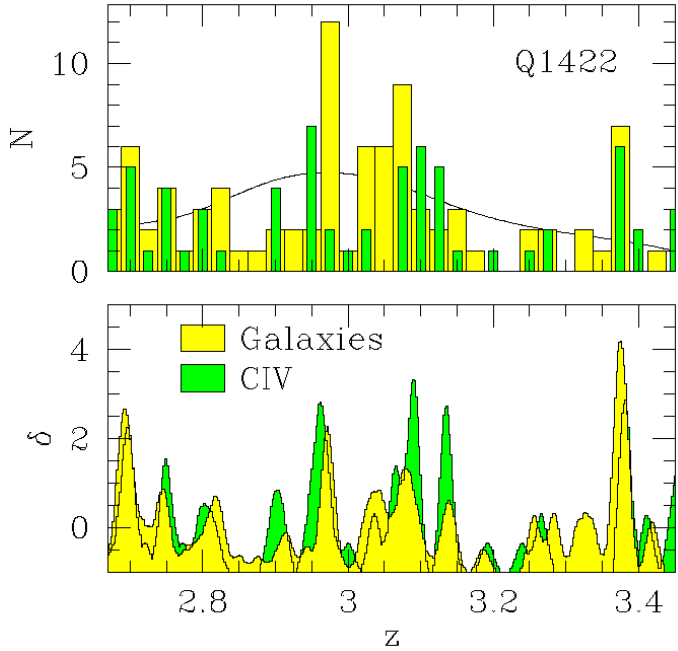}}
\figcaption[f7.eps]{The redshift distributions of galaxies and CIV absorbers in the field of Q1422+2309.
The top panel shows the number of objects observed at each redshift.  
The bottom panel shows
the implied overdensity as a function of redshift after smoothing
our raw redshifts by a Gaussian of width $\sigma_z=0.008$.
The good correspondence of features in the lower panel shows
that CIV systems are preferentially found within galaxy overdensities.
\label{fig:q1422_gal_civ}
}
\end{figure}

The notion that the intergalactic
material within galaxy overdensities may have already been preheated receives
some support from its observed 
metal content.  The association of
Lyman-break galaxies and metals can be quantified in any number of ways,
and will receive more attention below; but the qualitative point we wish
to make here is adequately illustrated by figure~\ref{fig:q1422_gal_civ}.
This figure shows the redshifts of the Lyman-break galaxies and CIV
systems in the field of Q1422+2309, the QSO with the best spectrum in our sample.
The top panel shows the number of galaxies and CIV systems in redshifts bins of width $\Delta z=0.025$.
The bottom panel shows the result of smoothing the raw redshifts by a Gaussian with
width $\sigma_z=0.008$ ($\sim 5.4h^{-1}$ comoving Mpc for $\Omega_M=0.3$, $\Omega_\Lambda=0.7$)
and then dividing by the selection function.  The selection function
used for the galaxies, shown in the top panel of the figure, is a spline fit to
a coarsely binned histogram of every Lyman-break galaxy redshift in our sample.
The selection function for CIV systems was approximated as constant with redshift.
Though obscured somewhat by shot noise, the connection between galaxy density and
CIV density is in fact surprisingly strong;  Pearson's correlation coefficient
between the two curves in the bottom panel of figure~\ref{fig:q1422_gal_civ}
is 0.61. It is interesting that the amplitudes of the CIV
and galaxy fluctuations are comparable.  The figure suggests that
a volume of the universe which is overdense in galaxies by a factor
of $1+\delta\sim 3$ (e.g.) will tend to be overdense in CIV systems
by a factor of $\sim 3$ as well.  But because Lyman-break galaxies are biased
tracers of the matter distribution (e.g., Adelberger et al. 1998), 
the baryonic overdensity of the same volume
will be significantly smaller.  This shows that there is more detectable
CIV absorption per baryon in galaxy overdensities than elsewhere.  One of
several possible interpretations is that the intergalactic metallicity
is enhanced near galaxies at $z\sim 3$.  A similar trend
might also result from the density-dependence of carbon's ionization
state.

\subsection{HeII}
\begin{figure}[htb]
\centerline{\epsfxsize=9cm\epsffile{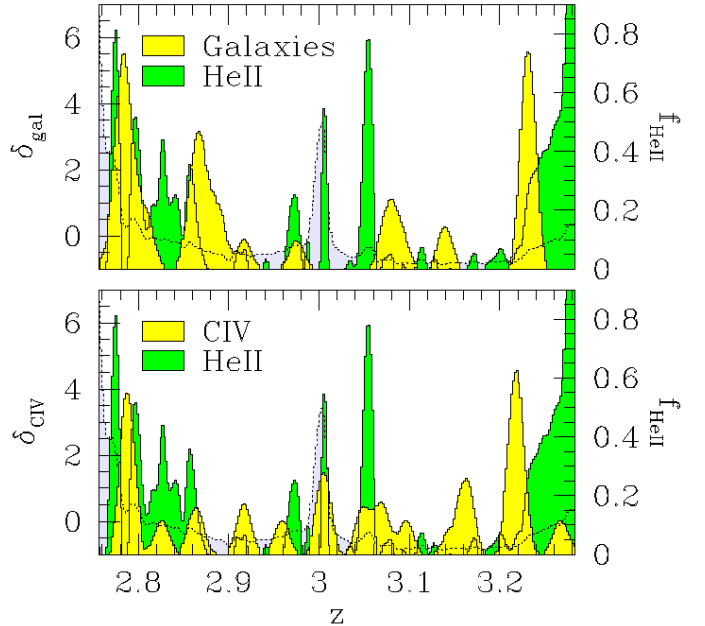}}
\figcaption[f8.eps]{The correlation between galaxies, CIV systems, 
and intergalactic HeII transmissivity in the field of Q0302-0019.  
The curves in
the foreground show the overdensity of galaxies and CIV
systems as a function of redshift; they were produced by
smoothing the list of redshifts with a Gaussian
of width $\sigma_z=0.008$.  The shaded curve
in the background shows the HeII Lyman-$\alpha$ absorption seen
in the spectrum of Q0302-0019 (Heap et al. 2000)
in units of $10^{-16}$ erg s$^{-1}$ cm$^{-2}$ \AA$^{-1}$
(refer to right axis).  The dotted line is the error spectrum.
The HeII content of the IGM is detectably reduced from its
mean value in regions where the spectrum significantly exceeds
its uncertainty.  These regions often (but not always) coincide
with galaxy and CIV overdensities.
\label{fig:q0302galcivheii}
}
\end{figure}

Heap et al. (2000) obtained a spectrum at wavelengths 
$1140\simlt\lambda_{\rm obs}\simlt 1300$\AA\ of one of the QSOs
in our sample, Q0302-0019.  This spectrum
revealed that the HeII Lyman-$\alpha$ 
($\lambda_{\rm rest}\simeq 304$\AA) optical depth 
of the intergalactic
medium in front of the QSO was large and
surprisingly variable.  The observed range of
optical depths in the spectrum, $\tau_{\rm HeII}\sim 1$
to $\tau_{\rm HeII}\simgt 5$, apparently implies that
the hardness of the ionizing background
and the ratio of HI to HeII number density must
vary significantly in the intergalactic medium at $z\sim 3$.
Figure~\ref{fig:q0302galcivheii} shows that 
the observed variations in intergalactic
HeII opacity appear to be spatially correlated with the locations of
galaxies and CIV systems.  Excluding the region
$z>3.24$, which is presumably affected by radiation from Q0302-0019
itself (e.g. Hogan et al. 1997), and the region
$2.987<z<3.016$, which is contaminated by geocoronal Lyman-$\alpha$
emission, we calculate $r_s\sim 0.21$, 0.27 for the
value of Spearman's rank correlation coefficient
between the smoothly varying galaxy, CIV overdensities shown
in figure~\ref{fig:q0302galcivheii} 
and the HeII absorption spectrum.
The galaxy and CIV overdensities were calculated as described above, near
figure~\ref{fig:q1422_gal_civ}.
Correlation coefficients of this size, $r_s\sim 0.2$--$0.3$, do not imply perfect
correspondence, and readers will easily find
examples in Figure~\ref{fig:q0302galcivheii}
where gaps in HeII opacity have few associated galaxies
or metals in our sample (e.g., at $z=2.83$ and $z=3.05$).
Nevertheless the statistical correlations visible in the figure
are moderately significant.
A correlation strength $r_s>0.21$ was found for roughly 13\% of 
randomized galaxy catalogs that we correlated with the HeII spectrum;
$r_s>0.27$ was found for roughly 6\% of the randomized CIV catalogs.
It is easy to think of reasons that the HeII opacity
of the intergalactic medium might decrease near galaxies
or CIV systems.  The most obvious is that
the reionization of HeII should happen first in the dense regions
where galaxies, metal-line absorbers, and AGN reside.
But this is unlikely to be the full explanation;
the HeII optical
depth at the mean density at $z\sim 3$ would be of order 1000 if
HeII were the dominant ionization state,  
and so the small but significant fraction of the QSO's 
light that is detectable at $\lambda\simlt 1310$\AA\
shows that HeII must already be highly reionized almost everywhere.  
A more likely explanation may be that the spatial clustering bias 
of galaxies and AGN causes the number of HeII-ionizing photons per baryon
to be largest in overdensities where galaxies and AGN tend
to be found.  
Because the mean free path of 4Ryd photons at $z\sim 3$
is likely to be $\sim 1500$ km s$^{-1}$ (Miralda-Escud\'e, Haehnelt,
\& Rees 2000), or roughly the size of independent bins in
figure~\ref{fig:q0302galcivheii}, we might reasonably expect to
see this effect in the figure.  We should also add that much of the
evidence for a correlation between HeII transmissivity and galaxy or CIV density
comes from redshifts $z\simlt 2.9$.  If the decrease in HeII opacity
at these redshifts reflects a global change in the hardness of
the ionizing background due to the growing dominance of AGN
(e.g. Songaila 1998; cf. Boksenberg, Sargent, \& Rauch 1998),
then the correspondence of high HeII transmissivity with
the observed galaxy and CIV overdensities would be only a coincidence,
and much of our evidence for a connection between galaxies and HeII
would be removed.  

\subsection{Damped Lyman-$\alpha$ systems}
\label{sec:largedla}
\begin{figure}[htb]
\centerline{\epsfxsize=9cm\epsffile{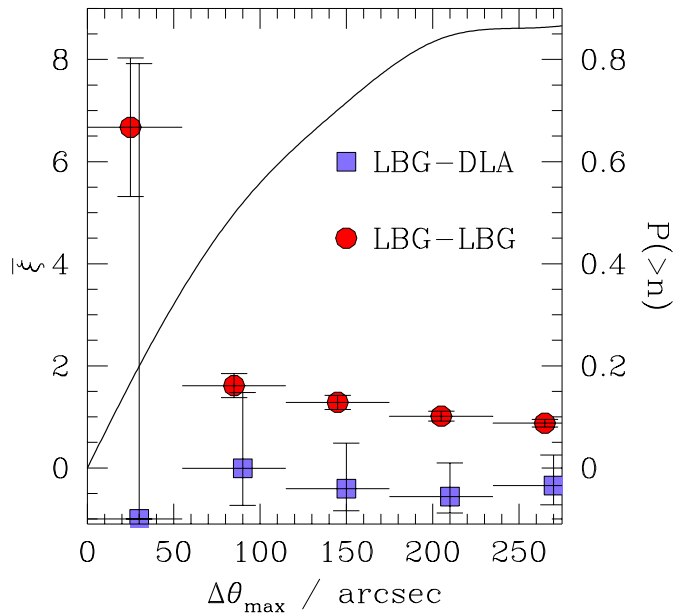}}
\figcaption[f9.eps]{The spatial association of Lyman-break galaxies and damped Lyman-$\alpha$
systems.  Square points show the mean overdensity of Lyman-break galaxies
in cylinders of constant depth $\Delta z=0.025$ ($\sim 17h^{-1}$ comoving Mpc
for $\Omega_M=0.3$, $\Omega_\Lambda=0.7$) and variable radius
$\Delta\theta$ surrounding the four damped systems in our survey.  
$250''$ corresponds to roughly $5.4h^{-1}$ comoving Mpc in the same
cosmology.
Circular points show the mean overdensity
of Lyman-break galaxies in similar cylinders centered on other Lyman-break
galaxies.  
A $5''$ offset was added to the true abscissae of the DLA points.
All error bars assume Poisson statistics.  Refer to left axis for scale.
The curved line shows how frequently randomly chosen Lyman-break galaxies
with similar redshifts to the two DLAs in our primary sample have more galaxy
neighbors than the DLAs.  Refer to right axis.  Damped systems do not appear
to cluster as strongly as Lyman-break galaxies.
\label{fig:dgaldlya}
}
\end{figure}

For completeness
we will briefly describe the association of Lyman-break galaxies
with damped Lyman-$\alpha$ systems (table~\ref{tab:dlas}).
This topic has been recently discussed by Gawiser et al. (2001).
Figure~\ref{fig:dgaldlya} compares the mean overdensity
of Lyman-break galaxies within cells centered on damped systems
to the mean overdensity of Lyman-break galaxies within cells
centered on other Lyman-break galaxies.  Each cell was a cylinder
with height $\Delta z= 0.025$ ($\sim 16.7h^{-1}$ comoving Mpc for $\Omega_M=0.3$,
$\Omega_\Lambda=0.7$) and radius equal to
value of $\Delta\theta$ shown on the figure's abscissa.  Even at the largest value
of $\Delta\theta$ ($265''$, or $r\sim 5.7h^{-1}$ Mpc comoving), 
each cylinder's diameter is significantly smaller than its height.
This helped ensure that
galaxies correlated with the central object would fall into
the surrounding cell even in the presence of substantial redshift errors.
If Lyman-break galaxies and damped systems were similar objects, the
mean density of Lyman-break galaxies around damped systems would be similar
to the mean density of Lyman-break galaxies around other
Lyman-break galaxies, but this is not the case.  The observed
number of Lyman-break galaxies close to the damped systems
in our sample is instead roughly what one would expect if the two populations
were independently distributed; we see no evidence for an excess
of Lyman-break galaxies near damped systems.  In contrast
the overdensity of Lyman-break galaxies near other Lyman-break
galaxies is large, as the filled circles in figure~\ref{fig:dgaldlya} show.
Table~\ref{tab:dlas} compares the observed number of Lyman-break galaxies
near each damped system with the number one would have expected
if damped systems and Lyman-break galaxies were the same objects.
In this case, we would have expected to find 5.96 Lyman-break
galaxies within $\Delta\theta=265''$, $\Delta z=0.0125$ of the damped systems.
Instead we found 2.  A Poisson distribution with true mean 5.96 will yield
2 or fewer counts about 6.4\% of the time, so the significance of this
result is slightly better than 90\%.  The significance can be assessed
in a more empirical way by exploiting the fact that our spectroscopic sampling
density in Q0933+2841 and SSA22D13 is very similar to the density
in the rest of the Lyman-break galaxy sample.
We selected at random from our Lyman-break galaxy catalogs 
many sets of two galaxies, with one galaxy at roughly the redshift of
the damped system in Q0933+2841, the other at roughly the redshift
of the damped system in SSA22D13.  We then counted the number of
other Lyman-break galaxies in cylindrical cells surrounding the two galaxies
in each set, and compared to the number of Lyman-break galaxies in
cells surrounding the damped systems.  The curve in
figure~\ref{fig:dgaldlya} shows the frequency $P(>n)$ with which 
a set of two random galaxies had more galaxy neighbors than the
set of two damped systems.  The lack of Lyman-break galaxies
within $\sim 5.7h^{-1}$ comoving Mpc ($\Delta\theta<265''$) of
these two damped systems lets us conclude with $\sim 90$\% confidence
again that bright Lyman-break galaxies and damped systems do not reside
in similar parts of the universe.
Taken together, the data in figure~\ref{fig:dgaldlya}
suggest that the statistical association
between damped systems and Lyman-break galaxies is weak;
the data are consistent with the idea that damped systems
tend to reside in small potential wells that are much more
uniformly distributed in space than the massive wells that presumably
host Lyman-break galaxies.  Others
have reached a similar conclusion from different starting points
(e.g., Fynbo, M\/oller, \& Warren 1999; Mo, Mao, \& White 1999;
Haehnelt, Steinmetz, \& Rauch 2000).

\begin{deluxetable}{llcccc
}\tablewidth{0pc}
\scriptsize
\tablecaption{Damped Ly-$\alpha$ systems [$+1$]}
\tablehead{
	\colhead{QSO} &
	\colhead{$z$} &
	\colhead{log($N_{HI}/{\rm cm}^{-2}$)} &
	\colhead{log($N_{CIV}/{\rm cm}^{-2}$)} &
	\colhead{$N_{\rm obs}$\tablenotemark{a}} &
	\colhead{$N_{\rm exp}$\tablenotemark{b}} 
}
\startdata
Q0000-2620	& 3.3902 & 21.3 & 14.7 & 1 & 0.67 \\
Q0201+1120	& 3.3864 & 21.3 & 13.9 & 0 & 0.45 \\
Q0933+2845	& 3.2352 & 20.3 & 12.8 & 1 & 2.23 \\
SSA22D13	& 2.9408 & 20.7 & 13.1 & 0 & 2.61 \\
SSA22D13\tablenotemark{c} & 2.7417 & 15.1 & 14.4 &  &  \\
\enddata
\tablenotetext{a}{Observed number of Lyman-break galaxies with $\Delta\theta<265''$, $\Delta z<0.0125$}
\tablenotetext{b}{Expected number of Lyman-break galaxies if the DLA-LBG cross-correlation function
were the same as the LBG-LBG correlation function}
\tablenotetext{c}{Not a damped system.  This is the gas at a distance $\Delta\theta=17''$ 
($90h^{-1}$ proper kpc), $\Delta z=0.003$ from a Lyman-break galaxy in SSA22 
(figure~\ref{fig:page1to3}).
Its HI and CIV column densities are listed for comparison.}
\label{tab:dlas}
\end{deluxetable}

\section{GALAXIES AND INTERGALACTIC GAS AT SMALL SEPARATIONS}
\label{sec:gpe}
Any influence of a galaxy's supernovae on its surroundings should
be most pronounced near to the galaxy itself.  This makes
it especially interesting to study the contents of the intergalactic
medium near Lyman-break galaxies.
Figure~\ref{fig:page1to3}
shows the distribution of HI and CIV absorption along the QSO sightline
segments that approach Lyman-break galaxies most closely.
The angular separation of the galaxy from the sightline
is marked in each panel.  At $z=3$ one arcminute corresponds to
roughly $1.3h^{-1}$ comoving Mpc for $\Omega_M=0.3$, $\Omega_\Lambda=0.7$,
and so at closest approach these sightlines probe the intergalactic
medium at $\sim 0.3$ to $1.3h^{-1}$ comoving Mpc from the galaxy.
The shaded curves show the Lyman-$\alpha$ forest transmissivity.
The symbol $\beta$ appears next to the impact parameter in each panel
if Lyman-$\beta$ absorption from gas at higher redshifts could have affected
the appearance of the Lyman-$\alpha$ forest at the galaxy's redshift.
The horizontal line marks the mean transmissivity at the galaxy's
redshift, estimated as described in the following paragraph.
Circles mark the locations of detectable CIV absorption.  The
size of each circle is related to the CIV column density; a tripling
of a circle's area corresponds to a factor of ten increase in
column density.  Due to significant gaps in our echelle spectrum
of Q0201, we did not attempt to make a catalog of the CIV systems in this field.
Our low quality spectrum of Q1422+2309b did not allow us to detect a significant
number of CIV systems; panels for this QSO show the locations of CIV
absorption in Q1422+2309 itself, a QSO which is $\sim 1'$ away.
The short and tall vertical bars mark the
observed redshifts of interstellar absorption and Lyman-$\alpha$ emission,
respectively, in each galaxy's spectrum.  Wide shaded boxes
mark the $1\sigma$ confidence
interval on each galaxy's systemic redshift, estimated from
the appropriate equation among~\ref{eq:dv_vs_ew}, \ref{eq:dv_eamidew},
and~\ref{eq:dv_abs}.  

\begin{figure*}
\centerline{\epsfxsize=18cm\epsffile{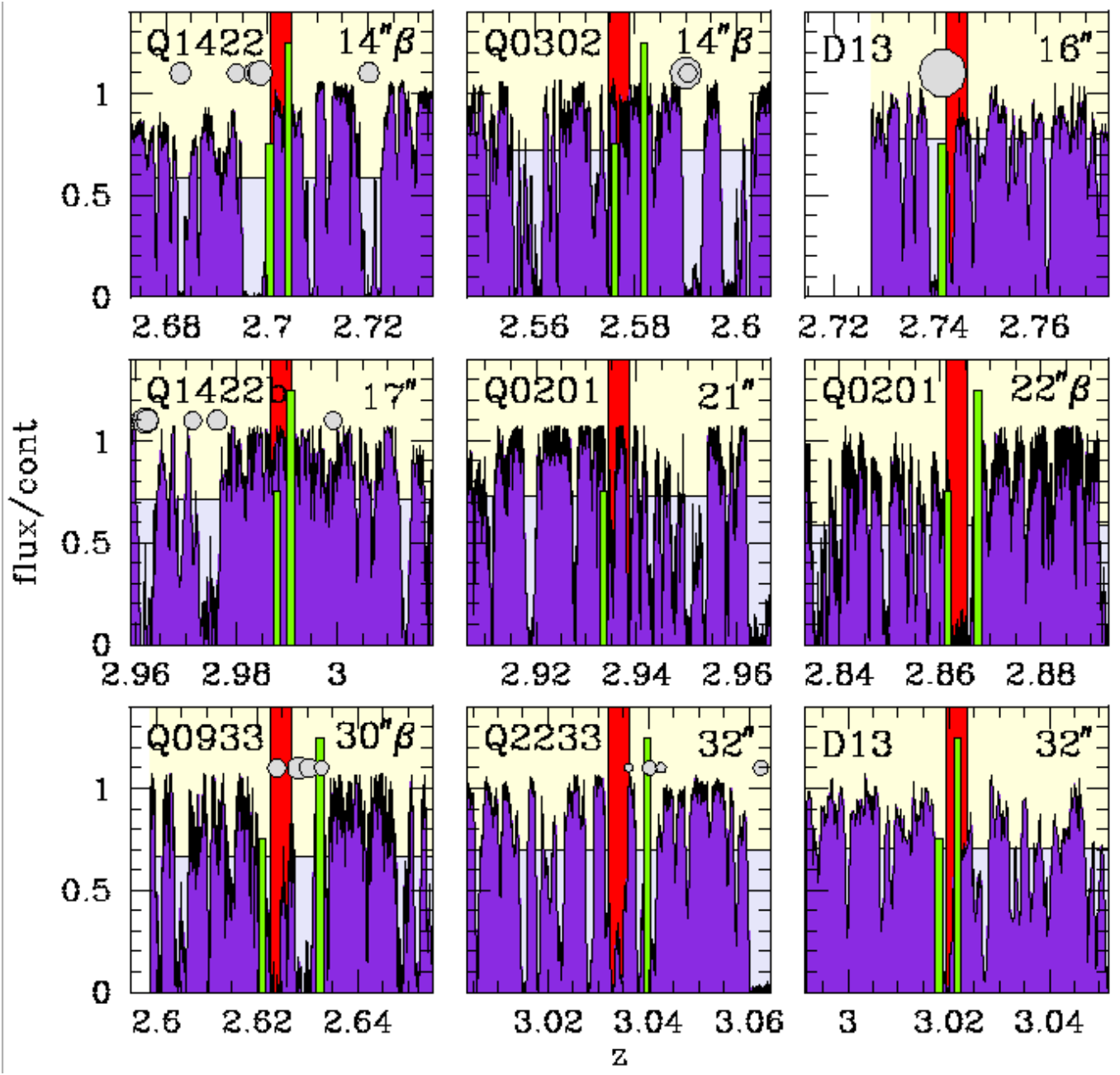}}
\figcaption[f10.eps]{The distribution of neutral hydrogen and CIV absorption along the
segments of the QSO spectra that pass closest to a Lyman-break galaxy.
The shaded curve shows the Lyman-$\alpha$ forest.  The horizontal line
marks the mean transmissivity at this redshift in the QSO's spectrum.
Circles mark the redshifts
of detectable CIV absorption.  Larger circles correspond to larger CIV column densities;
a tripling in the circle's area corresponds to a factor of 10 increase in column density.
Numerical values for the HI and CIV column densities of the absorbing
gas near to one of these galaxies can be found in the last entry
of table~\ref{tab:dlas}.
The wide vertical region shows our estimated redshift ($\pm 1\sigma$) for each galaxy.  Narrower
vertical bars mark the redshifts of Lyman-$\alpha$ (tall bar) or interstellar
absorption (short bar) in each galaxy's spectrum.  The distance from each galaxy to
the QSO sightline is indicated.  $10''$ corresponds to roughly $200h^{-1}$ comoving kpc.
A $\beta$ next to the distance indicates that the Lyman-$\alpha$ forest at this redshift
may be contaminated by Lyman-$\beta$ (or higher) absorption from gas at larger redshifts.
\label{fig:page1to3}
}
\end{figure*}
\begin{figure*}
\addtocounter{figure}{-1}
\centerline{\epsfxsize=18cm\epsffile{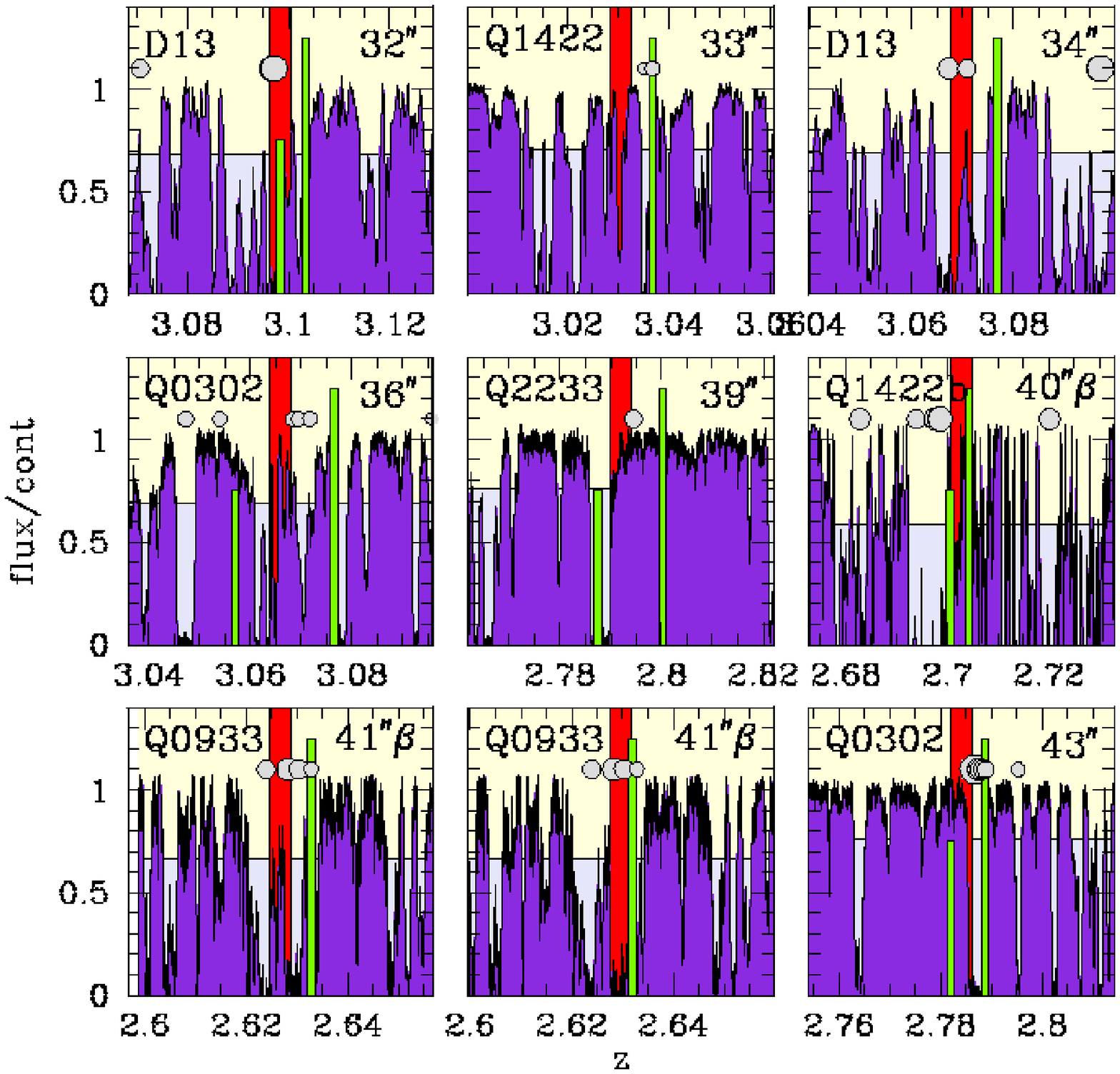}}
\figcaption[f10b.eps]{ continued.\label{fig:page1to3}}
\end{figure*}
\begin{figure*}
\addtocounter{figure}{-1}
\centerline{\epsfxsize=18cm\epsffile{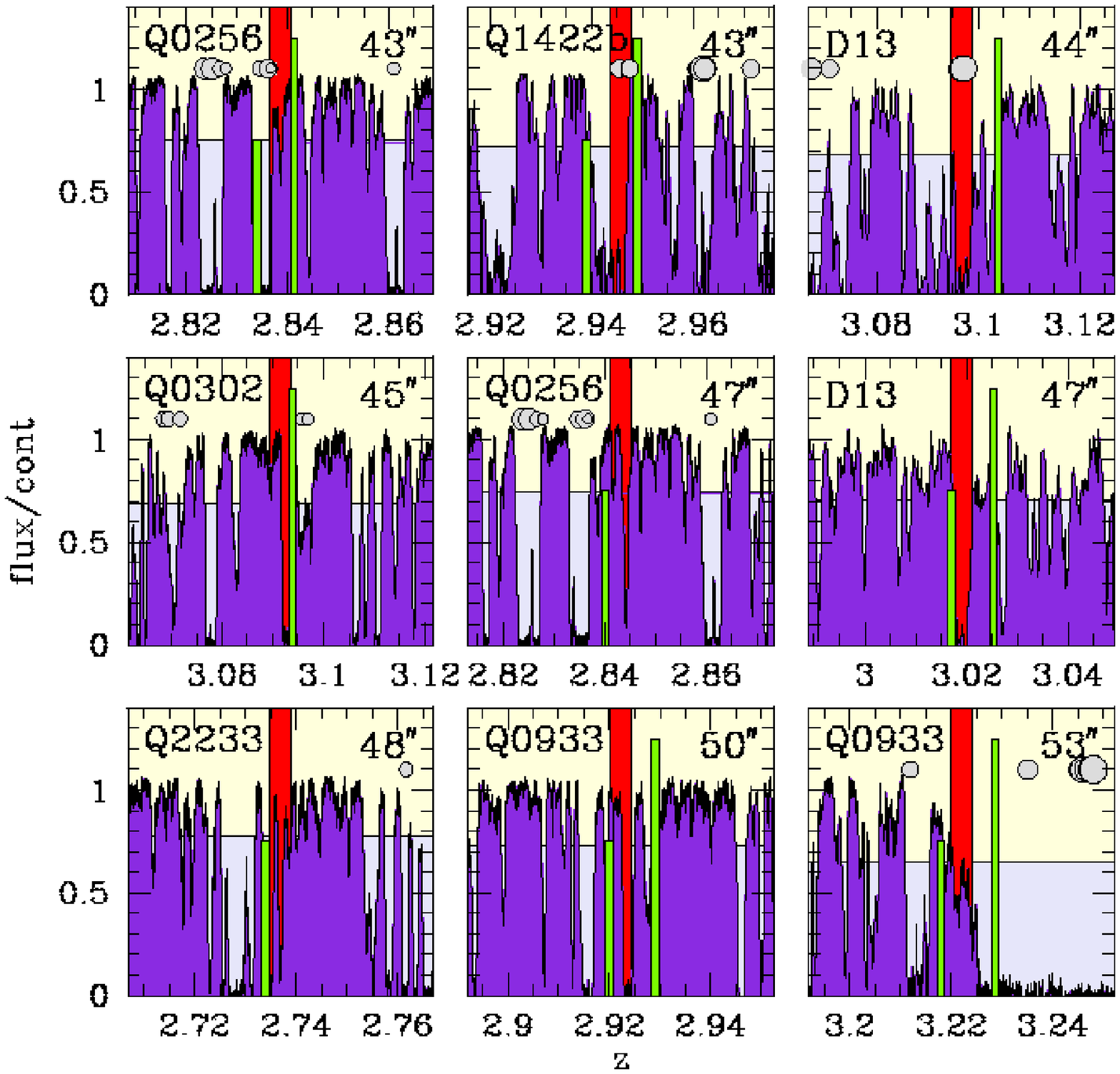}}
\figcaption[f10c.eps]{ continued.\label{fig:page1to3}}
\end{figure*}

The values of the mean intergalactic Lyman-$\alpha$ transmissivity
came from the following formulae.
When the forest was not contaminated by Lyman-$\beta$ absorption,
its mean transmissivity was taken to be
\begin{equation}
\bar f(z) = 0.676 - 0.220 (z-3),
\label{eq:fbarvsz}
\end{equation}
a fit to the relationship between mean transmissivity
$\bar f$ and redshift presented by McDonald et al. (2000).  
The mean transmissivity of the forest contaminated by Lyman-$\beta$ and higher lines
is roughly independent of redshift because the increased absorption in the blue
due to high Lyman-series lines is largely canceled out by the gradual thinning
of the forest towards lower redshifts.  The mean transmissivity in this case
was found to roughly obey the formula
\begin{equation}
\bar f = 0.633-0.40(z_{\rm QSO}-3.5)
\label{eq:fbarvszb}
\end{equation}
where $z_{\rm QSO}$ is the QSO's redshift.
These formulae are reasonable approximations only at the
redshifts $z\sim 3$ that are of interest to us; they should
not be used at other redshifts.

\subsection{A lack of HI near Lyman-break galaxies}
\label{sec:smallhi}
\begin{figure}[htb]
\centerline{\epsfxsize=9cm\epsffile{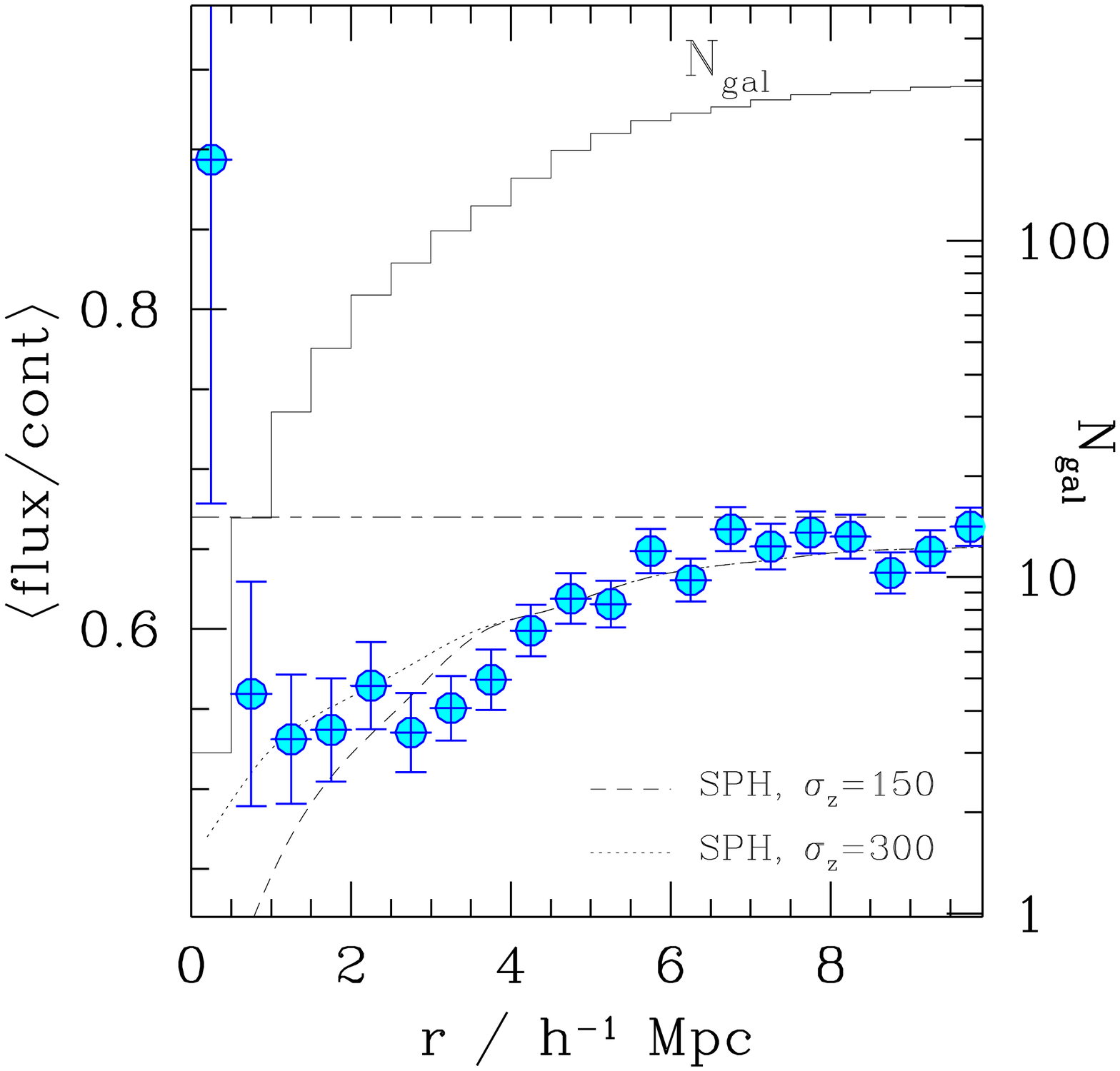}}
\figcaption[f11.eps]{Mean intergalactic Lyman-$\alpha$ transmissivity as a function of comoving distance
from Lyman-break galaxies.  Points with error bars mark our measurements; 
refer to the left axis for the scale.  If galaxies were distributed randomly
with respect to the Lyman-$\alpha$ forest, we would have measured 
$\langle {\rm flux/cont}\rangle\simeq 0.67$ at each radius (horizontal line).
Instead it appears that Lyman-break galaxies are associated with
neutral hydrogen overdensities (i.e., flux decrements) on scales extending to
several Mpc, and with HI underdensities on the smallest spatial scales.  
This behavior, especially on small scales, contrasts with predictions from numerical
smoothed-particle hydrodynamic simulations of $\Lambda$CDM universes where galaxies
have little impact on surrounding intergalactic matter.  In these simulations
the mean transmissivity drops increasingly rapidly near galaxies.  Examples
for different assumed galaxy redshift errors (as labeled, in km s$^{-1}$)
are shown with curved lines
(data from figure 9a of Croft et al. 2002, and scaled arbitrarily in $y$ to match roughly
our data on large spatial scales).
The number of galaxies in our sample within each radius
is shown by the arced staircase (refer to right axis).  
\label{fig:twopanelgpea}
}
\end{figure}

We would like to draw readers' attention to
an interesting aspect of figure~\ref{fig:page1to3}.  Although the intergalactic
medium within $\sim 10h^{-1}$ comoving Mpc of galaxy overdensities
appears to contain large amounts of neutral hydrogen (\S\ref{sec:largehi}), often
little neutral hydrogen is observed along the small segments of
the QSO sightline that pass within $\sim 0.5h^{-1}$ comoving Mpc
of a Lyman-break galaxy.  This is illustrated more clearly
in figure~\ref{fig:twopanelgpea}, which shows
the mean Lyman-$\alpha$ transmissivity of the intergalactic medium
as a function of apparent comoving distance from a Lyman-break 
galaxy\footnote{More precisely, the figure shows $(1+\bar\xi(r))\bar f$,
where $\bar f\equiv 0.67$ is approximately the mean transmissivity
at $z=3$ and $\bar\xi$ is an annular average of the two-dimensional
galaxy-flux correlation function $\xi(r_\theta, r_z)$ calculated as described in
\S~\ref{sec:corrfns} below.  The annular average was weighted
by $\ell(r_\theta,r_z)$, the total path length of our QSO spectra at a distance
$r_\theta$, $r_z$ from any galaxy in our sample. 
Scaling from the correlation function
helps reduce systematic errors from the geometry of our survey
and from the change in mean transmissivity with redshift.}.
We discarded parts
of the QSO spectra that were contaminated by Lyman-$\beta$ absorption
from gas at higher redshifts or by damped Lyman-$\alpha$ systems,
and normalized the result so that the mean transmissivity
at all radii would have been precisely 0.67 if the galaxies in our surveyed
volumes were randomly distributed and infinitely numerous.
Rough error bars were calculated by generating a large
ensemble of fake data sets, each containing the same QSO spectra
and same galaxies as our actual sample, but with each galaxy's
redshift modified by a Gaussian deviate
with $\sigma_z=0.075$ ($\sim 50h^{-1}$ comoving Mpc for $\Omega_M=0.3$,
$\Omega_\Lambda=0.7$).
The error bars in the figure 
show the $1\sigma$ scatter about the mean among these
fake data sets.  They may be underestimates of the true uncertainty,
as discussed below.

\begin{figure}[htb]
\centerline{\epsfxsize=9cm\epsffile{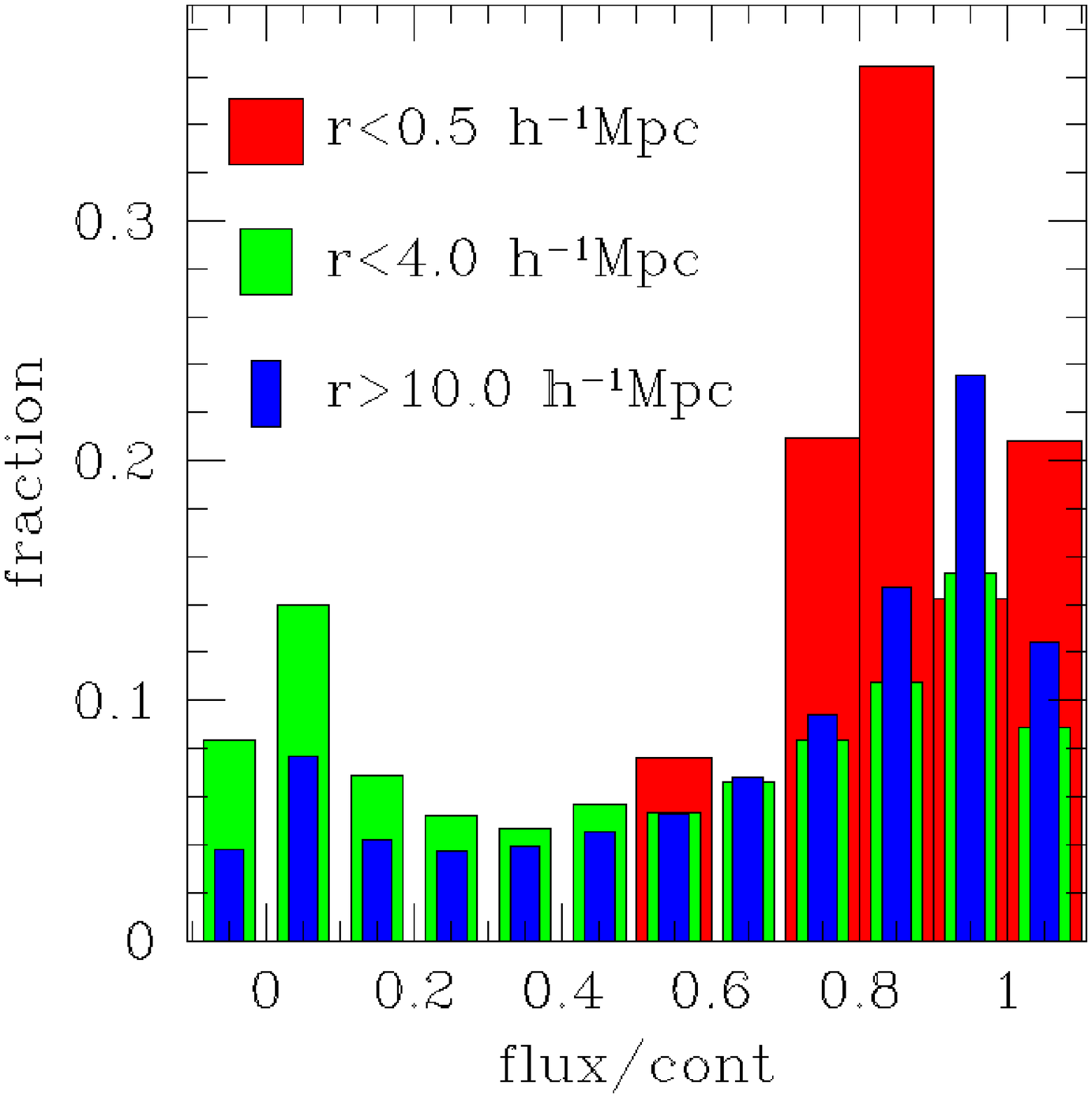}}
\figcaption[f12.eps]{
Histogram of pixel transmissivities at different distances from Lyman-break
galaxies.  The shape of figure~\ref{fig:twopanelgpea}
derives from a relative lack of
low transmissivity pixels within $r\sim 0.5h^{-1}$ comoving Mpc of 
Lyman-break galaxies 
and an excess with $1\simlt r\simlt 5h^{-1}$ comoving Mpc.
\label{fig:twopanelgpeb}
}
\end{figure}

Figure~\ref{fig:twopanelgpea} shows again the result discussed above (see
figures~\ref{fig:clustvoid} and~\ref{fig:dgdflux}):
on scales of $\sim 1$ to $\sim 5h^{-1}$ comoving Mpc Lyman-break
galaxies are associated with an excess of neutral hydrogen.
But on the smallest scales the trend appears to reverse.
Little neutral hydrogen is found within $\sim 0.5h^{-1}$ comoving Mpc
of the galaxies.  
Figure~\ref{fig:twopanelgpeb} shows that the change in mean
transmissivity with distance derives from spatial variations in the relative
proportion of lightly and heavily obscured pixels in the Lyman-$\alpha$ forest spectra:  few
pixels with transmissivities less than 0.5 are found
within $\sim 1h^{-1}$ comoving Mpc of a Lyman-break galaxy,
while a significant excess of saturated pixels with transmissivity $\sim 0$
are found within $\sim 4h^{-1}$ Mpc of Lyman-break galaxies.

The apparent lack of HI near galaxies is particularly interesting because it seems inconsistent
with currently favored numerical models of the intergalactic medium
at high redshift.  In these models the density of baryons closely traces
the density of dark matter and the $\rho^2$ 
dependence of the volumetric recombination rate
causes the density of neutral hydrogen to be highest where the density
of dark matter is high.  But galaxies also form where the density of dark matter
is high, and as a result galaxies in the simulations tend
to be surrounded by large amounts of neutral hydrogen.
In the simplest models the density of HI increases rapidly at small galactocentric radii.
This is illustrated by the curved lines in figure~\ref{fig:twopanelgpea},
which show the mean transmissivity around galaxies with baryonic mass
$M_b>10h^{-1} M_\odot$ 
in the $\Omega_M=0.3$, $\Omega_\Lambda=0.7$ smoothed-particle hydrodynamical (SPH) simulation of
Croft et al. (2002; taken from their figure 9a).  

The differences between the simulation results and our data are not
hard to discern.  Rather than a continued increase in the HI density
at smaller radii, we find that the HI density appears to level off
at radii $r\simlt 4h^{-1}$ comoving Mpc and then decrease at the
smallest radii $r\simlt 0.5h^{-1}$ Mpc.  Although these simulations
included many physical processes---gravitational
and hydrodynamical interactions, radiative cooling, and radiative heating due to a uniform ionized
background---they did not include star formation and supernova feedback
in a way that allowed galaxies to have much influence on the nearby intergalactic medium.
It is tempting to conclude from figure~\ref{fig:twopanelgpea}
that the actual impact of galaxies on their surroundings is considerably
larger than in the simulations shown.  Much of the rest of this section
will consider the idea in more detail (see also Croft et al. 2002).
But first we would like to discuss the statistical significance
of the result.

Our measurement of the mean transmissivity very close to galaxies ($\simlt 1h^{-1}$
comoving Mpc) should be viewed
with some skepticism.
Recall the large and relatively uncertain offsets that were
required to estimate galaxies' systemic redshifts from the
redshifts of the absorption and emission lines
in their optical spectra (\S~\ref{sec:dvsys}).  One need not contemplate
figure~\ref{fig:page1to3} for long to realize that even
minor changes to our estimated systemic redshifts could drastically
alter the inferred Lyman-$\alpha$ transmissivity of the intergalactic
medium close to Lyman-break galaxies.  A few judiciously applied
redshift adjustments of $\delta z\sim 0.003$ could easily erase
the inflection from figure~\ref{fig:twopanelgpea}, for example, and
$\delta z=0.003$ is hardly a large adjustment.
It is small compared to the range of redshifts seen
in most Lyman-break galaxy spectra.  It barely
exceeds our optimistic estimates of the $1\sigma$ redshift uncertainty
from \S~\ref{sec:dvsys}. 

\begin{figure}[htb]
\centerline{\epsfxsize=9cm\epsffile{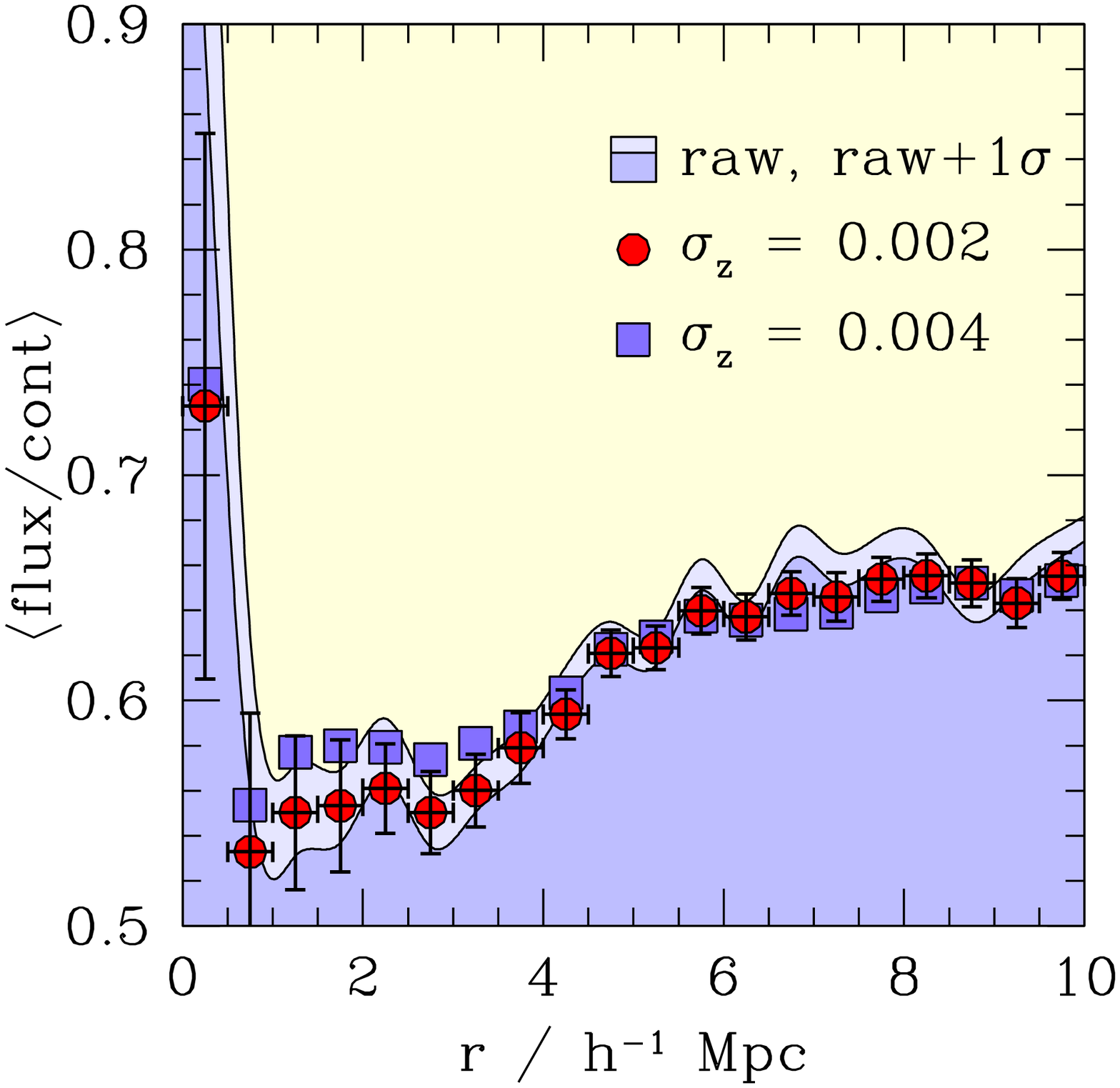}}
\figcaption[f13.eps]{The result of figure~\ref{fig:twopanelgpea} recalculated
after modifying each galaxy's redshift by an amount similar
to its uncertainty.  Numerous fake data sets were generated by
adding to each galaxy's redshift  a Gaussian deviate with the
standard deviation $\sigma_z$ shown. Points show the average
value of the transmissivity among the fake sets; error bars
show the $1\sigma$ spread.  
The error bars for $\sigma_z=0.004$,
suppressed for clarity, are similar to those for $\sigma_z=0.002$.
The smooth curves in the background
are a spline fit to the data of figure~\ref{fig:twopanelgpea}
and their $+1\sigma$ upper envelope.
If $\sigma_z$ were equal
to the uncertainty in each galaxy's redshift, then each of the fake
data sets would be roughly as compatible with our observations as the actual
data set, and any large differences between the mean transmissivities
in the fake and real sets would suggest that the mean transmissivity
in the real set may have been compromised by
an unusual combination of redshift errors.
\label{fig:fbarsigz}
}
\end{figure}

One way to assess how badly redshift errors might have compromised
our estimate of the mean intergalactic transmissivity at different
distances from Lyman-break galaxies is to change each of our redshifts
by an amount similar to its uncertainty, then recalculate the
implied mean transmissivity.  Figure~\ref{fig:fbarsigz} shows the
result.  Circles mark the average estimated transmissivity 
at each distance after adding a Gaussian
deviate with $\sigma_z=0.002$ to each of our redshifts; vertical
error bars show the $1\sigma$ range observed when we repeated the
exercise many times.  Squares show the result when the standard
deviation of the Gaussian deviate is increased to $\sigma_z=0.004$.
Since $\sigma_z=0.002$ is roughly the redshift
uncertainty that follows from the analysis of \S~\ref{sec:dvsys},
we conclude that
the true redshifts of our galaxies
could lie anywhere within our $\sim 2\sigma$ confidence intervals without
much affecting our conclusion that the Lyman-$\alpha$ opacity 
of the intergalactic medium decreases near galaxies.  

The small number of galaxies close to the QSO
sight-lines may be more worrying than our redshift uncertainties.
Only three galaxies contribute to the measurement of mean transmissivity
at $r<0.5h^{-1}$ comoving Mpc that is shown in figure~\ref{fig:twopanelgpea}.
The situation is not quite as dire as the figure
suggests, because three additional galaxies in our sample lie
within $r=0.5h^{-1}$ comoving Mpc of a QSO sight-line. See figure~\ref{fig:page1to3}.  
They were
excluded from the calculation shown in figure~\ref{fig:twopanelgpea}, along
with many other galaxies at larger impact parameters, because their low redshifts
corresponded to parts of their QSO's Lyman-$\alpha$
forest that were contaminated by Lyman-$\beta$ absorption
and correcting for Lyman-$\beta$ was a challenge that
we chose
to dodge.  
But since Lyman-$\beta$ absorption from gas
at higher redshifts can only decrease the transmissivity in the Lyman-$\alpha$ forest,
there is no reason not to use these galaxies  in a conservative search
for increases in transmissivity near galaxies.  The mean transmissivities within
$0.5h^{-1}$ comoving Mpc of the two galaxies in figure~\ref{fig:page1to3} at impact parameter
$14''$ ($\sim 0.28h^{-1}$ Mpc for $\Omega_M=0.3$, $\Omega_\Lambda$=0.7) 
are $\langle {\rm flux}/{\rm cont}\rangle = 0.92$, $0.97$; the mean transmissivity within
$0.5h^{-1}$ comoving Mpc of the galaxy at impact parameter $22''$ ($\sim 0.46h^{-1}$ Mpc)
is $\langle {\rm flux}/{\rm cont}\rangle = 0.04$.  Weighting
by the length of QSO spectrum that passes within $0.5h^{-1}$ comoving Mpc of each galaxy,
this implies a mean transmissivity of 0.78 for the three---an answer
encouragingly consistent with the result of figure~\ref{fig:twopanelgpea}
that was derived from three different galaxies.

Still, six galaxies with impact parameter $r_\theta<0.5h^{-1}$ comoving Mpc
is not a large sample.  The evidence for a decrease
in HI density within $0.5h^{-1}$ comoving Mpc of Lyman-break galaxies
has a significance of less than $2\sigma$; the difference between
our observations and the simulations in the same bin would
have a significance of about $2\sigma$ if we pessimistically assumed
that the accuracy of our redshifts was only $\sigma_z=300 {\rm km s}^{-1}$.
One might argue that even without the data at $r_\theta<0.5h^{-1}$
our measurements would not resemble the simulations perfectly, but it
may be more fruitful
to consider instead what physical processes could lead to
a lack of neutral hydrogen within $\sim 0.5h^{-1}$ comoving Mpc of
a galaxy and look for other signatures of their existence.

\subsection{Physical origin}
\subsubsection{Ionizing radiation}
Reduced Lyman-$\alpha$ absorption is observed in the intergalactic medium
surrounding high-redshift QSOs, a fact that
presumably reflects a decrease in the hydrogen
neutral fraction in regions where a quasar's radiation
overwhelms the ambient ionizing field
(see, e.g., Weymann, Carswell, \& Smith 1981; Murdoch et al. 1986; 
Bajtlik, Duncan, \& Ostriker 1988; Scott et al. 2000).
Could a similar physical cause account for the lack of HI absorption
in the vicinity of Lyman-break galaxies?  A rough argument
suggests that this is unlikely to be the case.
Consider the mean transmissivity within $0.5h^{-1}$ comoving Mpc of
an LBG, $\langle {\rm flux}/{\rm cont}\rangle \simgt 0.8$, which is
substantially higher than the mean transmissivity 
$\langle {\rm flux}/{\rm cont}\rangle \sim 0.5$ we might
expect in the absence of a proximity effect (Fig.~\ref{fig:twopanelgpea}).
In the $z=3$ output of the
SPH simulation described in White, Hernquist, \& Springel (2001),
an increase in the ionizing background by a factor of $\sim 8$
is required to increase the mean transmissivity of a random section of
the Lyman-$\alpha$ forest from 0.5 to 0.8, and so we might guess that
the ionizing flux within $0.5h^{-1}$ Mpc of a Lyman-break galaxy would have
to be $\sim 8$ times higher than its universal mean if ionizing radiation
from the galaxies were responsible for the observed proximity effect.
The Lyman-break galaxies in our sample could not easily produce
an ionizing flux so large, as the following crude calculation shows.
If our sample can be approximated as a uniform population 
of galaxies with ionizing luminosity $L$ (energy time$^{-1}$ frequency$^{-1}$)
and comoving number
density $\rho$
that produces a fraction $f_{\rm bg}$ of the total ionizing background 
$J_\nu^{\rm tot}$ (energy time$^{-1}$ frequency$^{-1}$ area$^{-1}$ 
steradian$^{-1}$) at $z\sim 3$,
then the contribution to the ionizing background from a single
galaxy at radius $r$ is
\begin{equation}
J_\nu^1 = L / (4\pi r)^2
\label{eq:jnu1}
\end{equation}
while the net contribution to the ionizing background from the population as a whole
is roughly
\begin{equation}
J_\nu^{\rm LBG} \sim L\rho (d/c) (c/4\pi) \equiv f_{\rm bg} J_\nu^{\rm tot}
\label{eq:jnuu}
\end{equation}
where $d$ is the average distance traveled by an ionizing photon
before it is absorbed.  The ionizing flux will therefore be more than $n$ times higher
than its universal average $J_\nu^{\rm tot}$ within a radius
\begin{equation}
r_n \sim [4\pi\rho (n-1)d/f_{\rm bg}]^{-1/2}
\label{eq:r_gpe}
\end{equation}
of a Lyman-break galaxy.  Assuming an $\Omega_M=0.3$, $\Omega_\Lambda=0.7$
cosmology, and
substituting 
$d\sim 117 h^{-1}{\rm Mpc}$ (i.e., $\Delta z\sim 0.18$) for the
comoving effective absorption distance (Madau, Haardt, \& Rees 1999) 
and
$\rho\sim 4\times 10^{-3} h^3 {\rm Mpc}^{-3}$
for the comoving number density of Lyman-break galaxies with magnitude ${\cal R}< 25.5$,
we find
\begin{equation}
r_n \sim \sqrt{f_{\rm bg}/(n-1)}\times 0.4 h^{-1} {\rm comoving\,\, Mpc}.
\label{eq:r_gpe_nums}
\end{equation}
In the extreme case where all of the hydrogen-ionizing background is produced by
Lyman-break galaxies with ${\cal R}<27$ (see, e.g., Steidel, Pettini, \& Adelberger 2001),
only $\sim 50$\% of it will be produced by Lyman-break galaxies with ${\cal R}<25.5$, provided
that the ratio $f_\lambda($1500\AA$)/f_\lambda($900\AA$)$ is independent of luminosity and that the
rest-frame 1500\AA\ luminosity distribution of Lyman-break galaxies is a Schechter function
with knee ${\cal R}_*=24.5$ and faint-end slope $\alpha=-1.6$ (Adelberger \& Steidel 2000),
and so we can take 0.5 as a rough upper limit on $f_{\rm bg}$.
This implies an upper limit of $r_{\rm max} \simlt 0.1 h^{-1}$ comoving Mpc
for the radius within which ionizing radiation from Lyman-break
galaxies could raise the ionizing flux to $\sim 8$ times its universal
value and increase the mean transmissivity from 0.5 to 0.8.
The argument is crude in a number of ways, but
the observed radius with $\langle {\rm flux}/{\rm cont}\rangle\sim 0.8$ is $\sim 5$
times larger than $r_{\rm max}$ and $r_{\rm max}$
does not depend strongly on any of its parameters $f$, $n$, $d$, or $\rho$.
The change in mean transmissivity near Lyman-break galaxies appears unlikely
to be produced solely by the Lyman-continuum radiation they emit.

\subsubsection{Galactic superwinds}
Could winds driven by the combined force of numerous supernova explosions
within Lyman-break galaxies be responsible instead?
Strong winds with velocities exceeding the
escape velocity have been observed around
a large fraction of starburst galaxies in the local universe (e.g. Heckman et al. 2000).
Similar outflows are seen in Lyman-break galaxies as well
(\S~\ref{sec:dvsys}; Pettini et al. 2001; Pettini et al. 2002).
Though the typical velocity of a Lyman-break galaxy's interstellar lines relative to
its nebular lines is only $\sim 150 {\rm km\,s}^{-1}$, the velocity
exceeds $300 {\rm km\,s}^{-1}$ for roughly one third of the galaxies
in the sample of~\S~\ref{sec:dvsys}.  Moreover the interstellar material
within an individual Lyman-break galaxy does not all flow outward at a single
rate.  Instead a reasonable fraction of the
absorbing interstellar material has been
accelerated to velocities significantly higher the mean interstellar velocity.
Pettini et al. (2002) found absorbing interstellar gas with blueshifts
of up to $\sim 750$ km s$^{-1}$ in MS1512-cB58, for example, a galaxy
with a mean interstellar blueshift of $\sim 250$ km s$^{-1}$.  The large
velocity widths of most Lyman-break galaxies 
($\sigma_v \sim$ 180--320 km s$^{-1}$; Steidel et al. 1996) 
show that this situation must be common.  The typical Lyman-break galaxy apparently
contains absorbing material flowing outwards with a range of velocities
$0\simlt v\simlt 600$ km s$^{-1}$.

In local galaxies, where similar outflows are observed, a distribution
of velocities from 0 to $v_{\rm max}$ is generally interpreted
as evidence that winds from supernovae are stripping material from
interstellar clouds and accelerating it to the velocity $v_{\rm max}$
(e.g. Heckman et al. 2000).  If Lyman-break galaxies' absorption spectra
were interpreted in the same way, one would conclude that most of
the absorbing material will eventually be accelerated to
a velocity of $\sim 600$ km s$^{-1}$.  In \paperthree\ we consider
the implications of 600 km s$^{-1}$ outflows from Lyman-break galaxies
in some detail.
We argue that $\sim 600$ km s$^{-1}$ winds
should be able to escape potential wells as deep as those that presumably
surround Lyman-break galaxies, 
show that the release of supernova
energy implied by the galaxies' star-formation rates and stellar masses
should be sufficient to set massive $\sim 600$ km s$^{-1}$ winds
in motion, and find that these winds would likely
travel a distance comparable the observed radius
of the galaxy proximity effect
during the typical $\sim 300$Myr star-formation
time-scale of Lyman-break galaxies. 
The upshot is that these winds may plausibly have driven 
intergalactic material from a cavity of
radius $r\sim 0.5$ comoving Mpc surrounding each galaxy,
producing the observed lack of neutral hydrogen 
near the galaxies.  In the remainder of this section we will
look for other evidence that this might be the case.

\subsection{Metals}
\label{sec:smallciv}
If Lyman-break galaxies drive winds into their surroundings,
one might expect to see an increase in the number density
of metal-line absorption systems near them.
The discussion above and in \paperthree\ suggests that
any material ejected by a Lyman-break galaxy would be likely
to lie within $\sim 0.5$ comoving Mpc 
and to have a redshift difference
$\simlt 600$ km s$^{-1}$.  These numbers are uncertain.
Aside from the crude spherical model of \paperthree\ and its poorly constrained
parameters (e.g., the fraction of supernova energy that 
is imparted to the winds),
there are the complications that winds from different galaxies
will have advanced to different radii, that winds slow as they advance,
and that much of a wind's velocity may be directed perpendicular to the
sightline.  Nevertheless the numbers above tell us roughly where
we should search for metals that might have been ejected from
Lyman-break galaxies.  

\begin{figure}[htb]
\centerline{\epsfxsize=9cm\epsffile{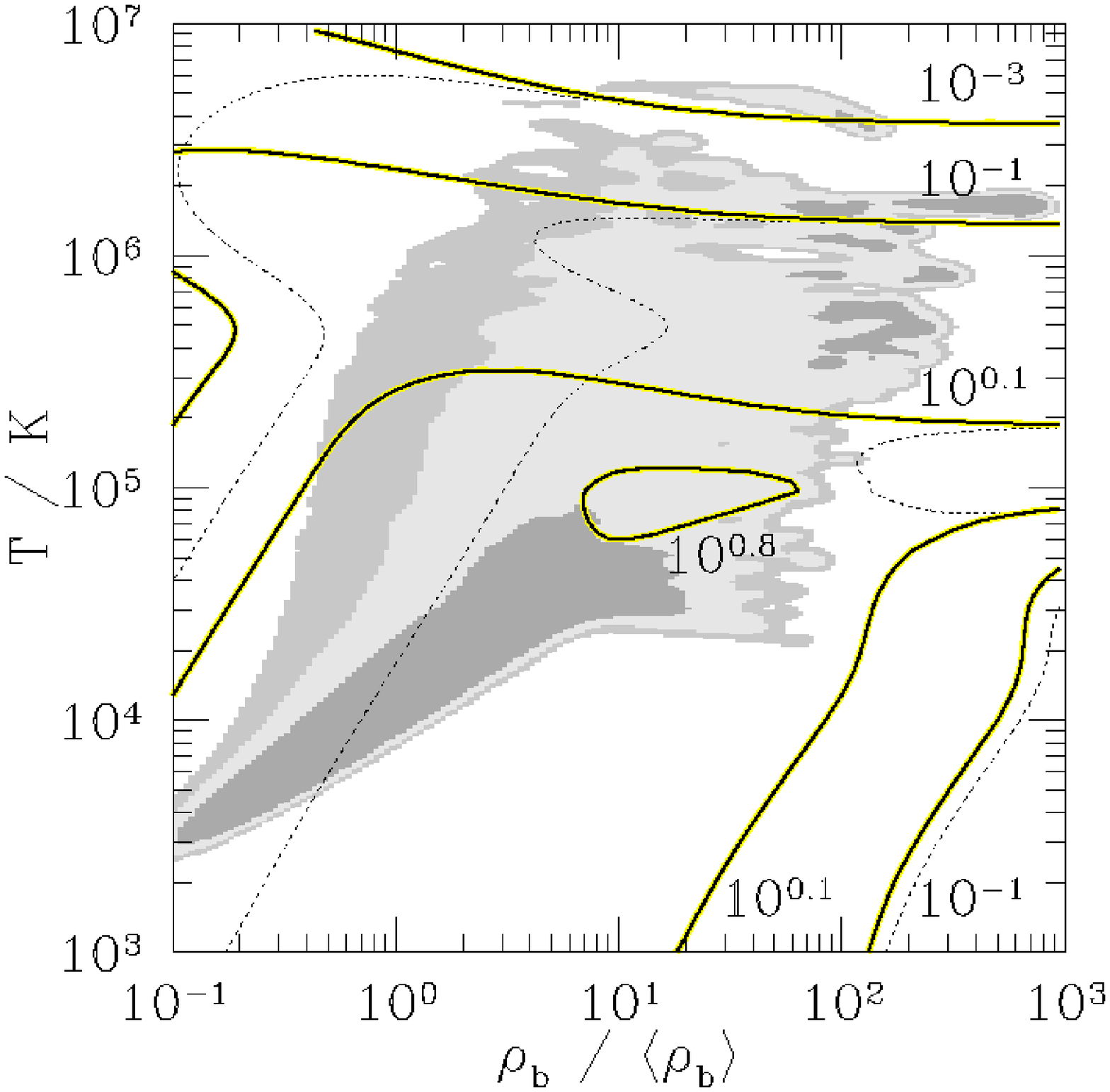}}
\figcaption[f14.eps]{
The ratio $n_{\rm CIV}/n_{\rm HI}$ for a gas with solar carbon
abundance in ionization equilibrium with a uniform radiation
field.  Densities are expressed in terms of $\langle\rho_b\rangle$,
the mean proper baryon density at $z=3$.  Thin dashed lines are appropriate for
the background radiation field 
$J_\nu = 1.2(h\nu/13.6{\rm eV})^{-1.8}\times 10^{-21}$
erg s$^{-1}$ cm$^{-2}$ Hz$^{-1}$ sr$^{-1}$;
thick solid lines for the same background except with a factor
of 10 reduction for $h\nu>4$ Ryd.
The shaded regions show the temperature and density
of 68\%, 90\%, and 95\% of the baryonic mass at $z=3$
in the SPH simulation described in White et al. 2001.
The upper limit $n_{\rm CIV}/n_{\rm HI}\simlt 6$
for solar carbon abundance shows that gas at the mean
intergalactic metallicity $[{\rm C}/{\rm H}]\sim -2.5$ should not have
a ratio $n_{\rm CIV}/n_{\rm HI}$ greater than $\sim 0.02$.
The measured ratio in gas near to Lyman-break galaxies can
exceed this maximum by a large amount, showing that the
gas is enriched well beyond the intergalactic mean.
\label{fig:civohi}
}
\end{figure}

The QSO absorption spectra available to us clearly cannot provide
perfect information about the distribution of metals in 
the intergalactic medium.  In most cases we can
detect only the absorption due to CIV, and this is what we will
be forced to rely on in the analysis below.  Although we will
often assume glibly that CIV absorption systems are found
where the metallicity is highest, we should remind readers
at the outset that the density of CIV is not related in a trivial
way to the density of metals: the fraction of carbon that
is in the third ionization state can be strongly affected
by the local gas density and by the 
intensity and shape of the ambient radiation field.
Still in at least some cases we can be reasonably
confident that the presence of CIV absorption near Lyman-break
galaxies signals the presence of significantly enriched
gas.  Our confidence derives
from Figure~\ref{fig:civohi}, which
shows the ratio $\eta\equiv n_{\rm CIV}/n_{\rm HI}$ in
ionization equilibrium
for a (76\% H, 24\% He) gas with solar carbon abundance.
This ratio depends sensitively on the possible presence
of a break at 4Ryd in the background radiation field,
since reducing the radiation at $h\nu>4$Ryd makes it
harder for CIV to be ionized to CV but does not
much affect the density of HI.
Nevertheless the plot suggests that most intergalactic gas would likely
have $0.1\simlt \eta \simlt 3$, though the ratio
could be driven as high as $\sim 6$ if the temperature
and density were carefully
chosen and if the intensity of the background radiation
decreased sharply at $h\nu>4$Ryd.  If the actual abundance of carbon in the gas
were $[{\rm C}/{\rm H}]\sim -2.5$, or $\sim$ one $300$th solar,
characteristic of the intergalactic medium at $z\sim 3$ (e.g., Dav\'e et al. 1998),
one would expect $\eta$ typically to be $\sim 10^{-3}$--$10^{-2}$
and never to exceed $\eta^{\rm max}_{-2.5}\sim 2\times 10^{-2}$.  The fact that
the observed
ratio $n_{\rm CIV}/n_{\rm HI}$ in gas close to Lyman-break galaxies sometimes
exceeds $\eta^{\rm max}_{-2.5}$ by an order of magnitude (see, e.g., the
entry in table~\ref{tab:dlas} for the $z=2.7417$ system along
the SSA22D13 sight-line)
shows that the metallicity of this gas must be significantly higher
than the mean $[{\rm C}/{\rm H}]\sim -2.5$.

We constructed catalogs of the CIV absorption systems 
in each QSO's spectrum 
by scanning by eye for double absorption lines with
the correct spacing and relative strengths for the redshifted
CIV$\lambda 1549$ doublet.  Rough column densities for each
system were estimated with the equation
\begin{equation}
\tau_\lambda^{\rm peak} = N_{\rm CIV} \frac{\pi e^2}{m_e c} f_{\rm CIV} \frac{\lambda_{\rm CIV}}{\sigma_v \sqrt{2\pi}},
\end{equation}
for the peak optical depth in terms of CIV's column density
$N_{\rm CIV}$, oscillator strength $f_{\rm CIV}$,
and (Gaussian) velocity dispersion $\sigma_v$,
after fitting $e^{-\tau_\lambda}$, with
$\tau_\lambda$ a Gaussian, to both components
of the doublet.  CIV systems with similar redshifts were treated
as independent systems if their velocity differences
exceeded twice the quadrature sum of their velocity full widths.
In the case of Q1422+2309, our CIV catalog
agrees well with the more carefully constructed catalog
of Ellison et al. (2000), though it differs somewhat in the arbitrary
grouping of neighboring CIV systems into single absorption complexes.

\begin{figure*}
\centerline{\epsfxsize=18cm\epsffile{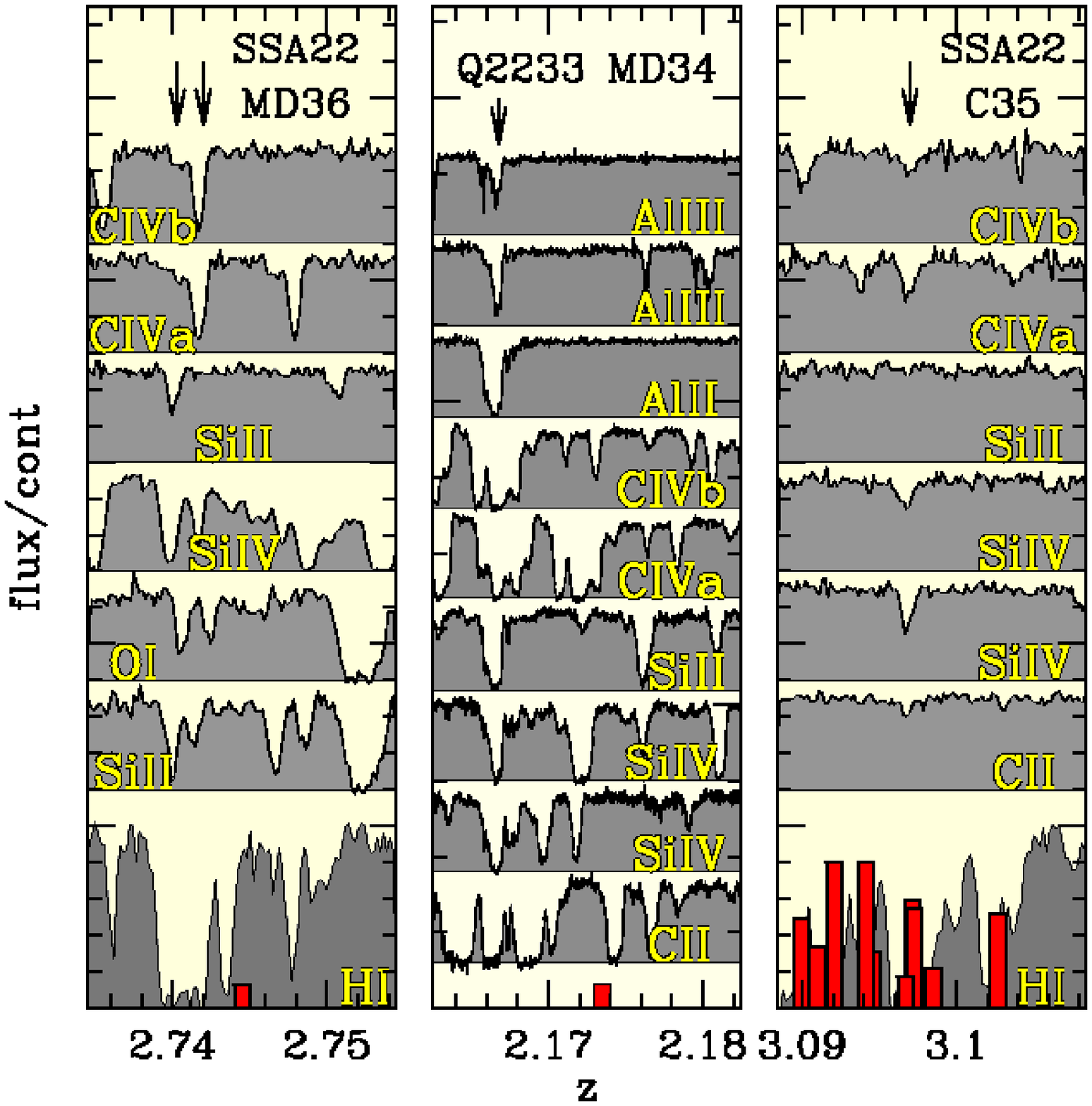}}
\figcaption[f15.eps]{ The Lyman-$\alpha$ and metal absorption along three QSO sightlines
than pass within (from left to right) $17''$, $31''$, and~$33''$ 
(90, 180, 170 proper $h^{-1}$ kpc)
of known $z\sim 3$ Lyman-break galaxies.
Sections of the QSO spectrum corresponding to different species were extracted
and aligned.  The bottom spectral segment shows the HI absorption; upper segments
show the associated metal-line absorption as indicated.  
Bluer lines are on the bottom.  Redshifts
with detected metal lines are marked by arrows. Absorption in the extracted
spectral segments can be produced by other metal lines
at other redshifts, and (in some cases) intergalactic Lyman-$\alpha$ at higher redshift,
and the observed absorption can believably be associated with metals at the
galaxy's redshift only when it is observed in more than one line.  Galaxy locations
are marked with vertical bars, as in Fig.~\ref{fig:clustvoid}; the galaxy closest
to the sightline lies in the middle of the extracted spectrum.  The velocity
offsets between the absorption in the QSO spectrum and the galaxy itself
($\sim 100-600$ km s$^{-1}$) are comparable to the measured outflow velocities
of the galaxies (e.g. Pettini et al. 2001).
\label{fig:galmetals}
}
\end{figure*}

Ten galaxies in our primary sample lie within $\Delta\theta = 35''$ 
($\sim 0.75h^{-1}$ comoving Mpc for $\Omega_M=0.3$, $\Omega_\Lambda=0.7$) of
the QSO sightline.  In nine of the ten cases there is detectable
CIV absorption in the QSO spectrum within 600 km s$^{-1}$ of the
galaxy's redshift.  The lone exception lies near the sightline
to SSA22D13, a faint QSO whose moderate resolution spectrum 
reveals only the strongest CIV systems.  In three cases, shown
in figure~\ref{fig:galmetals}, the CIV absorption is especially
strong and absorption due to many species is evident.  
The galaxies shown in the figure, SSA22-MD36, Q2233-MD34, and SSA22-C35,
lie~$17''$, $31''$, and~$33''$, respectively, from their background QSO's
sightline; of all the galaxies in our sample, their angular separations to
the sightline are the 3rd, 6th, and 8th closest.  
The metal-line system close to Q2233-MD34 has the largest CIV column density
of any in our QSO spectra; the system close to SSA22-MD36 
has the third largest.  Does this correspondence of metal-line systems with
galaxies near the sightline imply some physical connection
between the two, or could it be a coincidence?
Should we be surprised that 2 of the 3 strongest CIV systems in our
sample lie within $\Delta z=600 {\rm km\,s}^{-1}$ and $\Delta\theta=35''$
of a Lyman-break galaxy?

To address this question, we need some estimate of the observed
overdensity of Lyman-break galaxies near CIV systems
relative to what would be expected if they
were distributed independently.
We can estimate the number of Lyman-break
galaxies that would lie so close to the 3 strongest CIV systems
if galaxies and metal systems were independently distributed by
generating a large ensemble of fake data sets where each
galaxy within $\Delta\theta=35''$ of the sightline is assigned
a redshift drawn at random from our selection function.  
See appendix~\ref{sec:apdensest}.  Among these
fake data sets,
the mean number of galaxies
that lie within $\Delta z=600 {\rm km\,s}^{-1}$ of
one of the three strongest CIV systems is $0.049$.
Since the observed number is $2$, the implied overdensity
of Lyman-break galaxies is $\bar\xi_{gc}\sim 40$.  Only 1 time in 1000 will
sampling from a Poisson distribution with a true mean of $0.049$ 
yield a value of $\geq 2$, so we can conclude with 
confidence that strong CIV systems and Lyman-break galaxies are not
distributed independently.

But this does not show that the observed metals were driven
out of the galaxy by a superwind.  The number density of Lyman-break
galaxies in cells centered on other Lyman-break galaxies is much
higher than the number density in randomly placed cells, for example,
yet few believe that one galaxy ejected the next.
Is it possible that galaxies and CIV-systems tend to fall near 
each other for the same reason that galaxies fall near each other,
because they all trace the same large-scale structure?
A simple way to address this issue is to use the 
observation of Quashnock \& Vanden Berk (1998) that
the correlation function of CIV systems at $z\sim 3$ is similar
to the correlation function of Lyman-break galaxies, a fact compatible
with the idea that CIV systems and Lyman-break galaxies are similar
objects.  Suppose CIV systems and Lyman-break galaxies were in fact the
same objects, but the CIV associated with each galaxy extended only
to radii small compared to the smallest galaxy-QSO impact parameter
in our sample, so that the detected CIV absorption could never
have been produced by a galaxy we observed.  In this case
we would
expect the overdensity of galaxies within $\Delta\theta=35''$
and $\Delta z=600 {\rm km\,s}^{-1}$ of a CIV system to be
roughly equal to the mean overdensity of Lyman-break galaxies within
the same distance of another Lyman-break galaxy.  Among the 697
Lyman-break galaxies with the most certain redshifts in 13 fields
of our survey,
31 unique pairs have a separation $\Delta\theta<35''$
and $\Delta z<600 {\rm km\,s}^{-1}$ while 3.92 would have been
expected if the galaxies were distributed uniformly.
The implied galaxy-galaxy overdensity $\bar\xi_{gg}=6.9$ is significantly
smaller than the measured galaxy-CIV overdensity $\bar\xi_{gc}=40$.
This suggests the spatial coincidence of Lyman-break galaxies
and CIV systems may be too strong to be explained by arguing that both
trace the same large scale structure, that both tend to
reside in the same clusters and shun the same voids.  
A more direct connection between the observed CIV systems and galaxies
appears to be required.

This sort of argument can be formalized with a statistical
inequality derived in appendix~\ref{sec:apgenlim}.  If $f$ is a discrete
(i.e. Poisson) realization of the continuous function $f'$,
$g$ is a discrete realization of the continuous function $g'$,
and $\bar\xi_{fg}$ denotes the mean value within some volume of the
cross-correlation function $\xi_{fg}$ between $f$ and $g$,
then $\bar\xi_{fg}$ can exceed $(\bar\xi_{ff}\bar\xi_{gg})^{1/2}$
only if the locations of particles in $g$ are influenced by
where particles happen to lie in $f$ or vice versa.
This would be the case if (for example)
$g'$ were equal to $f'$ and the particles in $g$ were a random
subset of those in $f$, or if each particle in $f$ were surrounded
by a cloud of particles in $g$.  The statement holds provided
the correlation functions are volume averaged with one of a class
of 3D weighting functions that includes the Gaussian.
If we could show that the Gaussian-weighted cross-correlation
function of galaxies and CIV systems $\bar\xi_{gc}$
exceeded the root-product of the Gaussian-weighted autocorrelation functions
$(\bar\xi_{gg}\bar\xi_{cc})^{1/2}$, we would have evidence
that the number density of galaxies in a random volume
directly affected (or was directly affected by) 
the number density of CIV systems in the same volume.

The mean overdensity of Lyman-break galaxies in
Gaussian ellipsoids with $\sigma_z=350 {\rm km\,s}^{-1}$
and $\sigma_\theta=20''$ that are centered on the
strongest three CIV systems in our sample
is $\bar\xi_{gc}=21$.  The mean overdensity of Lyman-break
galaxies in similar ellipsoids centered on other Lyman-break
galaxies is $\bar\xi_{gg}=7$.  The mean overdensity of CIV systems
in similar ellipsoids centered on other CIV systems will be comparable,
$\bar\xi_{cc}\sim 7$, if Quashnock \& Vanden Berk's (1998)
estimate of the CIV correlation function remains accurate on 
small spatial scales.  $\bar\xi_{gc}$ evidently exceeds
$(\bar\xi_{gg}\bar\xi_{cc})^{1/2}$ by a large amount.  This may
be an aberration. The sample is small.
But if the result holds in a much larger sample
of galaxies and CIV systems, we would have solid evidence
for a physical link between galaxies and the strong CIV absorption observed
$\sim 0.5h^{-1}$ comoving Mpc away.

\begin{figure}[htb]
\centerline{\epsfxsize=9cm\epsffile{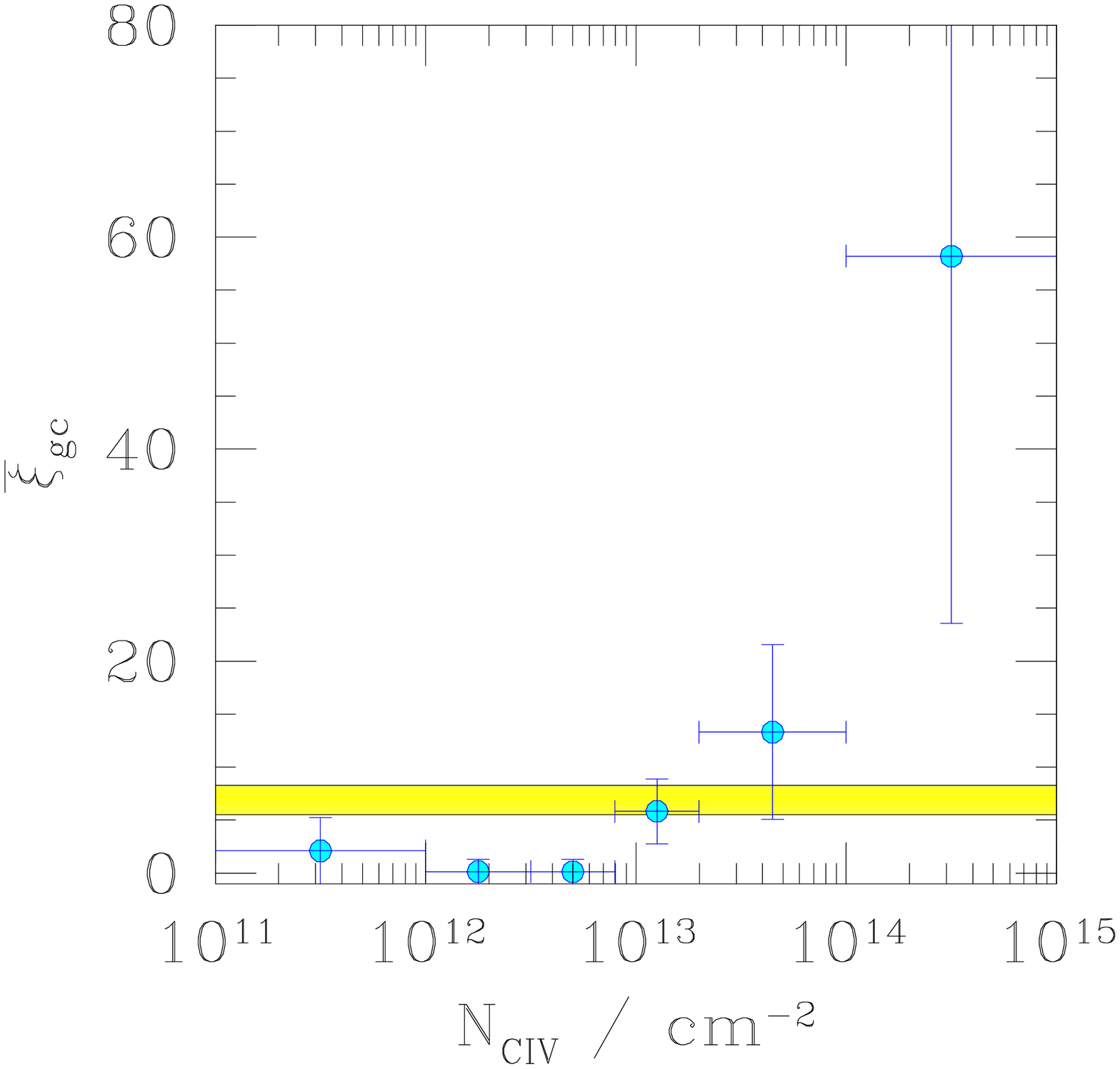}}
\figcaption[f16.eps]{
The observed overdensity of CIV systems within $\Delta z=600$ km s$^{-1}$
and $\Delta\theta=35''$ of a Lyman-break galaxy as a function of CIV
column density.  The spatial association of Lyman-break galaxies and CIV
systems at these separations 
grows rapidly stronger as the CIV column density increases.  
The shaded horizontal region
shows the overdensity of Lyman-break galaxies within a similar distance
of another Lyman-break galaxy.  As explained in \S~\ref{sec:smallciv}, 
the presence of CIV points above this region suggests strong CIV systems
and Lyman-break galaxies may be identical objects.
\label{fig:civcolumn}
}
\end{figure}

In any case, the connection between Lyman-break galaxies and
CIV systems clearly weakens as the column density of the CIV
systems is reduced.  If we take
the 20 strongest CIV systems, for example, rather than the
3 strongest, the mean overdensity of Lyman-break galaxies
in similar Gaussian ellipsoids surrounding the CIV systems
is $\bar\xi_{gc}=9.0$. This still exceeds 
$\bar\xi_{gg}=7\sim\bar\xi_{cc}$, but probably not by a significant amount.
The mean overdensity in Gaussian ellipsoids surrounding every CIV system
in our sample is a paltry $\bar\xi_{gc}=2.8$, showing that there
is little evidence that the weakest CIV systems
are directly associated with the observed galaxies.
The dependence of the cross-correlation strength $\bar\xi_{gc}$
on column-density is shown in figure~\ref{fig:civcolumn}.
Also shown is $\bar\xi_{gg}$ and its uncertainty.  Cylindrical
cells with $\Delta\theta<35''$
and $\Delta z<600 {\rm km\,s}^{-1}$ were used when calculating
the mean overdensities $\bar\xi$, rather than the ellipsoidal
Gaussians discussed above. This simplified the calculation of the
uncertainties in $\bar\xi$, which we optimistically took to be Poisson,
but it means that $\bar\xi_{gc}$ does not formally need
to satisfy inequality~\ref{eq:apbarxiineqnoshot}.  But similar
conclusions about the column densities where \ref{eq:apbarxiineqnoshot}
is and is not satisfied would follow had we averaged in Gaussian 
ellipsoids instead.

\begin{figure}[htb]
\centerline{\epsfxsize=9cm\epsffile{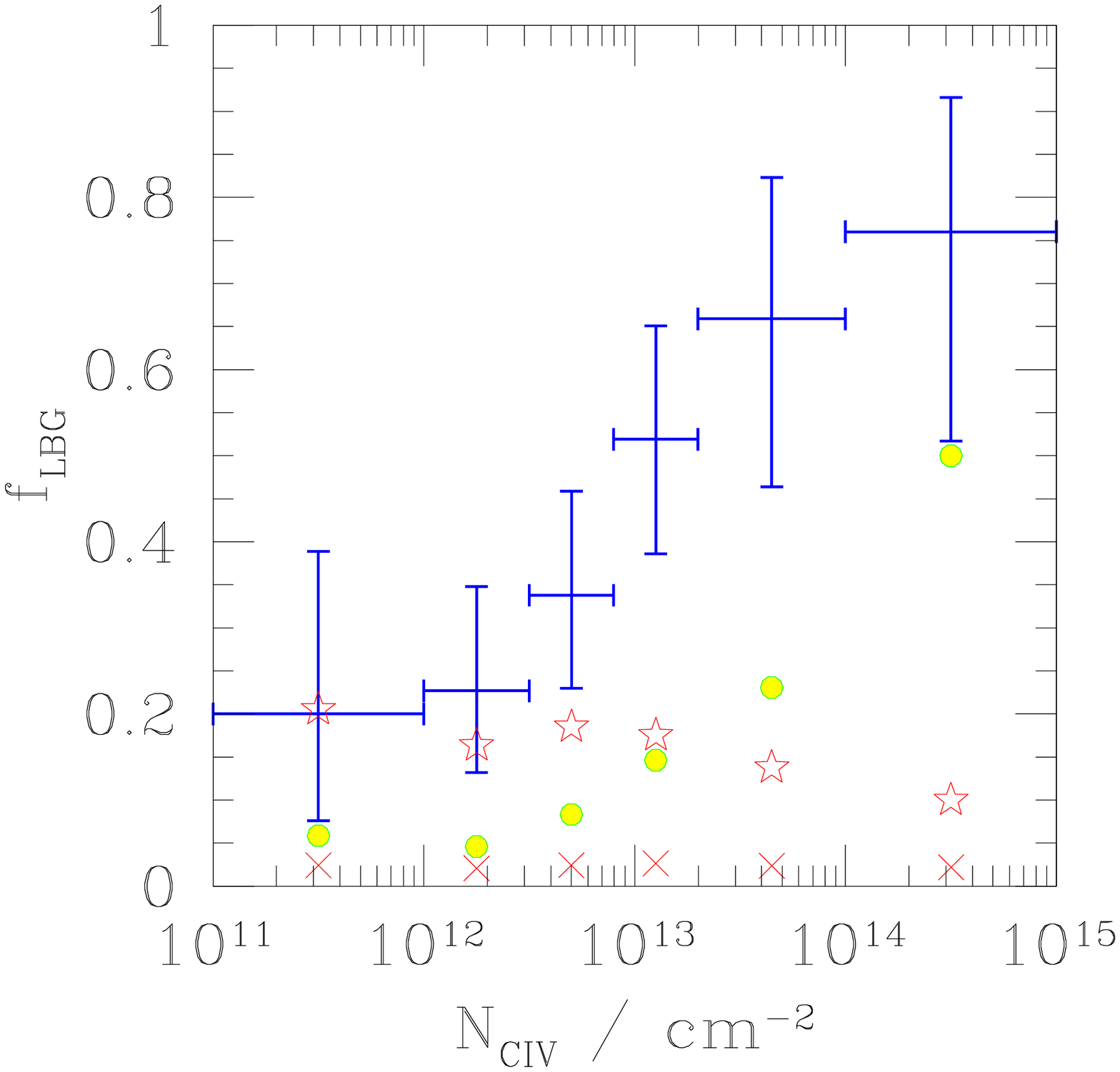}}
\figcaption[f17.eps]{
The fraction $f_{\rm LBG}$ of detected CIV systems that lie within
$\Delta z=600$ km s$^{-1}$ and $\Delta\theta=35''$ of a Lyman-break galaxy
as a function of CIV column density.  Circles mark the fraction that
lie within this distance of a galaxy in our spectroscopic sample.
Stars mark the expected fraction if each CIV system had one galaxy
within this distance.  Crosses mark the expected fraction if CIV
systems and Lyman-break galaxies were independently distributed. 
The low value of $f_{\rm LBG}$ for the stars reflects the severe
incompleteness of our spectroscopic sample; simulations suggest
that only $\sim$ one fifth
of Lyman-break galaxies with redshifts similar to those
of the CIV systems would satisfy our photometric selection
criteria and be included in our final spectroscopic catalogs.
Error bars mark the estimated fraction of CIV systems that
lie within $\Delta z=600$ km s$^{-1}$ and $\Delta\theta=35''$
of a Lyman-break galaxy once our selection effects are corrected.
\label{fig:galinrange}
}
\end{figure}

It is interesting to address the association of galaxies and metals
in a slightly different way.  What fraction of detectable CIV absorption
is produced by gas within $\Delta\theta<35''$, $\Delta z<600 {\rm km\,s}^{-1}$
of a Lyman-break galaxy?  Because only a small fraction of the
Lyman-break galaxies\footnote{i.e., the galaxies with
optical magnitude ${\cal R}\simlt 25.5$ and dust reddening
$0<E(B-V)\simlt 0.5$ that have a non-zero probability of satisfying
our photometric selection criteria; see Steidel et al. (1999).}
in any field at $z\sim 3$ are included in our spectroscopic sample,
we would not expect most CIV systems to lie near a galaxy in
the sample even if every CIV system had a Lyman-break galaxy nearby.
Limited observing time allowed us to obtain spectra for fewer than
half of the photometrically detected Lyman-break galaxies in each field,
for example, and Monte-Carlo simulations (Steidel et al. 1999) suggest
that only $\sim 50$\% of all Lyman-break galaxies
are included in our photometric sample even at $z=3.0$
where our selection is most efficient.  We know our sample is seriously incomplete;
but we can attempt to correct for this and estimate what fraction
of CIV systems would have been found to lie within 
$\Delta\theta<35''$, $\Delta z<600 {\rm km\,s}^{-1}$ of a Lyman-break galaxy
if we had been able to measure the redshifts of every Lyman-break galaxy
our fields.  The result is shown in figure~\ref{fig:galinrange}.
Circles mark the fraction $f_{\rm LBG}$ of detected CIV systems with different column densities
that lie within $\Delta\theta<35''$, $\Delta z<600 {\rm km\,s}^{-1}$
of a galaxy in our spectroscopic sample.
Stars mark the expected fraction if every CIV system lay within
this distance of one (and only one) Lyman-break galaxy.  Crosses
mark the expected fraction if CIV systems and Lyman-break galaxies
were distributed independently.  At column densities
$N_{\rm CIV}\simgt 10^{13.5} {\rm cm}^{-2}$ the observed number of
Lyman-break galaxies close to CIV systems significantly exceeds
the number expected if only one Lyman-break galaxy lay
within $\Delta\theta<35''$, $\Delta z<600 {\rm km\,s}^{-1}$ of
each CIV system.  This apparently shows that the strongest
CIV absorption is produced in gas with several Lyman-break galaxies
nearby, an interesting observation for which we have no ready
explanation.  Confidence limits on the fraction of CIV systems
that have at least one Lyman-break galaxy within $\Delta\theta<35''$
and $\Delta z<600 {\rm km\,s}^{-1}$ can be crudely derived by estimating
the mean number of Lyman-break galaxies within this distance of
a CIV system, $\bar N_{\rm LBG}$, then using Poisson statistics to work out how frequently
sampling from a distribution with this mean 
would yield one or more galaxies ($1-e^{-\bar N_{\rm LBG}}$).  Averaging
over the range of $\bar N_{\rm LBG}$ compatible with the data
leads to the rough confidence intervals
shown in figure~\ref{fig:galinrange}.  The data evidently
suggest that the majority of CIV absorption with $N_{\rm CIV}\simgt 10^{13} {\rm cm}^{-2}$
is produced by gas that lies no farther than $\Delta\theta<35''$
and $\Delta z<600 {\rm km\,s}^{-1}$ from a Lyman-break galaxy,
i.e., by gas that could plausibly have been ejected from the
galaxies in an outflow.

How might weaker metal-line absorption systems with $N_{\rm CIV}\simlt 10^{13} {\rm cm}^{-2}$
be associated with Lyman-break galaxies?  This question can be straightforwardly
addressed with results presented in \S~\ref{sec:corrfns} below.  The numerous weak CIV absorbers
dominate the cross-correlation function of galaxies and CIV systems, which (as we will show)
is roughly a power-law of the form $\xi_{\rm gc}=(r/r_{\rm gc})^{-\gamma_{\rm gc}}$
with $r_{\rm gc}\simeq 3.2h^{-1}$ comoving Mpc and $\gamma_{\rm gc}=1.6$.
The mean number of Lyman-break galaxies within a distance $r$ of a randomly
chosen CIV system is therefore
\begin{equation}
N_{\rm LBG} \simeq \frac{4}{3}\pi r^3 n_{\rm LBG} \biggl(1+\frac{3r^{-\gamma_{\rm gc}}}{(3-\gamma_{\rm gc})r_{\rm gc}^{-\gamma_{\rm gc}}}\biggr),
\end{equation}
where $n_{\rm LBG}\simeq 4\times 10^{-3}h^3$ Mpc$^{-3}$ is the comoving number density
of Lyman-break galaxies with magnitude ${\cal R}<25.5$.
This shows that the numerous weak CIV systems will typically have $\sim 1$ Lyman-break galaxy
within a comoving distance of $\sim 2.4h^{-1}$ Mpc.  If they were randomly distributed,
the nearest Lyman-break galaxy would lie $\sim 3.9h^{-1}$ Mpc away.  Weak CIV systems also tend to lie
close to Lyman-break galaxies, though not as close as their higher column-density counterparts.

In summary, it appears that metals in the intergalactic medium are closely connected to
the star-forming galaxies we observe.  Most CIV systems lie within $\sim 2.4h^{-1}$ Mpc
of a Lyman-break galaxy; higher column density systems tend to lie closer still; and
the highest column density systems are so strongly correlated with Lyman-break galaxies
that we may be forced to conclude that they are the same objects, a result
that is remarkable because the absorbing gas is typically $\sim 0.5h^{-1}$ comoving Mpc
from the galaxies' stars.

We are aware of two objections to the implied link between intergalactic
metals and Lyman-break galaxies' winds.  The large HI column densities
of the metal-line systems significantly exceed naive expectations 
based on the high expected post-shock temperature of fast-moving
winds (\paperthreep).  The constant observed intergalactic CIV density
at redshifts $1.5<z<5.5$ may imply that metal enrichment of the intergalactic
medium occurred at $z>5.5$ and not at the lower redshifts we have
observed (Songaila 2001).  Both objections are reasonable; neither is damning.
Cooling due to a plausibly high metallicity in the outflows could account
for the large observed HI column densities.  Metals ejected by winds might later be
driven back into collapsed objects by the continued action of gravitational
instability (\paperthreep); in this case the constant intergalactic CIV density
at $1.5<z<5.5$ would primarily reflect the constant comoving star-formation
density (e.g., Steidel et al. 1999) at similar redshifts.

\subsection{Interstellar absorption lines}
\label{sec:ism}
We conclude this section with a brief discussion of the gas that produces
the blueshifted absorption lines in the spectra of Lyman-break galaxies.
These absorption lines are often called ``interstellar,'' because
they resemble the absorption lines produced by interstellar material
in local galaxies.  But is it necessarily true that the absorbing material
lies between the galaxy's stars?  Or might it instead lie far from the
stars, perhaps at the front of a shock that has advanced $\sim 0.5h^{-1}$ comoving Mpc?

\begin{figure}[htb]
\centerline{\epsfxsize=9cm\epsffile{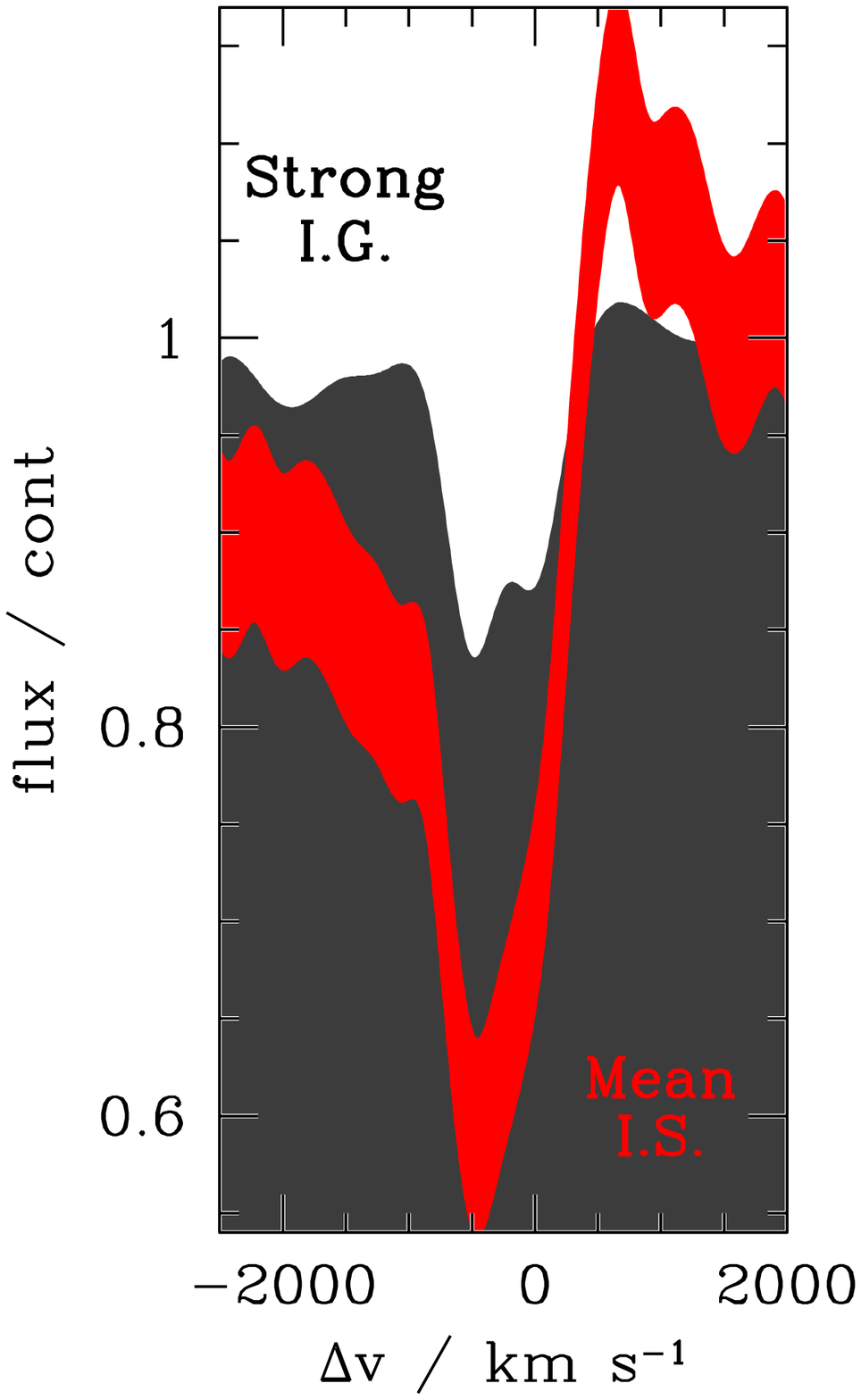}}
\figcaption[f18.eps]{Intergalactic and interstellar CIV absorption.  The thick line
in the foreground shows the mean CIV absorption observed in the
spectra of Lyman-break galaxies (Shapley et al. 2003, in preparation).
The equivalent widths implied by the upper and lower edges of
this line differ by $\pm 3\sigma$ from the mean equivalent
width reported by Shapley et al.
The darker shaded region in the background shows the CIV absorption
produced by the intergalactic gas that lies $\sim 0.35h^{-1}$
comoving Mpc from the Lyman-break galaxy
SSA22-MD36; it is a small piece of our spectrum of a background QSO 
that is offset from SSA22-MD36 by $17''$.  The QSO spectrum was smoothed
by a Gaussian with $\sigma=200{\rm km s}^{-1}$ to approximate the
resolution of the galaxy spectrum.
The intergalactic CIV absorption system is among the strongest in our entire
sample, yet its equivalent width is small compared to the CIV equivalent widths
observed in the spectra of Lyman-break galaxies themselves.  Intergalactic
CIV absorption may sometimes be produced by material ejected from Lyman-break
galaxies, but
not by the same material that produces the galaxies' own interstellar absorption lines.
\label{fig:igis}
}
\end{figure}

It is relatively easy to rule out the idea that the intergalactic CIV absorption
observed within $\sim 0.5h^{-1}$ Mpc of Lyman-break galaxies is produced by the same gas
responsible for the galaxies' own absorption lines.
The intergalactic equivalent widths are too small.  This is illustrated
by Figure~\ref{fig:igis}, which compares the mean interstellar CIV absorption 
in a sample of $\sim 800$ Lyman-break galaxies (Shapley et al. 2003)
to the CIV absorption produced by the intergalactic gas that lies $\sim 0.35h^{-1}$ comoving Mpc
from the Lyman-break galaxy SSA22-MD36. With a column density
$N_{\rm CIV}\simeq 10^{14.4}$ (see table~\ref{tab:dlas}), this intergalactic
CIV absorption is among the strongest in any of our QSO spectra.  But even though
its column density is $\sim 2$ orders of magnitude larger than  
the typical column density in our intergalactic sample, the resulting absorption
equivalent width is far smaller than observed in the average Lyman-break galaxy's spectrum.
We have argued that winds from Lyman-break galaxies may reach comoving
radii approaching $\sim 0.5h^{-1}$ Mpc, but we cannot claim
that most of the outflowing gas responsible for the galaxies' absorption
lines has traveled so far.

\begin{figure}[htb]
\centerline{\epsfxsize=9cm\epsffile{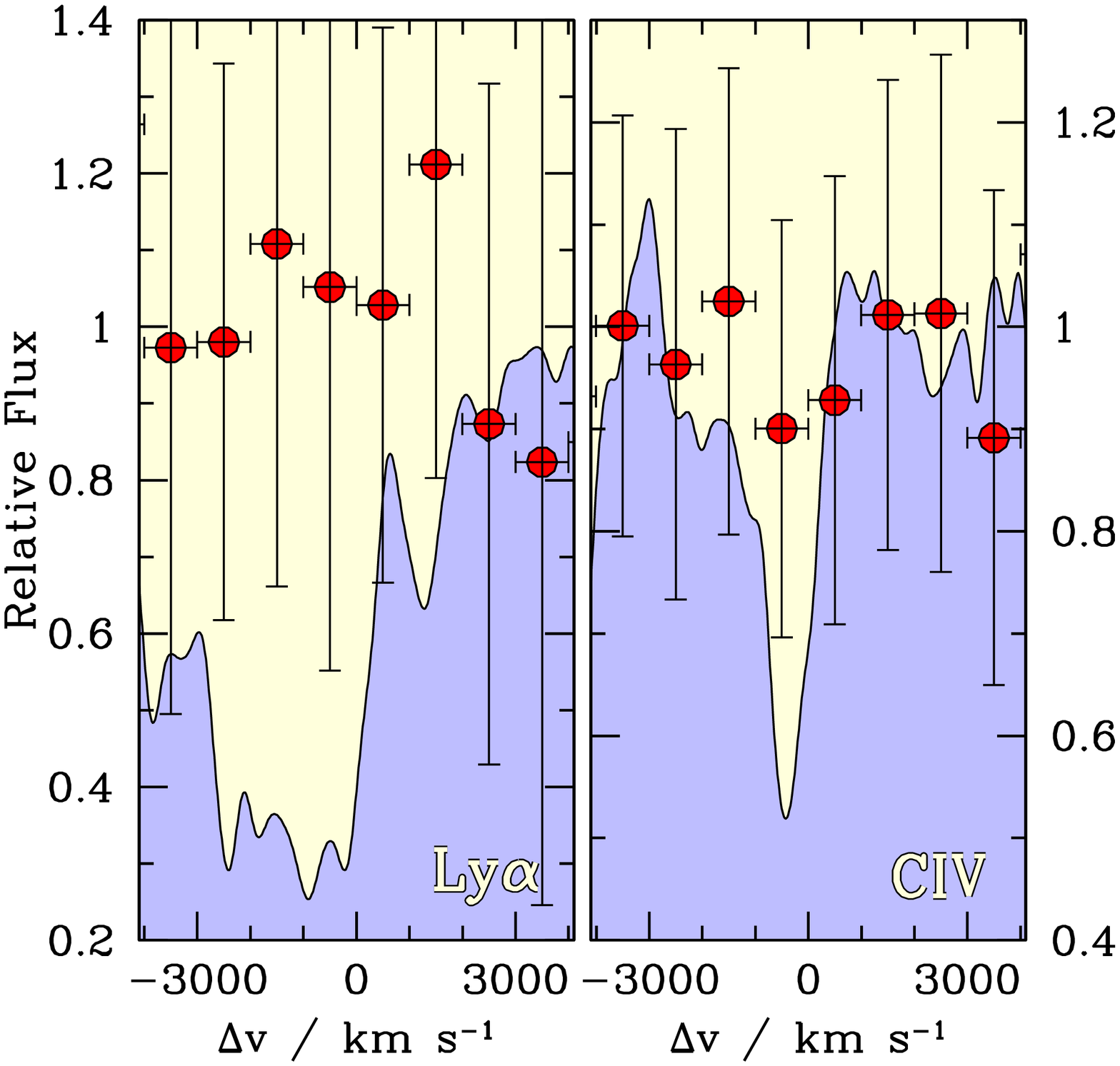}}
\figcaption[f19.eps]{Similar to Figure~\ref{fig:igis}, except this plot compares
Lyman-break galaxies' strong observed interstellar Ly-$\alpha$ and CIV
absorption (curves shaded beneath) to the undetected absorption that
foreground Lyman-break galaxies produce in the spectra of background
Lyman-break galaxies at small impact parameters $20<r_\theta<40h^{-1}$ proper kpc
(points).  Seventeen galaxy pairs have impact parameters so small.
The points' vertical error bars show the rms dispersion in
relative flux at each velocity offset among many sets of 17 galaxy pairs
drawn at random (with duplication allowed) from the 17 galaxy pairs
with $20<r_\theta<40h^{-1}$ proper kpc.
\label{fig:nightnday}
}
\end{figure}

The lack of galaxies close to the QSO sightlines prevents us from using
the QSO spectra to place tighter limits on the typical distance
between Lyman-break galaxies' stars and interstellar gas.
But a comparatively large number of Lyman-break galaxies lie
near sightlines towards other Lyman-break galaxies at higher redshift.
Since the spectra of the background galaxies are generally good enough
to reveal the background galaxies' own CIV absorption,
they should also be good enough to reveal CIV absorption
from a foreground Lyman-break galaxy---if the
foreground galaxy's CIV absorption were mainly produced
by gas at large impact parameters.
17 Lyman-break galaxies in our complete sample have a background
Lyman-break galaxy at an impact parameter of $r_\theta<0.16h^{-1}$ comoving Mpc.
Among these 17 pairs, the smallest and mean impact parameters are $0.07h^{-1}$ 
and $0.11h^{-1}$ comoving Mpc respectively.  The pairs are generally
chance alignments rather than physical associations; all but four
have redshifts that differ by more than $\Delta z=0.1$
($\sim 70h^{-1}$ comoving Mpc for $\Omega_M=0.3$, $\Omega_\Lambda=0.7$).
Figure~\ref{fig:nightnday} shows the mean Lyman-$\alpha$
and CIV absorption at the redshift of the foreground galaxy
that we observe in the spectra of the 17 background galaxies.  
In calculating the mean absorption, we discarded parts of the
spectra with $\lambda<4000$\AA, parts that were
contaminated by telluric absorption or by residuals from bright
sky lines, and parts that overlapped with any of the 
background galaxy's ten
strongest absorption lines.  Each spectrum was then multiplied
by a constant to give it a mean flux of unity
within $\pm 5000 {\rm km s}^{-1}$ of the foreground galaxy's absorption line.
Little absorption is apparent at the redshifts
of the foreground galaxies.  We conclude that the typical
Lyman-break galaxy's interstellar absorption lines arise
in gas that lies within $\sim 25h^{-1}$ proper
kpc of its stars.  This conclusion is consistent with
the idea that the interstellar absorption is produced
by pockets of cold gas entrained in a hot outflow that 
destroys them before they reach large radii (e.g., 
Klein, McKee, \& Colella 1994 \S 9.2).  It is equally consistent
with the idea that Lyman-break galaxies' winds rarely propagate 
so far.

\section{CORRELATION FUNCTIONS}
\label{sec:corrfns}
The previous sections were primarily concerned with spatial averages over the
auto- and cross-correlation functions of galaxies, CIV systems, and Lyman-$\alpha$
forest flux decrements.  These averages highlighted the aspects of the spatial
association of galaxies, metals, and gas that we found most interesting.
But readers may wonder how our conclusions would alter if we averaged the correlation
functions over different volumes, or may wish to know how galaxies and
intergalactic material are associated on spatial scales other than those
we considered.  This section presents the correlation functions with
no spatial averaging imposed.

\subsection{Two dimensional}
\begin{figure}[htb]
\centerline{\epsfxsize=9cm\epsffile{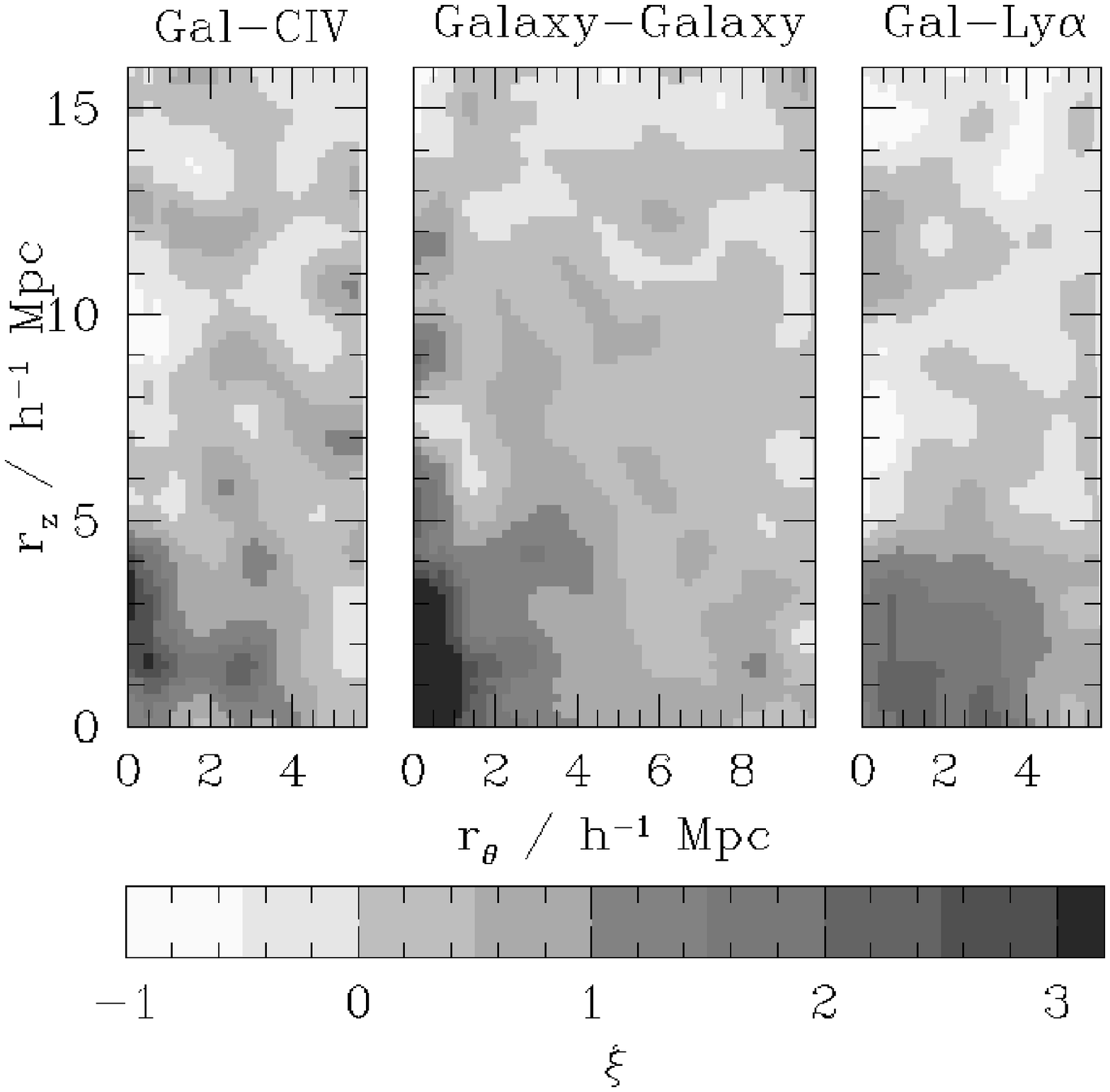}}
\figcaption[f20.eps]{
The two dimensional auto- and cross-correlation functions of galaxies
with galaxies, with CIV systems, and with intergalactic Lyman-$\alpha$
transmissivity.  The galaxy--transmissivity correlation function is
multiplied by $-10$.  Each correlation function was smoothed by
a two-dimensional Gaussian with $\sigma=0.5h^{-1}$ comoving Mpc.
\label{fig:corrfns}
}
\end{figure}

\begin{figure}[htb]
\centerline{\epsfxsize=9cm\epsffile{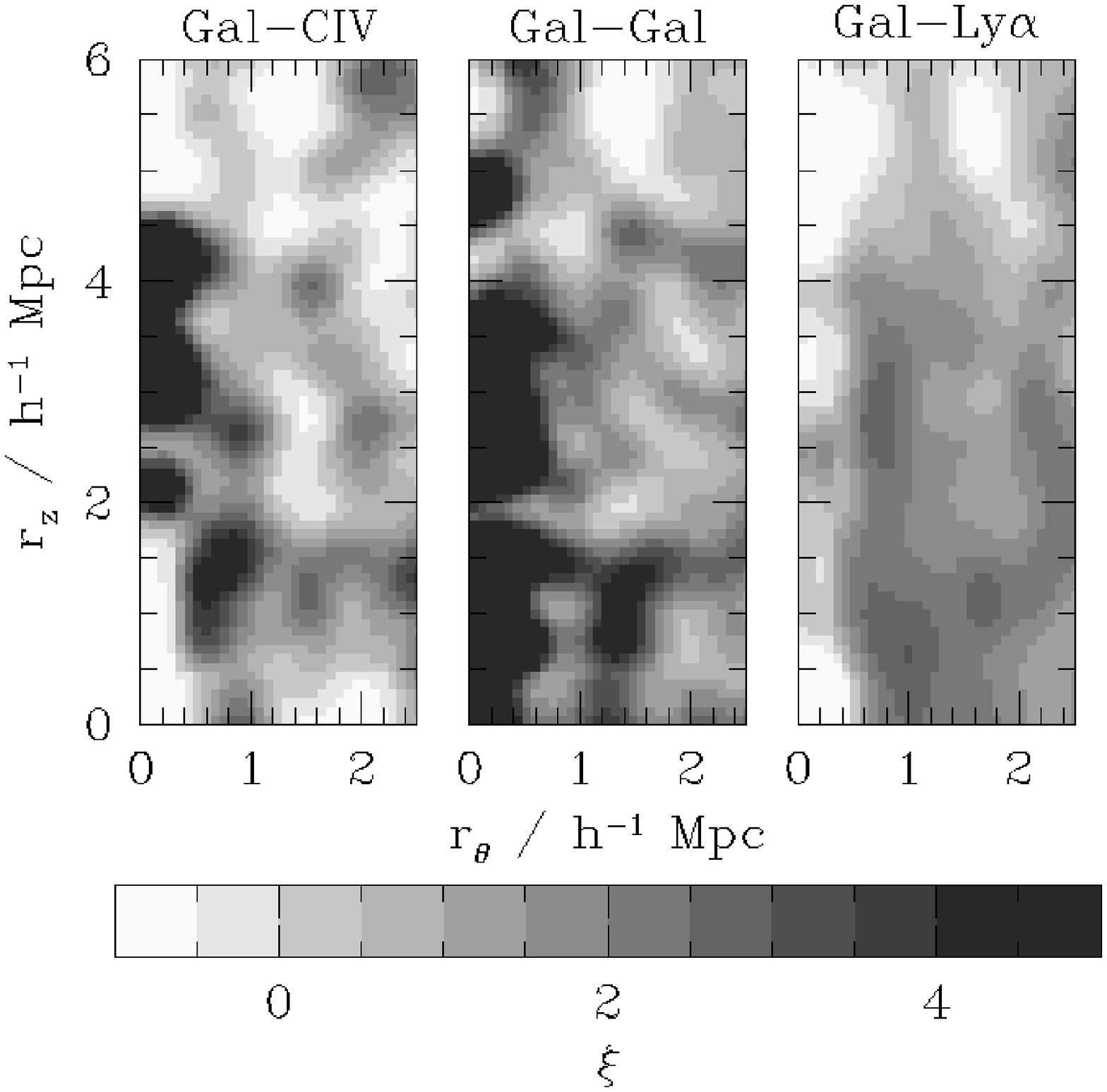}}
\figcaption[f21.eps]{
The correlation functions of figure~\ref{fig:corrfns} less heavily smoothed
($\sigma=0.2h^{-1}$ comoving Mpc).
\label{fig:corrfnszoom}
}
\end{figure}

Figures~\ref{fig:corrfns} and~\ref{fig:corrfnszoom}
show the two dimensional correlation functions
of galaxies with galaxies, of galaxies with detected CIV systems,
and of galaxies with Lyman-$\alpha$ forest flux decrements.  The
data used to generate the two figures were identical;
figure~\ref{fig:corrfnszoom}
is less heavily smoothed to help bring out the behavior of the correlation
functions on small spatial scales.  The
abscissae correspond to angular separations $r_\theta$ on the plane
of the sky; the ordinates correspond to redshift separations $r_z$.
Observed separations in arc-seconds and in redshift
were converted to the comoving distances
implied for
an $\Omega_M=0.3$, $\Omega_\Lambda=0.7$ cosmology.  The value of the correlation
function $\xi_{gc}$ between galaxies and CIV systems at separation $r_\theta$, $r_z$
was estimated with the statistic
\begin{equation}
\hat\xi_{gc} = (D_gD_c-D_gR_c-R_gD_c+R_gR_c) / R_gR_c
\label{eq:lsest}
\end{equation}
(e.g. Landy \& Szalay 1993) where $D_gD_c$ is the observed number of
galaxy--CIV-system pairs with separation $r_\theta, r_z$, $D_gR_c$ is
the number of pairs with the same separation between our galaxy catalog and
a random catalog of CIV systems, and $R_gR_c$ is the number of pairs with the
same separation between a random galaxy catalog and random CIV catalog.
The correlation functions $\xi_{gf}$ of galaxies with
Lyman-$\alpha$ flux decrements and $\xi_{gg}$ of galaxies with galaxies
were estimated similarly.  The random galaxy catalogs were generated by
assigning each galaxy in the true catalog 1000 redshifts drawn at random
from our selection function; the angular positions of galaxies in the
random and real catalogs were the same.  The random CIV catalogs were generated
by assigning each CIV system in the true catalog 1000 redshifts
drawn from a uniform distribution between the minimum and maximum redshifts
where our QSO spectra allowed us to detect CIV systems; the observed number density
of detected CIV systems among our QSO spectra at each redshift is closely
approximated by a constant once the different redshift selection ranges
are taken into account.
The random flux decrement
spectra were linear functions with the form of equation~\ref{eq:fbarvsz},
but scaled to match the observed
mean transmissivity of each QSO spectrum. 
The galaxy-galaxy correlation function was calculated using only data
in the redshift range $2.6<z<3.4$, to minimize the effect on our result
of the poorly determined wings of the selection function.  The calculation
of the galaxy-flux cross-correlation
function excluded data with $z<2.6$, with $z>3.4$,
with $z>z_{\rm QSO}-0.05$, and with $z< (1+z_{\rm QSO})\times 1026/1216 - 1$.
The final criterion excludes portions of the Lyman-$\alpha$ forest that
are contaminated by Lyman-$\beta$ absorption from material at higher redshifts.
Also excluded were data at the DLA-contaminated redshifts
 $2.91<z<2.98$ in SSA22 and $3.214<z<3.264$ in Q0933+2841.
The galaxy-CIV cross-correlation function calculation excluded
data with $z<2.6$, $z>3.4$, $z>z_{\rm QSO}-0.05$, or
$z<(1+z_{\rm QSO})\times 1216/1549 - 0.99$.  The last two criteria exclude CIV lines
that lie close to the QSO redshift or in the Lyman-$\alpha$ forest.
To help guide the eye, 
the resulting raw two-dimensional correlation functions 
were finally smoothed by two-dimensional Gaussians with $\sigma_\theta=\sigma_z=0.5h^{-1}$
(figure~\ref{fig:corrfns}) or $\sigma_\theta=\sigma_z=0.2h^{-1}$ (figure~\ref{fig:corrfnszoom})
Mpc comoving.


\begin{figure}[htb]
\centerline{\epsfxsize=9cm\epsffile{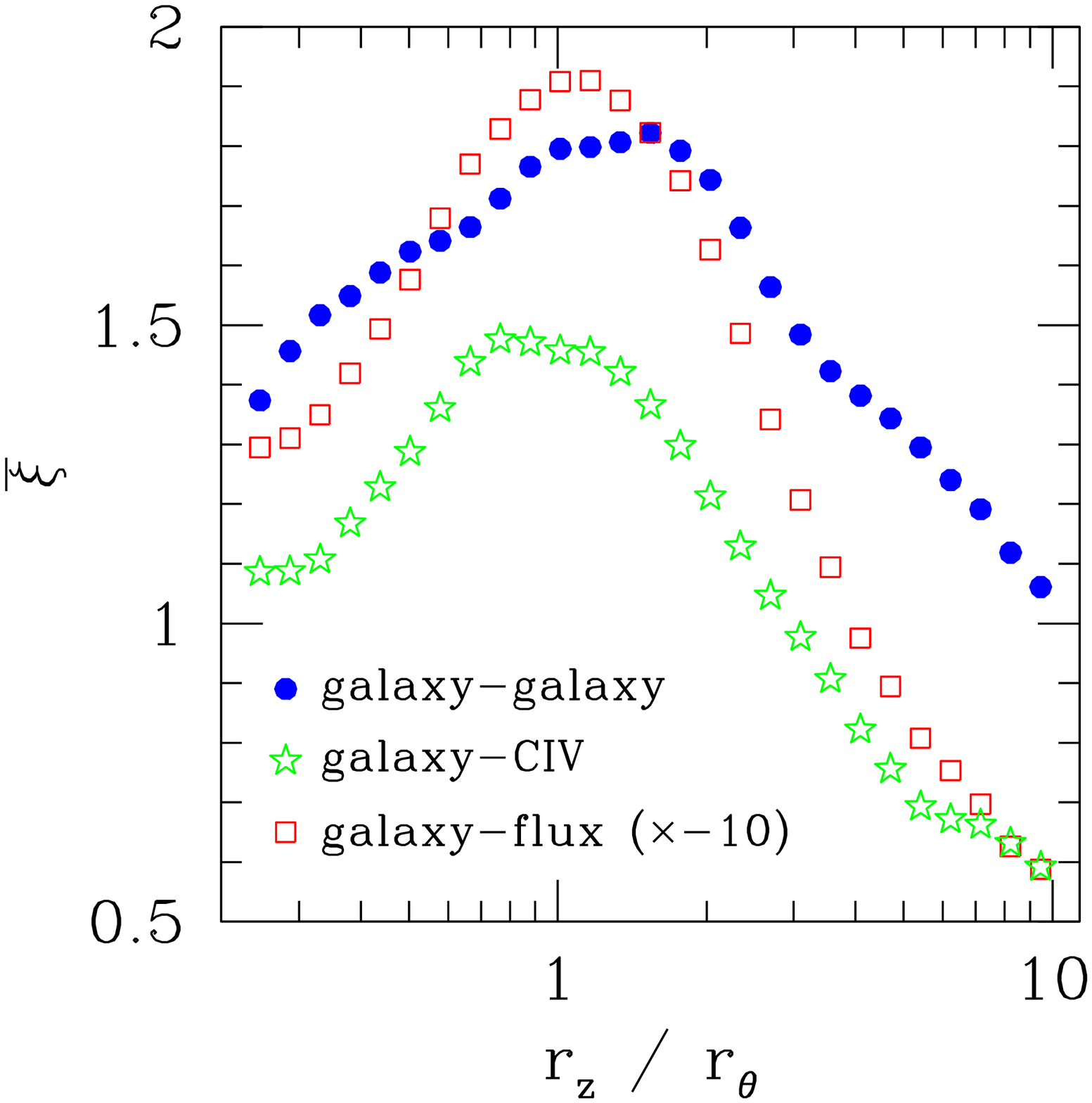}}
\figcaption[f22.eps]{
Average value $\bar\xi$ of the auto- and cross-correlation functions of figures~\ref{fig:corrfns}
and~\ref{fig:corrfnszoom} within ellipses of volume $4\pi(4h^{-1})^3/3$ comoving Mpc$^3$
as a function of the axial ratio $r_z/r_\theta$.  $\bar\xi$ reaches an extremum
close to $r_z/r_\theta=1$ for each correlation function, showing that they
are reasonably isotropic on these scales.
\label{fig:meanell}
}
\end{figure}

Aside from some elongation in the redshift direction at small separations,
which is at least partly due to errors in galaxy redshifts, these two-dimensional
correlation functions are reasonably isotropic.  This can be seen in 
figure~\ref{fig:meanell}.  The figure shows the average value of each correlation
function $\bar\xi$ in three-dimensional ellipses of comoving volume
$4\pi(4h^{-1}{\rm Mpc})^3/3$ as a function of axial ratio $r_z/r_\theta$.
The galaxy-galaxy correlation function is the most elongated in the redshift
direction, reaching a maximum of $\bar\xi$ at $r_z/r_\theta\sim 1$--$1.6$.
This is presumably because galaxy redshift errors have twice the impact and because close galaxy
pairs tend reside in dense parts of the universe where the pair-wise velocity
dispersion is highest.  But the two averaged cross-correlation functions both
hit an extremum at $r_z/r_\theta=1.0\pm 0.2$,
demonstrating their isotropy on $\sim 4h^{-1}$ Mpc scales.
If one were convinced that the cross-correlation functions should
be isotropic despite possible peculiar velocity distortions
and despite the significant errors in our galaxy redshifts,
figure~\ref{fig:meanell} would show that the Alcock-Paczynski (1979)
test at $z\sim 3$ favors a $\Lambda$ cosmology; if we had assumed
$\Omega_M=1.0$, $\Omega_\Lambda=0$ or $\Omega_M=0.3$, $\Omega_\Lambda=0$
instead of $\Omega_M=0.3$, $\Omega_\Lambda=0.7$, we would have found
that the averages hit their extrema at $r_z/r_\theta\sim 0.8$ not $r_z/r_\theta\sim 1.0$.
In any case, one unsurprising implication of the lack of redshift elongation in the
cross-correlation functions is that the bulk of
intergalactic absorbing material is not flowing outwards with respect to
nearby galaxies; the peculiar velocities of intergalactic material
relative to galaxies is not sufficient to make the galaxy--intergalactic material
cross-correlation functions be significantly more elongated than the
galaxy--galaxy correlation function.  There is no contradiction with the
galactic superwind hypothesis because a wind's evolution is dominated
by its stalling phase and because in any case only a small fraction of the intergalactic
volume would be affected by active winds at any time (e.g., \paperthreep).

The scale dependence of correlation functions is often easiest to apprehend when
they are presented as one-dimensional rather than two-dimensional functions.  It would
be trivial to average the correlation functions of figure~\ref{fig:corrfns} in circular
annuli, but because of peculiar velocities and redshift measurement errors, a more robust
estimate of the one-dimensional correlation functions will come from another approach.

\subsection{One dimensional}
\begin{figure}[htb]
\centerline{\epsfxsize=9cm\epsffile{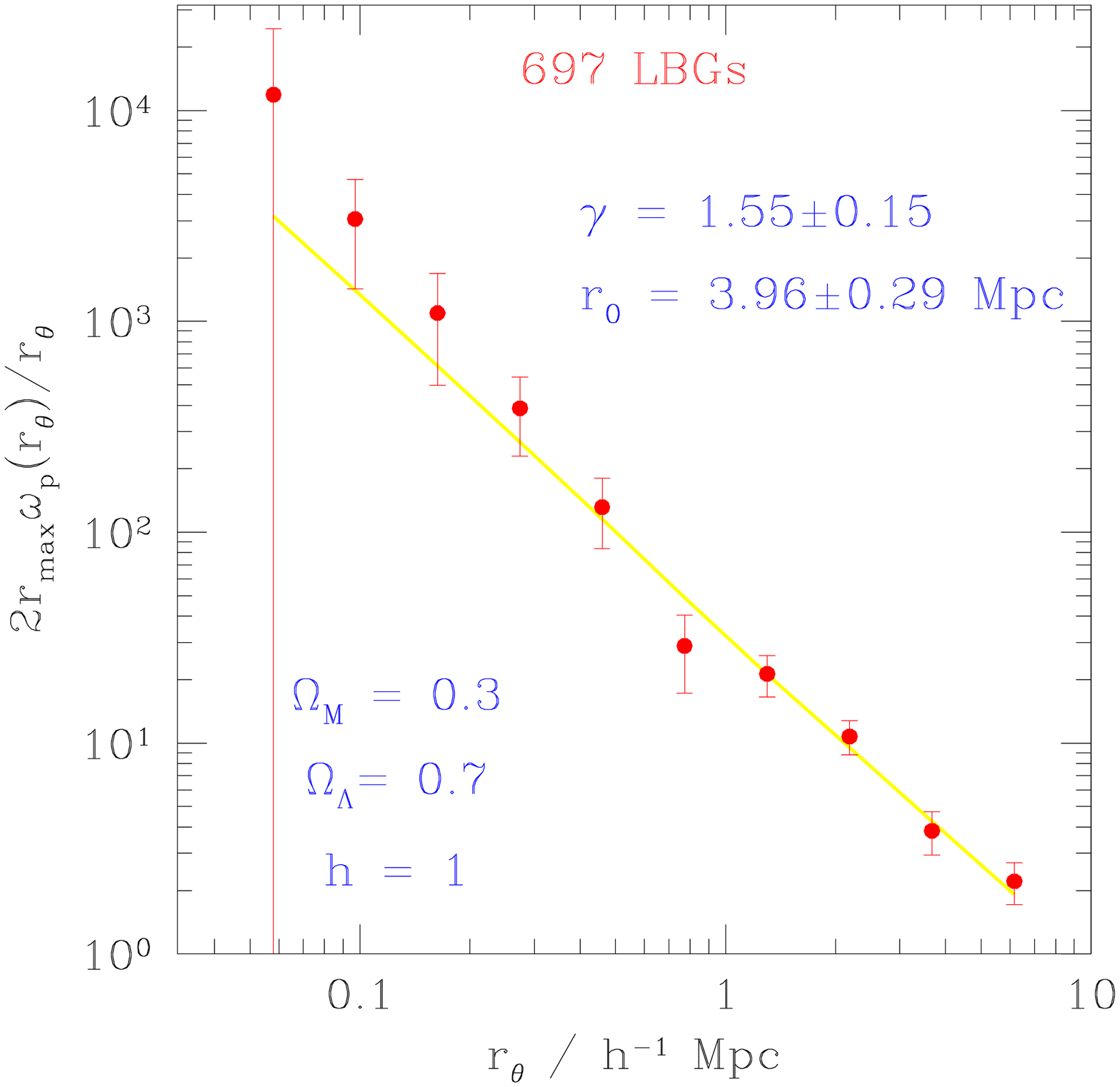}}
\figcaption[f23.eps]{
The projected correlation function of Lyman-break galaxies.  The parameters
of the best power-law fit to the three-dimensional correlation function
$\xi(r)=(r/r_0)^\gamma$ are shown.
\label{fig:apwp}
}
\end{figure}

\begin{figure}[htb]
\centerline{\epsfxsize=9cm\epsffile{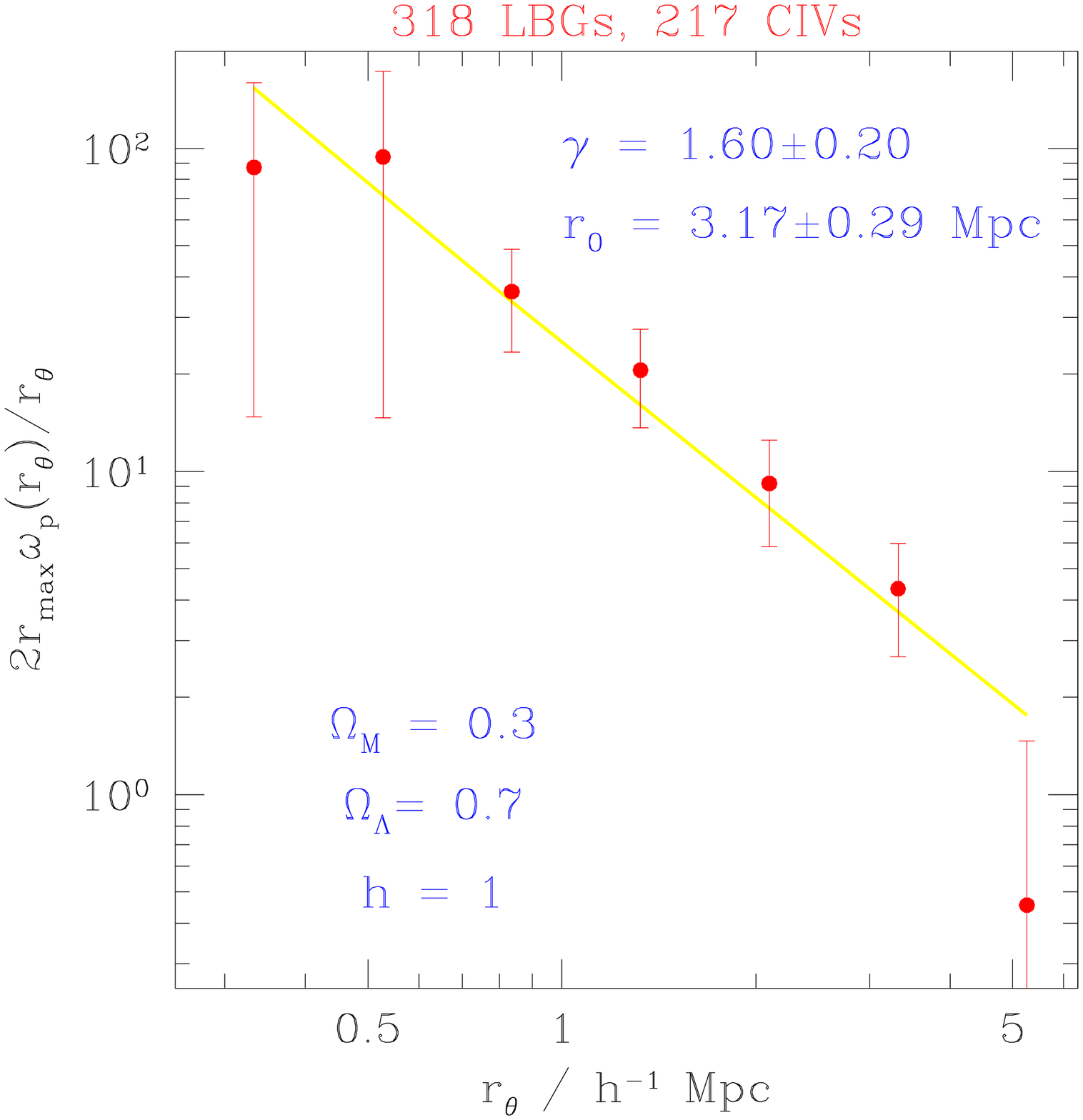}}
\figcaption[f24.eps]{
The projected cross-correlation function of CIV systems and Lyman-break galaxies.
\label{fig:wpcross}
}
\end{figure}

Appendix~\ref{sec:apwp} describes the approach adopted by Adelberger (2000)
to estimate the one-dimensional auto-correlation function 
of Lyman-break galaxies at $z\sim 3$.  Following Davis \& Peebles (1983),
we first estimated the projected correlation function, which is a marginalization
in the redshift direction of the two dimensional correlation function
(e.g., a marginalization of figure~\ref{fig:corrfns} or~\ref{fig:corrfnszoom}).
We then estimated the shape of the one-dimensional correlation function
from a power-law fit to the projected correlation function.
See appendix~\ref{sec:apwp}.  An identical approach can be used
to estimate the cross-correlation function of galaxies with CIV systems.
Figures~\ref{fig:apwp} and~\ref{fig:wpcross} show the auto-correlation function
of the $\sim 700$ galaxies with the most secure redshifts in the
Lyman-break sample of Steidel et al. (2003; in preparation) and
the cross-correlation function of galaxies and CIV systems in the sample
of this paper.  In both cases the data appear to be reasonably well
fit by a power-law of the form $\xi(r)=(r/r0)^{-\gamma}$,
justifying the method we have adopted.  The best-fit values and $1\sigma$ uncertainties
of $\gamma$ and $r_0$ are shown in the figures.


The galaxy-CIV cross-correlation function resembles the galaxy-galaxy
correlation function to a large extent.  The slopes are identical;
the correlation lengths differ by only $\sim 20$\%.  This suggests
that galaxies and CIV systems may be similar objects, a point
that has been made by Sargent, Steidel, \& Boksenberg (1988), by Quashnock \& Vanden Berk (1998),
and by us (\S~\ref{sec:smallciv} above) on slightly different grounds.

\begin{figure}[htb]
\centerline{\epsfxsize=9cm\epsffile{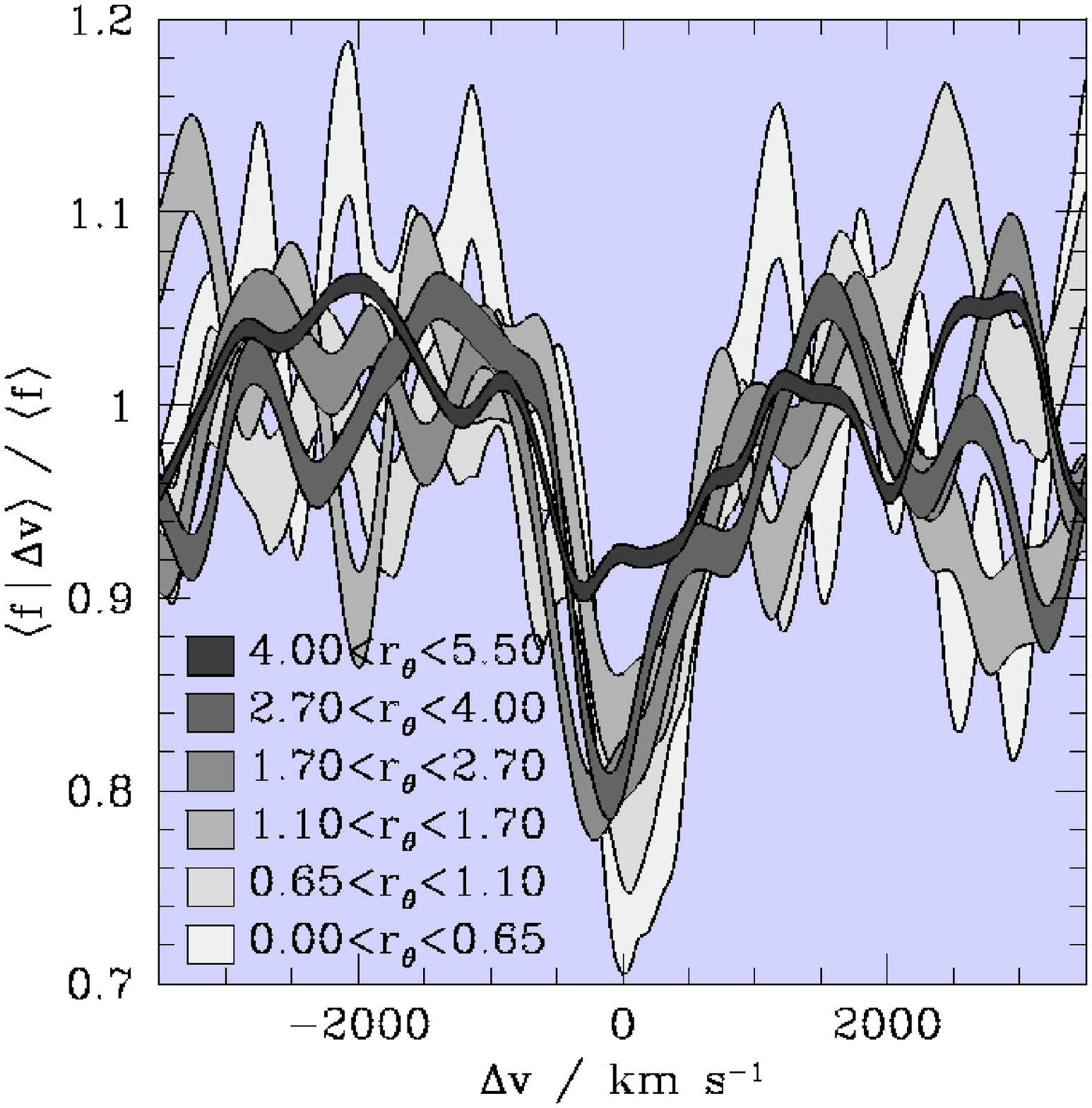}}
\figcaption[f25.eps]{
Cross-cuts through the galaxy-transmissivity cross-correlation
function (plus one) at different angular separations (as marked, in 
$h^{-1}$ comoving Mpc).  The line for $4<r_\theta<5.5h^{-1}$ Mpc (for example)
shows how the typical Lyman-$\alpha$ transmissivity of the intergalactic
medium varies along a line that passes a Lyman-break galaxy at
an impact parameter of $\sim 4$--$5$ Mpc.  The transmissivity is
near the global mean $\langle f\rangle$ at large redshift separations
from the galaxy, begins to fall within $\sim 1000 {\rm km s}^{-1}$ of
the galaxy, and reaches a minimum near the galaxy's redshift.  
Data for $r<0.65h^{-1}$ Mpc were calculated from the absorption that
foreground LBGs produce in the spectra of background LBGs; the galaxy
proximity effect should not be detected in this data because the
resolution of our galaxy spectra ($\sim 500{\rm km s}^{-1}$ FWHM,
or $\sim 4.5h^{-1}$ comoving Mpc) significantly exceeds the
$\simlt 1h^{-1}$ comoving Mpc diameter of the region around each galaxy that appears
to have a lowered HI content.
\label{fig:hi_vs_rthet}
}
\end{figure}

Unfortunately we were unable to estimate the cross-correlation
function between galaxies and Lyman-$\alpha$ transmissivity
with a similar approach.  The difficulty is that (after the first
$\sim$ Mpc) the HI content of the IGM appears to decline quite slowly
with increasing distance from a galaxy.
This can be guessed from figure~\ref{fig:twopanelgpea},
but is seen most robustly in figure~\ref{fig:hi_vs_rthet}, which shows
cuts along the redshift direction through the two-dimensional
galaxy-transmissivity cross-correlation function.  To improve 
the statistics, data
for $r_\theta<1.1h^{-1}$ Mpc were estimated from the absorption
that foreground LBGs produce in the spectra of background LBGs (see \S~\ref{sec:ism}
for more details);
at larger separations the
data are the same as in figure~\ref{fig:corrfns}.
One must advance to a impact parameter $r_\theta\simgt 4h^{-1}$ comoving Mpc
before the net Lyman-$\alpha$ absorption within velocity separations
$-600<\Delta v<600{\rm km s}^{-1}$ decreases to half its value
at the smallest impact parameters $r_\theta\simlt 1h^{-1}$ Mpc.
This implies that the slope $\gamma_{\rm gf}$ of the galaxy/transmissivity
correlation function lies precariously close to the values
$\gamma_{\rm gf}\leq 1$ that cause our statistical deprojection (equation~\ref{eq:apwp})
to break down entirely.
Although the best formal fit of a power-law correlation function to
the data in Figure~\ref{fig:hi_vs_rthet} (marginalized by $\pm 6h^{-1}$ comoving
Mpc, or $\sim\pm 670 {\rm km s}^{-1}$) has
$r_0\sim 0.49h^{-1}$ comoving Mpc, $\gamma\sim 1.15$, 
the data are almost equally compatible with $\gamma_{\rm gf}\leq 1$,
and that complicates our attempts to place confidence intervals on the
parameters.  In any case the best fit parameters
depend uncomfortably on the distance over which
we marginalize the data.  The problem is that the measured mean transmissivity
stays slightly lower than the global mean even at large distances
from galaxies (cf. figure~\ref{fig:twopanelgpea}), and it is unclear
whether this is part of the true correlation function or merely
a $\sim$ percent level systematic error due (e.g.) to our continuum
fitting. 
Fitting the two-dimensional data directly with
a power-law smoothed by a Gaussian in the redshift direction
does not help.  None of our fits filled us with complete confidence.
We shall defer a proper estimate
of the galaxy-transmissivity correlation function.  

\begin{figure}[htb]
\centerline{\epsfxsize=9cm\epsffile{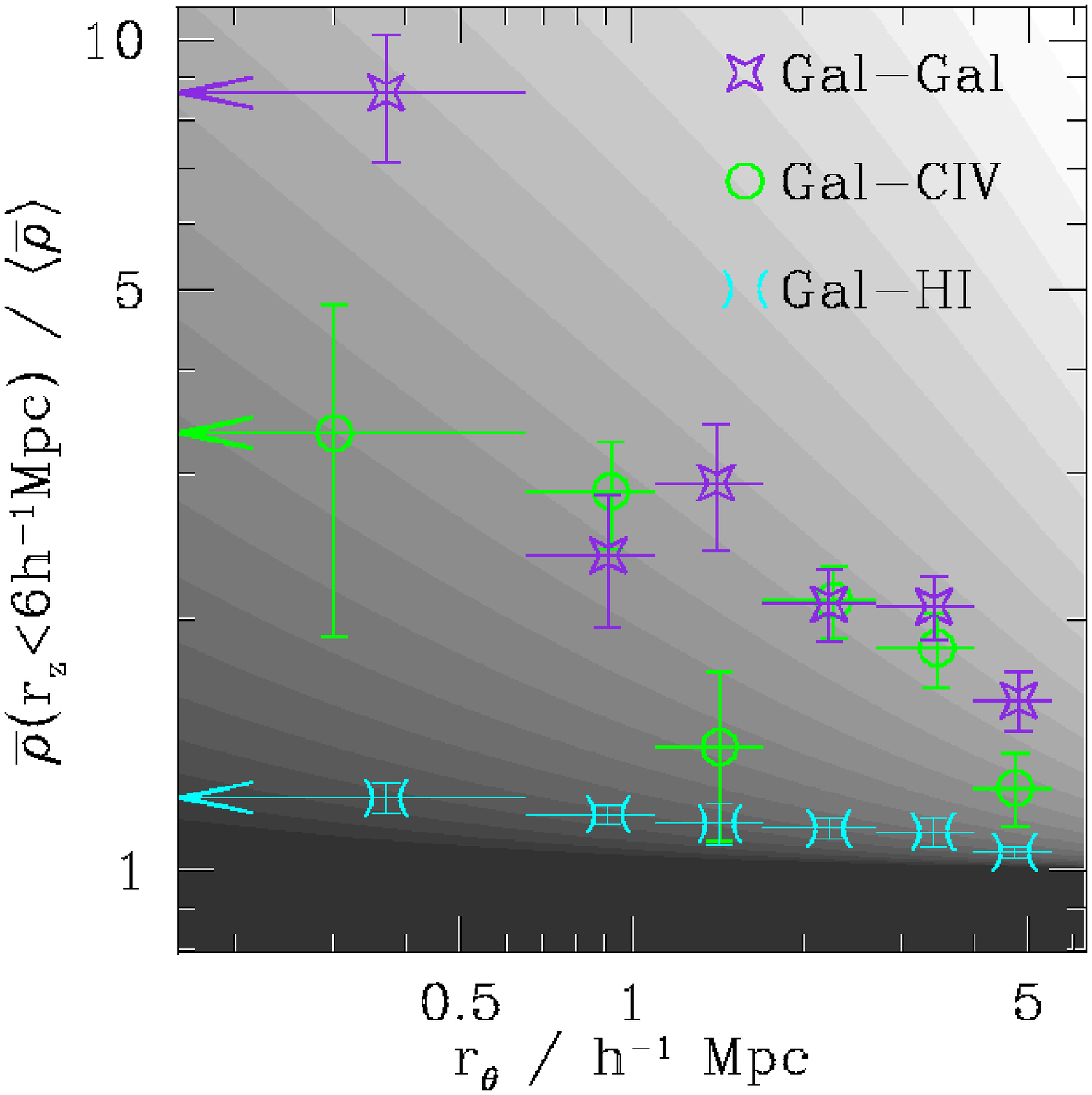}}
\figcaption[f26.eps]{
Marginalizations over $0<r_z<6h^{-1}$ comoving Mpc of the data shown
in figure~\ref{fig:corrfns}.  Points with error bars show the
marginalized galaxy-galaxy, galaxy-CIV system, and galaxy/Ly-$\alpha$
transmissivity correlation functions plus one.  The plot shows
(for example) that a cylinder of depth $r_z=\pm 6h^{-1}$ Mpc
and radius $r_\theta=0.65h^{-1}$ Mpc centered on a typical
LBG will contain $\sim 9$ times the
number of (other) LBGs and $\sim 3.5$ times the number of CIV
systems as a similar cylinder randomly placed.  Other bins
of $r_\theta$ on the plot correspond to volumes
that are cylindrical annuli centered on LBGs rather than cylinders.
To improve the statistics,
the galaxy-HI measurement at the smallest separation was estimated
from the absorption that foreground LBGs produce in the spectra
of background LBGs; see \S~\ref{sec:ism}.
Striations in the background show how this plot would appear
if the unmarginalized density had the form $\rho(r)\propto 1+(r/r_0)^{-1.6}$.
\label{fig:xigcgg}
}
\end{figure}

The uncertainty in the shape of the galaxy-flux correlation
function does not stop us from making one final observation:  the ratio
of CIV to HI in the intergalactic medium almost certainly increases
as one approaches a Lyman-break galaxy.  The mean Lyman-$\alpha$
transmissivity of the intergalactic medium does not change
much with distance from a Lyman-break galaxy, 
only by $\simlt 30$\% (cf. figure~\ref{fig:twopanelgpea}),
so it may be a reasonable first approximation to assume that
the density of intergalactic HI is roughly independent of distance
to a Lyman-break galaxy.  But the number density of CIV systems
depends very strongly on distance to a Lyman-break galaxy;
according to the power-law fit described above, the density of CIV
systems is $\sim 6.4$ times higher at $r=1h^{-1}$ comoving Mpc
than at $r=10h^{-1}$ Mpc.  Readers may wish to refer to
Figure~\ref{fig:xigcgg}, which shows marginalizations by $\pm 6h^{-1}$ comoving
Mpc in the redshift direction over the two-dimensional correlation
functions of Figure~\ref{fig:corrfns}.  The marginalization
reduces the differences between the correlation functions somewhat,
but nevertheless they are clear.
The significantly different strengths
of the galaxy-CIV and galaxy-transmissivity correlation functions make it
seem almost inevitable that the ratio of intergalactic CIV to HI density
increases dramatically close to Lyman-break galaxies---though
we should emphasize again that neither the mean Lyman-$\alpha$ transmissivity
nor the number density of CIV systems is related in a simple way to column density.
One may demonstrate through simple calculations, omitted at the referee's
request, that increasing the intensity of the radiation field
at $z\sim 3$ would lead to a decrease in the CIV content of
most of the Lyman-$\alpha$ forest.  This shows that radiation from
Lyman-break galaxies is unlikely to account for any enhancement in
CIV density near to them.  Figure~\ref{fig:civohi} shows
that plausible changes in the intergalactic density and temperature
near the galaxies would be unlikely to change $n_{\rm CIV}/n_{\rm HI}$
by a factor of $\sim 6$.  We conclude that the simplest explanation for the
(apparently) strong change of $n_{\rm CIV}/n_{\rm HI}$ with radius
may be that the intergalactic metallicity is
higher near the galaxies.  See Steidel et al. (2003; in preparation)
for a less superficial analysis.

\section{SUMMARY}
The goal of this paper was to compare the spatial distributions of galaxies,
metals, and neutral hydrogen at high redshift.  We conducted a redshift survey
of $z\sim 3$ Lyman-break galaxies in 6 primary fields containing
QSOs at $3.1\simlt z\simlt 3.6$ whose spectra revealed the locations
of HI and metals along sightlines through the galaxy distributions.
We quantified the relationship between galaxies and intergalactic material
by calculating two-point correlation functions or closely
related statistics.  Various averages over the correlation functions
were presented in \S\S~\ref{sec:large} and~\ref{sec:gpe};
the one-dimensional and two-dimensional correlation functions of
galaxies with galaxies, with CIV systems, and with Lyman-$\alpha$ forest
transmissivity were presented in \S~\ref{sec:corrfns}.  The most lasting
contribution of this paper may be its methodology, which
will be applied to a larger sample in the future.  Applying these
statistical methods to the present sample leads to the
following principal conclusions:

\smallskip

1.  The HI content of the intergalactic medium is closely correlated with
the positions of nearby galaxies.  The material within $r\sim 0.5h^{-1}$ comoving Mpc
of Lyman-break galaxies absorbs Lyman-$\alpha$ photons less heavily
than the material at random points
in the intergalactic medium (\S\S~\ref{sec:smallhi},~\ref{sec:gpe},
and~\ref{sec:corrfns}); the material
at slightly larger radii $1\simlt r\simlt 5h^{-1}$ Mpc absorbs Lyman-$\alpha$ photons
more heavily (\S\S~\ref{sec:largehi}
and~\ref{sec:corrfns}) than average.  Interested readers will find in Paper II
a more detailed discussion of the correlation
between galaxies and intergalactic HI on large spatial scales.

\smallskip

2.  The gas within the largest galaxy overdensities at $z\sim 3$, presumably
the young intracluster medium, is rich in HI and CIV 
(\S\S~\ref{sec:largehi},~\ref{sec:largeciv}, and~\ref{sec:corrfns}).

\smallskip

3.  The cross-correlation function of galaxies with CIV systems is similar to
the auto-correlation function of galaxies with galaxies. 
Both appear to be power-laws; the best fit parameters to a correlation
function of the form $\xi(r) = (r/r_0)^{-\gamma}$ are
$r_0=3.2\pm 0.3$, $4.0\pm 0.3$ comoving $h^{-1}$ Mpc and
$\gamma=1.60\pm 0.20$, $1.55\pm 0.15$ for the galaxy-CIV and galaxy-galaxy
correlation functions respectively ($1\sigma$; $\Omega_M=0.3$, $\Omega_\Lambda=0.7$).
This shows that
CIV systems and Lyman-break galaxies are found in similar parts of the universe
and suggests that they may be similar objects (\S~\ref{sec:corrfns}).

\smallskip

4.  The ratio of the number density of CIV systems 
to HI density (\S~\ref{sec:corrfns}) and to HII density (\S~\ref{sec:largeciv}) almost certainly
increases within a few comoving Mpc of Lyman-break galaxies.  Most detectable
CIV systems ($N_{\rm CIV}\simgt 10^{11.5}$ cm$^{-2}$) lie within
$2.4h^{-1}$ comoving Mpc of a Lyman-break galaxy (\S~\ref{sec:smallciv}).  Stronger
CIV systems tend to lie even closer.  The strongest
CIV systems ($N_{\rm CIV}\simgt 10^{13.5} {\rm cm}^{-2}$) appear to be so strongly
correlated with nearby ($r\simlt 0.5h^{-1}$ comoving Mpc) Lyman-break galaxies
that they should be considered the same object---despite the large distance
between the galaxy and CIV-absorbing gas (\S~\ref{sec:smallciv}).

\smallskip

5. Damped Lyman-$\alpha$ systems appear to reside in different environments than
Lyman-break galaxies (\S~\ref{sec:largedla}).  Though the sample is small
and the statistics are poor, the available data suggest that they should not be considered
similar objects (cf. Fynbo et al. 1999, Haehnelt et al. 2000, Gawiser et al. 2001).

\smallskip

6. One of our surveyed volumes surrounds the line of sight to Q0302-0019, a QSO
whose spectrum reveals patchy HeII absorption.
Comparing the HeII spectrum of this QSO with our galaxy and CIV data
suggests that
HeII tends to be absent in regions of higher
galaxy density and in regions with a higher incidence of CIV absorption (\S 3.3).  Presumably it has been ionized to HeIII.

\smallskip

Readers may find in these results some indications
that we have observed powerful superwinds emerging from Lyman-break
galaxies.  In particular, the lack of HI within $\sim 0.5h^{-1}$ Mpc
of the galaxies and excess of CIV at slightly larger radii
could be neatly explained if
the $\sim 600$ km s$^{-1}$ outflows common to Lyman-break galaxies
(\S~\ref{sec:data}; Pettini et al. 2001, 2002) are able to
escape the galaxies' potentials and maintain their large velocities
throughout the galaxies' $\sim 300$Myr (Shapley et al. 2001)
star-formation timescale.  This idea will be considered more
fully in subsequent papers of our series.  The present
paper is merely a first report.

\bigskip
\bigskip
Every paper benefits from the work of others.
Our debts are especially numerous.  The referee, M. Rauch,
read the paper closely, offered many perceptive comments,
and graciously tolerated the extended outburst of poor taste that appears
in \S~1.
R. Croft suggested making the plot that revealed the galaxy proximity
effect.  Conversations with T. Abel were
a phenomenal source of inspiration, education, and 
amusement during the middle stages
of this project.  They shaped this paper greatly.
C. Metzler patiently fielded numerous questions in the middle
of the night.  J. Prochaska provided information about the damped system
in Q0933+2845.  P. Molaro gave us a reduced spectrum of Q0000-2620.
M. Dickinson, D. Erb, M. Giavalisco, M. Hunt and M. Kellogg
helped take and reduce some of the data.
W. Sargent generously shared his QSO spectra with us.
Financial and enological support 
from the Harvard Society of Fellows sustained KLA during
his hapless lucubrations.
CCS and AES were supported by grant AST0070773
from the U.S. National Science Foundation and by
the David and Lucile Packard Foundation.
Our results relied in part on public data released
from UVES Commissioning observations at the VLT Kueyen telescope.
This research made use of the NASA/IPAC Extragalactic Database (NED),
which is operated by the Jet Propulsion Laboratory, California Institute
of Technology, under contract with the National Aeronautics and Space
Administration.
The authors wish to extend special thanks to those
of Hawaiian ancestry for graciously allowing
telescopes and astronomers upon their sacred mountaintop.
Without their indurable hospitality, few of the observations
presented here would have been possible. 

\onecolumn\appendix

\section{A LIMIT ON THE ASSOCIATION OF TWO RANDOM FIELDS}
\label{sec:apgenlim}
Consider two standard ways of picturing the relationship between
galaxies and intergalactic metals.  The metals might
lie only within some
radius of the galaxies---for example, within the galaxies' extended
gravitational potentials or within the largest radius reached
by the galaxies' winds.  Or they might be truly intergalactic
but still statistically correlated with galaxies because
galaxies are found in parts of the universe where the
density of other sorts of matter is high.
One can safely assert that the truth lies between these
two extremes.  But one need not stop with this meek conclusion.
Our goal is to illustrate that it is possible, at least in principle,
to tell which of the two pictures lies closer to the truth.
Various approaches could be taken; ours is to derive a statistical
inequality that would be satisfied by the cross-correlation function
of galaxies and CIV systems in the second picture but not necessarily
the first.  The inequality is violated by our data, suggesting
a direct connection between intergalactic metals and
nearby galaxies in at least some cases.

We begin by deriving the inequality.
Let $f({\mathbf r})$, $g({\mathbf r})$, $f_k({\mathbf k})$, and
$g_k({\mathbf k})$ be two arbitrary random fields and their
Fourier transforms.  $f({\mathbf r})$ and $g({\mathbf r})$
might represent the density of CIV systems and of galaxies at spatial
position ${\mathbf r}$.
The Cauchy-Schwarz inequality in the form
\begin{equation}
|\langle a^\ast b\rangle|^2 \leq \langle a^\ast a\rangle \langle b^\ast b\rangle
\label{eq:apcscomplex}
\end{equation}
leads to the following two inequalities:
\begin{equation}
\xi_{f\!g}^2({\mathbf r}) \leq \xi_{f\!f}(0)\xi_{gg}(0),
\label{eq:apcrosscorrlim}
\end{equation}
where $\xi_{f\!f}$ is the auto-correlation function of $f$ and
$\xi_{f\!g}$ is the cross-correlation function of $f$ with $g$, and
\begin{equation}
|P_{f\!g}({\mathbf k})|^2 \leq P_{f\!f}({\mathbf k}) P_{gg}({\mathbf k}),
\label{eq:apcrossspeclim}
\end{equation}
where $P_{f\!f}$ is the power-spectrum of $f$ and $P_{f\!g}$ is the cross-spectrum
of $f$ and $g$.  Equations~\ref{eq:apcrosscorrlim} and~\ref{eq:apcrossspeclim}
result from inserting $a=f({\mathbf r'})$, $b=g({\mathbf r}-{\mathbf r'})$
and $a=f_k({\mathbf k})$, $b=g_k({\mathbf k})$ into equation~\ref{eq:apcscomplex}.

Equation~\ref{eq:apcrosscorrlim} does not appear to be particularly
useful, because the RHS is independent of $r$ and depends
on the value of the correlation function
at zero lag, which is often difficult to estimate.  Fortunately the 
possible association of $f$ and $g$ in real space can be limited
by a large number of other inequalities.
Consider, for example, the mean value of the correlation function $\xi_{fg}$
weighted by an arbitrary function $W({\mathbf r})$,
\begin{equation}
\bar\xi_{f\!g} \equiv \int d^3r\, \xi_{f\!g}({\mathbf r}) W({\mathbf r}) \bigg/ \int d^3 r W({\mathbf r}).
\label{eq:apbarxidef}
\end{equation}
If $W$ were equal to a constant for $|{\mathbf r}|<r_{\rm sph}$ and to zero elsewhere,
$\bar\xi_{f\!g}$ would be the mean value of the correlation function $\xi_{f\!g}$ within a sphere of
radius $r_{\rm sph}$.  After writing $\xi_{f\!g}$ as the Fourier
transform of the cross-spectrum and applying Fubini's theorem, 
equation~\ref{eq:apbarxidef}
becomes
\begin{equation}
\bar\xi_{f\!g} \propto \int d^3k\, P_{f\!g}({\mathbf k}) W_k^\ast({\mathbf k})
\label{eq:apbarxifourier}
\end{equation}
or
\begin{equation}
\bar\xi_{f\!g}^2 \propto \bigg|\int d^3k\, \langle f_k({\mathbf k})g_k^\ast({\mathbf k})\rangle W_k^\ast({\mathbf k})\bigg|^2
\label{eq:apbarxifourier2}
\end{equation}
where $W_k({\mathbf k})$ is
the Fourier transform of $W({\mathbf r})$.  If $W_k({\mathbf k})$ is real and
positive, $W_k({\mathbf k})^{1/2}$ can be absorbed into each of $f_k$ and $g_k$
in equation~\ref{eq:apbarxifourier2}, leading through equation~\ref{eq:apcscomplex}
to the inequality
\begin{equation}
\bar\xi_{f\!g}^2 \leq \bar\xi_{f\!f}\bar\xi_{gg}
\label{eq:apbarxiineqnoshot}
\end{equation}
which holds only if the correlation functions are averaged in a suitably chosen
way.  Because the Fourier transform of a Gaussian centered on the origin
is real and positive, correlation functions averaged within a Gaussian volume
must obey inequality~\ref{eq:apbarxiineqnoshot}.

We have assumed so far that $f$ and $g$ are smoothly varying functions.
In fact galaxies and CIV systems are (nearly) discrete objects; if we
divided all space into small cells, we would find an occasional
cell that contained one galaxy and/or one CIV system but most cells
would be empty.  How does this affect inequality~\ref{eq:apbarxiineqnoshot}?
It depends.  
Imagine
underlying and unobserved continuous functions $f'$ and $g'$
defined such that the mean number of observed particles
in places where the continuous functions take on a given value
is equal to that value: $E(f|f')=f'$, $E(g|g')=g'$, with
$E(x|y)$ the expectation value of $x$ given $y$.
The continuous functions will clearly satisfy 
inequality~\ref{eq:apbarxiineqnoshot}.
Peebles (1980, \S33) has shown that the auto-correlation functions of
$f$ and $g$ are identical to the auto-correlation functions of $f'$ and $g'$.
Their cross-correlation function depends on $\langle fg\rangle$, which
may be written
\begin{equation}
\langle fg\rangle = \int df'\,dg'\,\,P(f'g') E(fg|f'g')
\label{eq:apexpfg}
\end{equation}
where $P(f'g')$ is the probability that the functions $f'$, $g'$
simultaneously take on the values $f'$, $g'$ and
$E(fg|f'g')$ is the expectation value of the product $fg$ given
the values of $f'$ and of $g'$.  If $P(fg|f'g')=P(f|f')P(g|g')$,
then $E(fg|f'g')=E(f|f')E(g|g')=f'g'$ and
substitution into equation~\ref{eq:apexpfg} shows that
$\langle fg\rangle = \langle f'g'\rangle$.  This implies that
the cross-correlation function of
$f$ and $g$ will be equal to the 
cross-correlation function of 
$f'$ and $g'$.  Since the auto-correlation functions
are equal as well, we conclude that particle distributions $f$ and $g$
related via this Poisson process to the continuous functions $f'$ and $g'$
will obey inequality~\ref{eq:apbarxiineqnoshot}
provided $P(fg|f'g')=P(f|f')P(g|g')$. 
It is worth giving some thought to the meaning of
this condition.  We introduced the continuous functions $f'$ and $g'$
as a mathematical aid to calculating the particle distributions'
auto- and cross-correlation functions, but
suppose momentarily 
that the particle distributions $f$ and $g$ were in fact generated
from the continuous functions $f'$ and $g'$ through some
stochastic process to be specified, for example by associating
an independently and identically distributed random number with
each volume element $dV$ and then placing a particle in $f$
in every volume element where the random number was less
than $f'({\mathbf r})dV$.  
If $P(fg|f'g')=P(f|f')P(g|g')$, then
the job of constructing the particle
realizations $f$ and $g$ of the continuous functions
$f'$ and $g'$ could be completed
in the two independent steps of generating $f$ from $f'$
and $g$ from $g'$.
This condition should be appropriate if metal-line absorption systems
were truly intergalactic, associated with galaxies only to the
extent that both traced the same large-scale structure, because
then one might imagine that the locations of galaxies and of
metal-line systems were determined primarily by the large scale
distribution of matter without much regard to each other.

It would not be appropriate if there were a stronger correspondence
of galaxies to metal-line absorbers.
Suppose we were asked to design the intergalactic medium
for a universe where most metal-line
absorption was produced by galaxies' superwinds, for example.
We clearly could not begin until we were told precisely where amid
the jumbled large-scale structure the galaxies happened to lie.
A counterpart of equation~\ref{eq:apbarxiineqnoshot} for
the general particle case with $P(fg|f'g')\neq P(f|f')P(g|g')$
can be derived as follows.
Our original derivation relied on the assumption
that the correlation function $\xi({\mathbf r})$ and power-spectrum
$P({\mathbf k})$ would be Fourier transforms of each other.
When a random field is composed of discrete objects,
however, the definition of the correlation function is customarily
modified to exclude the correlation
of an object with itself.  In this case the power-spectrum is
equal to the Fourier transform of the correlation function plus
an additive constant (see, e.g., Peebles 1980 \S 41), and 
equation~\ref{eq:apbarxiineqnoshot} accordingly becomes
\begin{equation}
\bar\xi_{gf}^2 \leq (\bar\xi_{ff} + 1/\bar n_f) (\bar\xi_{gg} + 1/\bar n_g),
\label{eq:apbarxiineqshot}
\end{equation}
where $n_f$ is the global number density of particles in $f$
and $\bar n_f\equiv n_f\int d^3r\,W({\mathbf r})/W(0)$,
rather than~\ref{eq:apbarxiineqnoshot}.  Equation~\ref{eq:apbarxiineqshot}
is the statistical inequality that must be obeyed by the cross-correlation
function of any two arbitrary particle distributions.
Since $\bar n_f>0$ and $\bar n_g>0$ for the weighting
functions $W({\mathbf r})$ that we are considering, it is less restrictive
than equation~\ref{eq:apbarxiineqnoshot}.  This reflects the obvious
fact that if the positions of particles in $f$ and $g$ are allowed
to influence each other, we may obtain a stronger cross-correlation
than otherwise.  

The difference between these two inequalities provides one way of
testing whether an observed association of galaxies with CIV absorbers
requires the observed galaxies to have influenced
the intergalactic CIV content\footnote{Or, equally likely
a priori, for the intergalactic metal content to have
influenced galaxy formation} in some way:  calculate the
mean value of the cross-correlation function within a Gaussian
volume and see if it violates equation~\ref{eq:apbarxiineqnoshot}.
This was the approach of \S~\ref{sec:smallciv}.
Perhaps an example more mundane than galaxies and metals
can help clarify the difference between particle distributions
whose cross-correlation satisfies~\ref{eq:apbarxiineqnoshot}
and those whose cross-correlation satisfies~\ref{eq:apbarxiineqshot}.
Although pedestrians,
eyeglasses, and fire hydrants all tend to be found in similar
places on the earth's surface, the cross-correlation function
of pedestrians and eyeglasses obeys~\ref{eq:apbarxiineqshot}
while (except in very unfortunate circumstances) the
cross-correlation function of pedestrians and fire hydrants
obeys~\ref{eq:apbarxiineqnoshot}.  Not all pedestrians
wear eyeglasses, and not all eyeglasses are on pedestrians,
but the locations of at least some eyeglasses
are determined by where pedestrians happen to be standing.
The locations of fire hydrants are not. 

We conclude with two obvious caveats.  
First, even if the global means satisfied the relationship
$E(fg|f'g')=E(f|f')E(g|g')$, the observed means for any small
sample might not.  Consequently statistical fluctuations will cause
violations of inequality~\ref{eq:apbarxiineqnoshot}
among small sub-samples of a population
that obeys the inequality as a whole.  Reasonably large samples, far
larger than the one in this paper, will be required to make this sort
of test convincing.  Second, even if it were shown that
the cross-correlation function of galaxies and metal-line
systems does not violate~\ref{eq:apbarxiineqnoshot}, this
would not justify the conclusion that the observed galaxies and
observed metals exerted no influence on each other.  It would
show only that the influence was not dominant on the
chosen spatial smoothing scale.

\section{ESTIMATING AN OVERDENSITY}
\label{sec:apdensest}
Suppose we would like to estimate the universal average overdensity
of Lyman-break galaxies in cells centered on CIV systems
from a measurement of $\{N_i\}$, the number
of Lyman-break galaxies in cells centered on each of the ${\cal N}_c$
CIV systems in our sample.  What estimator should
we use?  If an average of $\mu$ galaxies were expected
to lie in each CIV system's cell in the absence of a correlation between
galaxies and CIV systems, a natural choice for the estimator
would be
\begin{equation}
\bar\xi_{gc} = \frac{1}{{\cal N}_c}\sum_{i=1}^{{\cal N}_c} N_i/\mu\quad -\quad 1,
\end{equation}
the mean of the observed galaxy overdensity $(N_i-\mu)/\mu$ around each
of the $i$ CIV systems.  But in practice we would not expect
to observe $\mu$ Lyman-break galaxies around each CIV system
if galaxies and CIV systems were independently distributed; because
the selection function of our galaxy survey peaks around $z\simeq 3.0$, 
more Lyman-break galaxies should lie close to
a CIV system at $z=3.0$ than a CIV system at (say) $z=2.2$ or $z=3.5$.
The expected number of galaxy neighbors is not a constant but
instead depends on each CIV system's redshift, and we must
estimate the global mean overdensity not from $\{N_i\}$ and $\mu$
but from $\{N_i\}$ and $\{\mu_i\}$ where $\mu_i$ is the expected
overdensity around the $i$th CIV system given the system's redshift
and the shape of our selection function.  What estimator should
we use now?

Let $\bar\xi_{gc}$ be the
(unknown) 
global mean overdensity of galaxies in cells centered on CIV systems,
and let $n_i\equiv\mu_i(1+\bar\xi_{gc})$ be the expected number of galaxies
in a cell centered on a CIV system that lies at redshift $z_i$.
Approximating the probability of observing
$N_i$ galaxies when $n_i$ were expected with
the Poisson distribution, $P(N_i|n_i)=e^{-n_i}n_i^{N_i}/N_i!$,
and assuming that different cells are independent, one finds that
the probability of
observing the set of actual $\{N_i\}$ and expected $\{n_i\}$
galaxy numbers in cells surrounding CIV systems
is equal to the product of the individual probabilities:
\begin{equation}
P(\{N_i\}|\{n_i\}) = \exp\biggl[-(1+\bar\xi_{gc})\sum_i\mu_i\biggr] \\
        \biggl(1+\bar\xi_{gc}\biggr)^{\sum_iN_i} \prod_i\mu_i^{N_i}/N_i!.
\label{eq:poissonprob}
\end{equation}
The $\bar\xi_{gc}$ dependence of equation~\ref{eq:poissonprob}
is identical to the $\bar\xi_{gc}$ dependence of a single Poisson
distribution of the form $P(x|\bar x) = e^{-\bar x} {\bar x}^x/x!$
with $x\equiv \sum_iN_i$ and $\bar x\equiv(1+\bar\xi_{gc})\sum_i\mu_i$.
Since the Poisson likelihood $P(x|\bar x)$ is maximized for
$x=\bar x$, we can see immediately that we should adopt
\begin{equation}
\bar\xi_{gc} = \frac{\sum_iN_i}{\sum_i\mu_i} - 1
\label{eq:barxi_maxlik}
\end{equation}
as the maximum likelihood estimator of $\bar\xi_{gc}$.  In practice
the assumptions leading to equation~\ref{eq:barxi_maxlik}
are somewhat incorrect---different cells are not completely
independent, for example---and so the estimator is not optimal;
but it provides an attractively simple and reasonably good way
to correct for the complexities of our selection bias.

In the notation of equation~\ref{eq:lsest}, estimator~\ref{eq:barxi_maxlik}
would be written $D_cD_g/D_cR_g - 1$.  The point of this section
is not to advocate this particular estimator---in our more careful calculations
the superior estimator of equation~\ref{eq:lsest} (Landy \& Szalay 1993)
was adopted instead---but to justify our practice of 
estimating $\bar\xi$ by summing the number of observed and expected pairs
in our full sample and then dividing, rather than (e.g.) by
averaging together the values of $N_i/\mu_i - 1$ observed around
individual CIV systems or in individual fields.

\section{SPATIAL CLUSTERING OF LYMAN-BREAK GALAXIES}
\label{sec:apwp}
Because of peculiar velocities and redshift uncertainties, our
estimate of any 
Lyman-break galaxy's position along the line-of-sight
is imprecise.  This imprecision complicates our attempts to measure
the correlation function of galaxies at small separations.
With a redshift uncertainty of $\sigma_z\sim 0.0025$ ($\sim 1.8$ Mpc)
we cannot estimate the strength of the correlation function
at separations $r<1$ Mpc (say) by
counting the number of galaxies that lie within~1 Mpc
of each other and comparing to the expected number if galaxies
were uniformly distributed; the number of galaxies within~1 Mpc
of each other is something we do not know.

One way around this problem (e.g., Davis \& Peebles 1983) is to count not the
number of galaxies whose estimated redshifts place them within
a distance $r$ of each other, but instead the number of galaxies
with angular separation $r_\theta\pm dr_\theta$ and
redshift separation $|\Delta z|<r_z$.  If $r_z$ is significantly
larger than each galaxy's positional uncertainty $\Delta l$, 
the resulting function of $r_\theta$, $n(r_\theta,<r_z)$, will
be unaffected by the size of our redshift
errors.  $n(r_\theta,<r_z)$ can then be inverted to produce
an estimate of the correlation function $\xi(r)$ that is not
corrupted by redshift uncertainties.

Different schemes can be used to estimate $\xi(r)$
from $n(r_\theta,<r_z)$ (see, e.g., Davis \& Peebles 1983).
Here is our approach.  It is designed for the low signal-to-noise
regime where one can only hope to recover the gross features
of the correlation function.
The expected number of galaxy neighbors
with angular separation $r_\theta$ and radial separation
within $\pm r_z$ is
\begin{equation}
\langle n(r_\theta, <r_z)\rangle = \bar n \biggl(1 + \frac{1}{r_z}\int_0^{r_z}dl\,\xi(\sqrt{l^2+r_\theta^2})\biggr)
\label{eq:apnrtrz}
\end{equation}
where $\bar n$ is
the number of galaxies we would expect to observe along a similar line
randomly placed.
If the correlation function is a power law, $\xi(r)\equiv(r/r_0)^{-\gamma}$,
dull algebra shows that the expected excess number of
pairs is
\begin{equation}
\omega_p(r_\theta,<r_z)\equiv \frac{\langle n\rangle}{\bar n}-1 = \\
	\frac{r_o^\gamma r_\theta^{1-\gamma}}{2r_z} B\Bigl(1/2, (\gamma-1)/2\Bigr) I_x\Bigl(1/2, (\gamma-1)/2\Bigr)
\label{eq:apwp}
\end{equation}
where $B$ and $I_x$ are the beta function and incomplete beta function in
the convention of Press et al. (1992, \S 6.4) and 
$x\equiv r_z^2(r_z^2+r_\theta^2)^{-1}$.
We estimate the correlation function parameters $r_0$, $\gamma$
by fitting equation~\ref{eq:apwp} to observed number of galaxies with
similar redshifts that lie at angular separation $r_\theta$.

In principle any value of $r_z$ can be used, provided
$r_z\gg \sigma_z$, but in practice some
choices of $r_z$ are better than others.  If $r_z$ is too small
we will miss some correlated pairs. The incomplete beta function
in equation~\ref{eq:apwp} will correct for this only imperfectly.
If $r_z$ is too large our sample will include a needlessly large
number of uncorrelated pairs, and statistical fluctuations in the
number of uncorrelated pairs may obscure the number of correlated
pairs.  Our choice of $r_z$ was designed to fall between these
two extremes.  We took $r_z$ to be the greater of
$1000 {\rm km s}^{-1}(1+z)/H(z)$ and $7r_\theta$.  $1000 {\rm km s}^{-1}$
is several times larger than our redshift uncertainty, and so the
first lower limit helps ensure that redshift errors will not make
us fail to recognize correlated pairs.  The second limit ensures
that we will integrate far enough down the correlation function
to include at least 80\% of correlated pairs for $\gamma\simgt 1.6$.
The derived correlation function does not change significantly
if the $1000 {\rm km s}^{-1}$ is changed by a factor of 2 in either direction or
if we adopt (say) $15r_\theta$ rather than $7r_\theta$ as the lower
limit on $r_z$.

Figure~\ref{fig:apwp} shows the observed excess galaxy counts
as a function of angular separation $r_\theta$.  Overlaid
are the expected excess galaxy counts at each separation
for the best power-law fit to $\xi(r)$.  The data are reasonably
consistent with a power-law correlation function.
$\omega_p$ was estimated with
the Landy-Szalay (1993) estimator $(DD-2DR+RR)/RR$;
this nomenclature and method of generating random catalogs
are described in the main body of the text
below equation~\ref{eq:lsest}.  The only
change is that here
$DD$ (e.g.) is the observed number of galaxy pairs with angular
separation $r_\theta$ and redshift separation $|\Delta z|<r_z$,
rather than the number of pairs with separation $r_\theta$, $r_z$.
As described above, near equation~\ref{eq:lsest},
the angular locations of objects in the random and real catalogs
were the same. This eliminated any artificial clustering signal
due to angular variations in the fraction of Lyman-break galaxies
with measured redshifts, an inevitable result
of our sparse multislit spectroscopy, but also eliminated any
contribution from the true angular clustering of Lyman-break galaxies
to our estimate of $\xi(r)$.
As the angular correlation $\omega(\theta)$ of Lyman-break
galaxies is almost undetectably weak (Giavalisco \& Dickinson 2001), 
however, the resulting bias in $\xi(r)$ should be small.

The estimates $r_0=3.96\pm 0.29 h^{-1}$ comoving Mpc,
$\gamma=-1.55\pm 0.15$ follow from
fitting equation~\ref{eq:apwp} to the data by minimizing $\chi^2$
with Powell's direction set method (Press et al. 1992; \S 10.5).
The uncertainties were calculated by generating a large number
of fake realizations of $\omega_p$ by adding
to our estimated $\omega_p$ at each $r_\theta$ a Gaussian deviate
with standard deviation equal to the observed uncertainty,
then fitting these fake realizations of $\omega_p$ to equation~\ref{eq:apwp}
with Powell's method.  68.3\% of the fake realizations had best fit
parameter values in the range listed above.

We should emphasize that figure~\ref{fig:apwp} provides little support
for or against Porciani \& Giavalisco's (2002) claim that the correlation function
of Lyman-break galaxies becomes negative at very small separations.
Any small-scale anticorrelation would be washed out in the redshift direction
by our redshift uncertainties and by the integration of equation~\ref{eq:apnrtrz};
any anti-correlation in the angular positions of galaxies would be
missed in our analysis because of the way we generated our random catalogs.

In any case, the correlation function estimated here
agrees well with the clustering strength estimated from
a counts-in-cells analysis.  Using the statistic ${\cal S}$ of
Adelberger et al. (1998) on our current
full sample of Lyman-break galaxies, we estimate the relative variance of
galaxy number density in cubical cells of side-length $11.6h^{-1}$ Mpc
($\Omega_M=0.3$, $\Omega_\Lambda=0.7$) to be
$\sigma_{\rm gal}^2=0.8\pm 0.2$, which corresponds roughly to
a correlation length of $r_0=4.65\pm 0.75 h^{-1}$ comoving Mpc
for $\gamma=-1.55$ (see Adelberger et al. 1998).

\bigskip

\clearpage


\begin{references}

\reference{} Adelberger, K. L. \& Steidel, C. C. 2000, ApJ, 544, 218
\reference{} Adelberger, K. L., Steidel, C. C., Giavalisco, M., Dickinson, M.,
	Pettini, M., \& Kellogg, M. 1998, ApJ, 505, 18
\reference{} Adelberger, K. L. 2000, in Clustering at High Redshift, eds. A. Mazure, O. Le F\`evre,
	\& V. Le Brun, ASP, 200, 13
\reference{} Alcock, C. \& Paczynski, B. 1979, Nature, 281, 358
\reference{} Bajtlik, S., Duncan, R. C., \& Ostriker, J. P. 1988, ApJ, 327, 570
	304, 15
\reference{} Boksenberg, A., Sargent, W.L.W., \& Rauch, M. 1998, in The Birth of Galaxies, Proc. of the Xth Rencontres de Blois
		
\reference{} Cen, R. \& Bryan, G. L. 2001, ApJL, 546, 81


\reference{} Cole, S., Arag\'on-Salamanca, A., Frenk, C. S., Navarro, J. F.,
	\& Zepf, S. E. 1994, MNRAS, 271, 781
\reference{} Cowie, L. L., Songaila, A., Kim, T.-S., \& Hu, E. M. 1995, AJ, 109, 1522
	
\reference{} Croft, R.A.C., Hernquist, L., Springel, V., Westover, M., \& White, M., ApJ, in press
\reference{} Dav\'e, R., Hellsten, U., Hernquist, L., Katz, N., \& Weinberg, D.H. 1998, 
	ApJ, 509, 661
\reference{} Davis, M. \& Peebles, P. J. E. 1983, ApJ, 267, 465
\reference{} Dekel, A. \& Silk, J. 1986, ApJ, 303, 39
\reference{} Dekker, H., D'Odorico, S., Kaufer, A., Delabre, B., \& Kotzlowski, H. 2000,
	SPIE, 4008, 534
\reference{} Ellison, S.L., Songaila, A., Schaye, J., \& Pettini, M. 2000,
	AJ, 120, 1175
\reference{} Ellison, S.L., Pettini, M., Steidel, C.C., \& Shapley, A.E. 2001, ApJ, 549, 770
	
\reference{} Fynbo, J.U., M\/oller, P., \& Warren, S.J. 1999, MNRAS, 305, 849
\reference{} Gawiser, E., Wolfe, A.M., Prochaska, J.X., Lanzetta, K.M., Yahata, N., \&
	Quirrenbach, A. 2001, ApJ, 562, 628
\reference{} Giavalisco, M. \& Dickinson, M. 2001, ApJ, 550, 177
	Quinn, T., \& Stadel, J. 1998, Nature, 392, 359
\reference{} Haehnelt, M.G., Steinmetz, M., \& Rauch, M. 2000, ApJ, 534, 594
\reference{} Heap, S. R., Williger, G. M., Smette, A., Hubeny, I., Sahu, M. S., Jenkins, E. B.,
	Tripp, T. M., \& Winkler, J. N. 2000, ApJ, 534, 69
\reference{} Heckman, T. M., Lehnert, M. D., Strickland, D. K. \& Armus, L. 2000, ApJS,
	129, 493


\reference{} Hogan, C.J., Anderson, S.F., \& Rugers, M.H. 1997, AJ, 113, 1495

\reference{} Kaiser, N. 1984, ApJL, 284, 9
\reference{} Kaiser, N. 1991, ApJ, 383, 104
	1999, MNRAS, 303, 188
\reference{} Kells, W., Dressler, A., Sivaramakrishnan, A., Carr, D., Koch, E.,
	Epps, H., Hilyard, D., \& Pardeilhan, G. 1998, PASP, 110, 1489
\reference{} Klein, R.I., McKee, C.F., \& Colella, P. 1994, ApJ, 420, 213
\reference{} Landy \& Szalay 1993, ApJ, 412, 64


\reference{} Madau, P., Haardt, F., \& Rees, M. J. 1999, ApJ, 514, 648
\reference{} Madau, P., Ferrara, A., \& Rees, M. J. 2001, ApJ, 555, 92
\reference{} Mathews, W. G. \& Baker, J. C. 1971, ApJ, 170, 241

\reference{} McDonald, P., Miralda-Escud\'e, J., Rauch, M., Sargent, W. L. W., Barlow, T. A.,
	Cen, R., Ostriker, J. P. 2000, ApJ, 543, 1
\reference{} McLean, I.S. et al. 1998, Proc. SPIE, 3354, 566
\reference{} Miralda-Escud\'e, J., Haehnelt, M., \& Rees, M. J. 2000,
	ApJ, 530, 1
\reference{} Mo, H.J., Mao, S., \& White, S.D.M. 1999, MNRAS, 304, 175
\reference{} Murdoch, H.S., Hunstead, R.W., Pettini, M., \& Blades, J.C. 1986, ApJ, 309, 19 


\reference{} Mushotzky, R. F. \& Loewenstein, M. 1997, ApJL, 481, 63 
\reference{} Norman, C.A., Bowen, D.V., Heckman, T., Blades, C., \& Danly, L. 1996, ApJ, 472, 73
\reference{} Oke, J. B. et al. 1995, PASP, 107, 3750
\reference{} Ostriker, J. P. \& Cowie, L. L., 1981, ApJL, 243, 127
\reference{} Peebles, P.J.E. 1980, ``The Large-Scale Structure of the Universe'',
	(Princeton: Princeton University Press)
\reference{} Pen, U.-L. 1999, ApJL, 510, 1
\reference{} Pettini, M., Shapley, A. E., Steidel, C. C., Cuby, J.-G., Dickinson, M.,
	Moorwood, A. F. M., Adelberger, K. L., \& Giavalisco, M. 2001, ApJ, 554, 981

\reference{} Pettini, M., Rix, S.A., Steidel, C.C., Adelberger, K.L., Hunt, M.P., \& Shapley, A.E.
	2002, ApJ, 569, 742
\reference{} Ponman, T.J., Cannon, D.B., \& Navarro, J.F. 1999, Nature, 397, 135
\reference{} Porciani, M. \& Giavalisco, M. 2002, ApJ, 565, 24
\reference{} Press, W. H., Flannery, B. P., Teukolsky, S. A., \& Vetterling, W.
	T. 1992, ``Numerical Recipes in C'', (Cambridge:  Cambridge 
	University Press)
\reference{} Quashnock, J.M. \& Vanden Berk, D.E. 1998, ApJ, 500, 28
	Weinberg, D. H., Hernquist, L., Katz, N., Cen, R., \& Ostriker, J. P.
	1997, ApJ, 489, 7
\reference{} Sargent, W.L.W., Steidel, C.C., \& Boksenberg, A. 1988, ApJS, 68, 539
\reference{} Scott, J., Bechtold, J., Dobrzycki, A. \& Kulkarni, V., 2000, ApJS, 130, 67
\reference{} Shapley, A.E., Steidel, C.C., Adelberger, K.L., Dickinson, M., Giavalisco, M., \& Pettini, M. 2001, ApJ, 562, 95
\reference{} Sheinis, A.I., Miller, J.S., Bolte, M., \& Sutin, B.M. 2000, SPIE, 4008, 522
\reference{} Songaila, A. 1998, AJ, 115, 2184
\reference{} Songaila, A. 2001, ApJL, 561, 153
\reference{} Springel, V. \& Hernquist, L. 2002, MNRAS, in press
\reference{} Steidel, C. C., Giavalisco, M., Pettini, M., Dickinson, M., \&
	Adelberger, K. L. 1996, ApJL, 462, 17
\reference{} Steidel, C. C., Pettini, M., \& Adelberger, K. L., 2001, ApJ, 546, 665
\reference{} Steidel, C. C., Adelberger, K. L., Giavalisco, M., Dickinson, M.,
	\& Pettini, M. 1999, ApJ, 519, 1
\reference{} Steidel, C. C., Adelberger, K. L., Dickinson, M., Giavalisco, M., Pettini, M.,
	\& Kellogg, M. 1998, ApJ, 492, 428
\reference{} Tenorio-Tagle, G., Silich, S. A., Kunth, D., Terlevich, E.,
	\& Terlevich, R. 1999, MNRAS, 309, 332
\reference{} Theuns, T., Mo, H.-J., \& Schaye, J. 2001, MNRAS, 321, 450

\reference{} Vogt, S.S. et al. 1994, SPIE, 2198, 362
\reference{} Weil, M. L., Eke, V. R., \& Efstathiou, G. 1998, MNRAS, 300, 773
\reference{} Weinberg, D. H., Miralda-Escud\'e, J., Hernquist, L., \& Katz, N. 1997, ApJ, 490, 564


\reference{} Weymann, R. J., Carswell, R. F., \& Smith, M. G. 1981, ARAA, 19, 41
\reference{} White, M., Hernquist, L., \& Springel, V. 2001, ApJL, 550, 129
\reference{} White, S.D.M. \& Rees, M.J. 1978, MNRAS, 183, 341
\reference{} de Young, D. S. 1978, ApJ, 223, 47
	
\end{references}
\end{document}